\documentclass[prl,twocolumn,floatfix]{revtex4-1}

\usepackage{graphicx}
\usepackage{amssymb,amsmath}
\usepackage[colorlinks,citecolor=blue,urlcolor=blue]{hyperref}
\usepackage{dcolumn}

\begin{document}

\title{Emergent Coherent Lattice Behavior in Kondo Nanosystems}

\author{Marcin Raczkowski}
\author{Fakher F. Assaad}
\affiliation{Institut f\"ur Theoretische Physik und Astrophysik,
             Universit\"at W\"urzburg, Am Hubland, D-97074 W\"urzburg, Germany}

\date{\today}

\begin{abstract} 
How many magnetic moments periodically arranged on a metallic surface are needed to generate
a coherent Kondo lattice behavior? We investigate this fundamental issue within the particle-hole 
symmetric Kondo lattice model using quantum Monte Carlo simulations. Extra magnetic atoms 
forming closed shells around the initial impurity induce a fast splitting of the Kondo resonance 
at the inner shells which signals the formation of composite heavy-fermion bands. 
The onset of the hybridization gap matches well the enhancement of  antiferromagnetic spin correlations 
in the plane perpendicular to the applied magnetic field, a genuine feature of the coherent Kondo 
lattice. In contrast, the outermost shell remains dominated by a local Kondo physics with spectral 
features resembling the single-impurity behavior.

\end{abstract}

\maketitle


In the realm of condensed matter physics, the theory of wave propagation in periodic structures---the Bloch 
theorem~\cite{Bloch29}---forms the basis for calculating  the electronic band structure of solids. 
It is also a cornerstone for understanding low-temperature Fermi-liquid-like $\propto T^2$  transport properties 
of heavy-fermion  (HF) metals which arise from a coherent Kondo screening of periodically arranged $f$-shell moments 
by conduction electrons~\cite{Wirth16}. The quantum entanglement between localized $f$-moments 
and mobile conduction electrons leads to the enlargement of the Fermi surface, which determines, below the 
coherence scale $T_{\textrm{coh}}$, transport and thermodynamic properties of HF liquids.   
A coherent Kondo lattice behavior differs markedly from the single-impurity Kondo physics with a logarithmic 
increase, below the Kondo temperature $T_K$, of spin-flip scattering for conduction electrons~\cite{Hewson_book}. 
Elucidating the crossover between both regimes~\cite{Costi88,Schlottmann92,Miranda97,Affleck00,Riseboro03,Vojta07,Pruschke08,Ogata10,Kumar14,Moreno15,Wei17} 
as well as identifying the energy scale $T_{\textrm{coh}}$   associated with the formation of the coherent HF state is a long-standing 
issue~\cite{Martin82,Lacroix86,Tesa86,Continentino89,Jarrell98,Pruschke00,Burdin00,Costi02,Assaad04,Fisk08,Beach08,Raczkowski10,Tanaskovic11}.

Starting from an incoherent dilute limit of Kondo impurities, one possibility to study 
how the lattice effects come into play is to increase the concentration of magnetic 
ions~\cite{Steglich84,Onuki85,Sumiyama86,Lin87,Budko98,Fisk02,Pikul12,Budko15}.
However, the electronic structure of bulk HF materials is essentially three dimensional, and one may wonder 
how the coherence phenomena are affected either by reduced dimensionality or in spatially restricted geometries 
where both quantum fluctuations and correlations effect are enhanced~\cite{Groten01,Burdin18}. In this respect, the 
experimental realization of artificial $f$-electron superlattices~\cite{superlatt16} has opened a new avenue 
to investigate the onset of coherence in a two-dimensional (2D) regime followed up by theoretical 
studies~\cite{Tada13}. 

In recent years, new insight into the Kondo physics at the nanoscale has come from scanning tunneling microscopy (STM)~\cite{Ternes09,Morr17}. 
It allows one to probe Kondo screening at a single magnetic adatom~\cite{Madhavan98,Li98}, to study the composite nature of the HF quasiparticles 
in a lattice situation~\cite{Schmidt10,Wirth11,Hamidian11,Yazdani12}, and also to image a mutual RKKY interaction between magnetic 
impurities mediated by conduction electrons  as a function of the interatomic distances~\cite{Wiebe10}.
The possibility for a systematic and controlled study of the competition between different energy scales 
in Kondo nanostructures resulted in a resurgence of interest in the interplay between the Kondo effect and magnetic RKKY 
correlations in adatom dimers~\cite{Pruser14,Spinelli15,Carlos15,Bulla15}, trimers~\cite{Savkin05}, and multiple 
impurities~\cite{DiLullo12,Chudzinski12,Mitchell15}.


Given the distinct difference between a single-impurity Kondo physics and the coherent Kondo lattice behavior, 
it is natural to ask how many magnetic moments periodically arranged on a metallic surface 
are needed to resolve a crossover between both regimes? In this Letter, we address this question on the basis 
of the Kondo lattice model (KLM) at half-filling~\cite{Ueda97},
\begin{equation}
H_{\textrm{KLM}} = -t\sum_{\langle\pmb{i},\pmb{j}\rangle, \sigma } 
                  c^{\dagger}_{\pmb{i},\sigma} c^{}_{\pmb{j},\sigma} +
           \sum_{\pmb{i}  }  J_{\pmb{i}} \pmb{S}^{c}_{\pmb{i}} \cdot \pmb{S}^{f}_{\pmb{i}}, 
\label{KLM}
\end{equation}
where $\pmb{S}^{c}_{\pmb{i}} = \tfrac{1}{2} \sum_{\sigma,\sigma'}  c^{\dagger}_{\pmb{i},\sigma}
\pmb{\sigma}^{}_{\sigma,\sigma'} c^{}_{\pmb{i},\sigma'}$ are spin operators of conduction electrons and
$\pmb{S}^{f}_{\pmb{i}} = \tfrac{1}{2} \sum_{\sigma,\sigma'}  f^{\dagger}_{\pmb{i},\sigma}
\pmb{\sigma}^{}_{\sigma,\sigma'} f^{}_{\pmb{i},\sigma'}$ are localized spins with  $\pmb{\sigma} $ being the Pauli matrices.
The Hamiltonian (\ref{KLM}) describes localized  spin 1/2 magnetic moments coupled via site-dependent 
antiferromagnetic (AF)  exchange interaction $J_{\pmb i}$ to conduction electrons on a square lattice. 
By switching on and off the values of $J_{\pmb{i}}$,  we can investigate a crossover from the zero-dimensional Kondo effect 
to the 2D particle-hole symmetric Kondo lattice, where the coherence of individual Kondo screening clouds 
opens up a hybridization gap at the Fermi level and gives rise to the Kondo insulating phase.    
We study the most challenging regime $J/t=1.6$  close to the quantum critical point of the 2D KLM~\cite{Assaad99}, 
where the RKKY interaction and Kondo screening are of the same order of magnitude, and thus it is crucial to treat them on equal footing. 
To this end, we use a finite-temperature version of the auxiliary-field quantum Monte Carlo (QMC) algorithm, as implemented 
in Ref.~\cite{alf}.

\begin{figure}[t!]
\begin{center}
\includegraphics[width=0.45\textwidth]{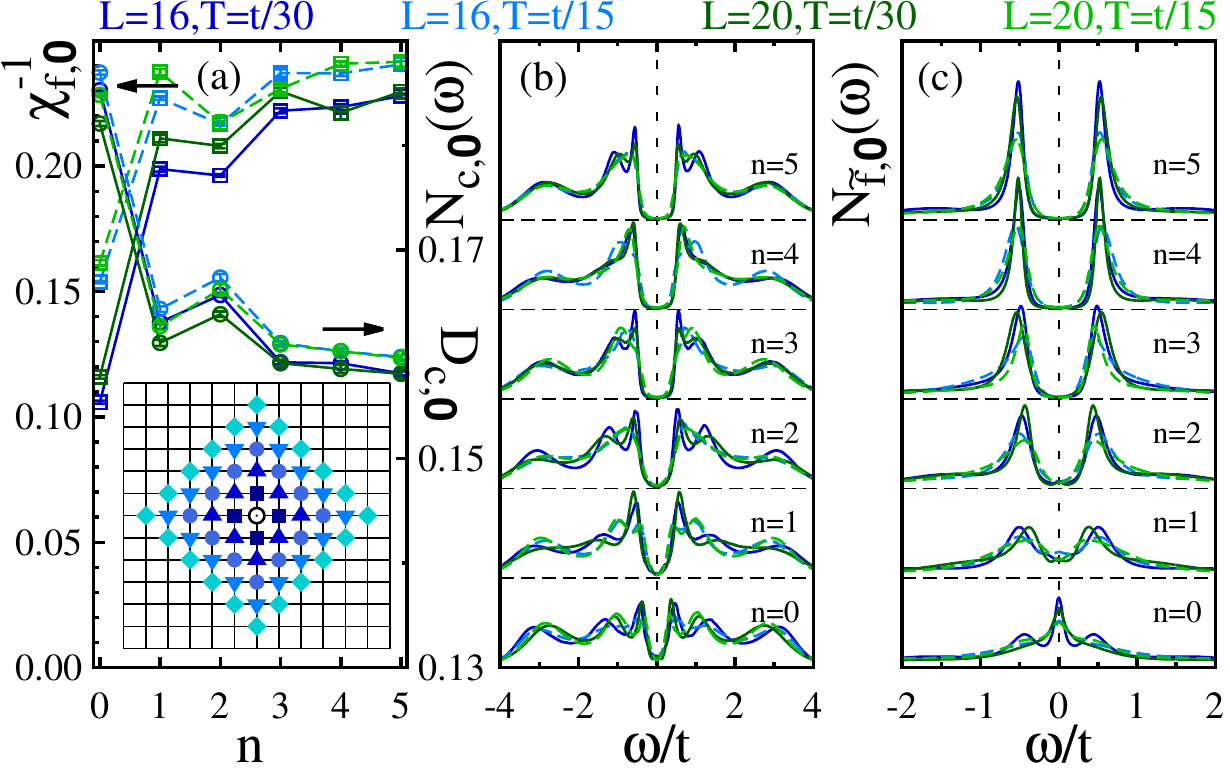}
\end{center} 
\caption
{
Inverse $f$-spin susceptibility $\chi_{f,\pmb{0}}^{-1}$ and $c$-electron double occupancy $D_{c,\pmb{0}}$ (a),  
$c$-electron LDOS $N_{c,\pmb{0}}(\omega)$ (b), and $\tilde{f}$-operator LDOS $N_{\tilde{f},\pmb{0}}(\omega)$ (c) 
at the central site $\pmb{r}=\pmb{0}$ of a Kondo superlattice (see inset) upon increasing the number 
of impurity shells $n$ coupled to a $L\times L$ lattice of $c$-electrons at temperature $T$ much below
the corresponding single-impurity Kondo scale $T_K\simeq t/8$~\cite{Assaad04}.
Inset: construction of the superlattice; extra impurity shells around the initial impurity 
(open circle) are indicated using different symbols.  
}
\label{local}
\end{figure}

We consider the situation depicted in the inset of Fig.~\ref{local}(a):
starting with a single magnetic impurity deposited in the middle part of the metallic surface, in consecutive steps 
we add  closed shells of magnetic moments around the initial impurity so as to keep the coordination number fixed. 
This line of research has explicit experimental relevance: atomically precise engineering 
with STM was used to study the evolution of local density of states (LDOS) from an isolated iron(II) 
phthalocyanine molecule to the 2D superlattice on a Au(111) surface~\cite{Tsukahara11}. 
Although the underlying physics is  complicated by the SU(4) Kondo effect~\cite{Lobos14}, Ref.~\cite{Tsukahara11} 
provides a novel route to interpolate between the physics of a single Kondo impurity and the 2D Kondo lattice behavior.


First, we focus on local properties at the central impurity $\pmb{r}=\pmb{0}$: $f$-spin susceptibility
$\chi_{f,\pmb{r}}=\int_0^{\beta}d\tau\langle S^z_{f,\pmb{r}}(\tau) S^z_{f,\pmb{r}}(0) \rangle $ and  
$c$-electron double occupancy $D_{c,\pmb{r}}=\langle n^{c}_{\pmb{r},\uparrow} n^{c}_{\pmb{r},\downarrow} \rangle $ 
with $n^{c}_{\pmb{r},\sigma}= c^{\dagger}_{\pmb{i},\sigma} c^{}_{\pmb{i},\sigma}$, and discuss their evolution 
upon increasing number of impurity shells $n$. 
In a single-impurity problem,  the susceptibility follows the Curie-Weiss law  $\chi_{f}\propto 1/(T+\Theta)$. 
As shown in Fig.~\ref{local}(a), $\chi_{f,\pmb{0}}^{-1}$  deviates rapidly from its single-impurity limit, 
shows an oscillating behavior, and finally starts to saturate in larger systems. Likewise, $D_{c,\pmb{0}}$ shows 
a strong initial reduction and displays noticeable fluctuations before reaching saturation.  
The behavior of both $\chi_{f,\pmb{0}}^{-1}$ and $D_{c,\pmb{0}}$ suggests a fast onset of lattice effects.

The corresponding evolution of $c$-electron LDOS $N_{c,\pmb{0}}(\omega)$ obtained from the imaginary-time-displaced Green's function
$G_{c,\pmb{0}}(\tau) = \sum_{\sigma}\langle c^{}_{\pmb{0},\sigma}(\tau) c^{\dagger}_{\pmb{0},\sigma}(0) \rangle$ 
by means of the stochastic analytic continuation method~\cite{Beach04a} is shown in Fig.~\ref{local}(b).
While a precise form of the spectra is difficult to assess  due to a complicated lineshape, 
the data reproducibly show that the narrow $\sim T_K$ dip at $\omega=0$ inherent to the single-impurity problem broadens 
and evolves into a full gap already in the $n=3$ system. A very natural account for the observed broadening is 
the RKKY interaction which is responsible for a broad quasiparticle gap $\sim J$ in the 2D KLM~\cite{Assaad99,Capponi01}.
Thus, our model QMC calculations rationalize the interpretation of spectroscopic characteristics in Ref.~\cite{Tsukahara11} 
in terms of the RKKY coupling between molecular spins in the central part of the superlattice.

The KLM forbids charge fluctuations on the $f$-orbitals. Instead, we examine a local spectral function 
$N_{\tilde{f},\pmb{0}}(\omega)$  extracted from the Green's function 
$G_{\tilde{f},\pmb{0}}(\tau) = \sum_{\sigma}\langle \tilde{f}^{}_{\pmb{0},\sigma}(\tau) \tilde{f}^{\dagger}_{\pmb{0},\sigma}(0) \rangle$,  
where the $\tilde{f}$-operator is defined as 
%
$ \tilde{f}^{\dagger}_{\pmb{0},\sigma'}  =  \sum_{\sigma}  \biglb(   
         c^{\dagger}_{\pmb{0},\sigma} f^{}_{\pmb{0},\sigma} f^{\dagger}_{\pmb{0},\sigma'}   +    
         f^{\dagger}_{\pmb{0},\sigma'}c^{\dagger}_{\pmb{0},\sigma} f^{}_{\pmb{0},\sigma} \bigrb) $.
%
The $\tilde{f}$-operator is derived from a single-impurity Anderson model (SIAM) using the Schrieffer-Wolff transformation~\cite{SW_transf,SM}  
and describes the so-called cotunelling process~\cite{Dzero09,Morr10,Wolfle10}: the tunelling of an electron from the STM tip  
into the conduction sea involves a spin-flip of the local magnetic $f$-moments. As is apparent, $N_{\tilde{f},\pmb{0}}(\omega)$ 
reproduces  the Abrikosov-Suhl resonance of the SIAM  in the Kondo model with a single impurity, see Fig.~\ref{local}(c). 
Upon increasing $n$, the resonance splits first, then is replaced by a pseudogap, and finally a full gap opens up at $n=3$, 
coinciding with the gap that appears in $N_{c,\pmb{0}}(\omega)$. 
While the splitting of the Abrikosov-Suhl resonance is already observed in the interimpurity spin-singlet state of the two-impurity 
Anderson model with a direct Heisenberg interaction between the $f$-spins~\cite{Zhu11,Grewe12,Wang15}, 
here we assume the absence of a direct overlap between the $f$-orbitals. Thus, the opening of the gap seems to signify  the onset of 
Kondo coherence. Moreover,  given that upon further increasing $n$ peaks on the flanks of the gap sharpen  and become 
reminiscent of those found in the lattice limit~\cite{Vekic95,Pruschke00,Logan03}, it is tempting to assume that they stem from 
nearly flat HF bands with predominantly $f$-character.

\begin{figure}[t!]
\begin{center}
\includegraphics[width=0.15\textwidth]{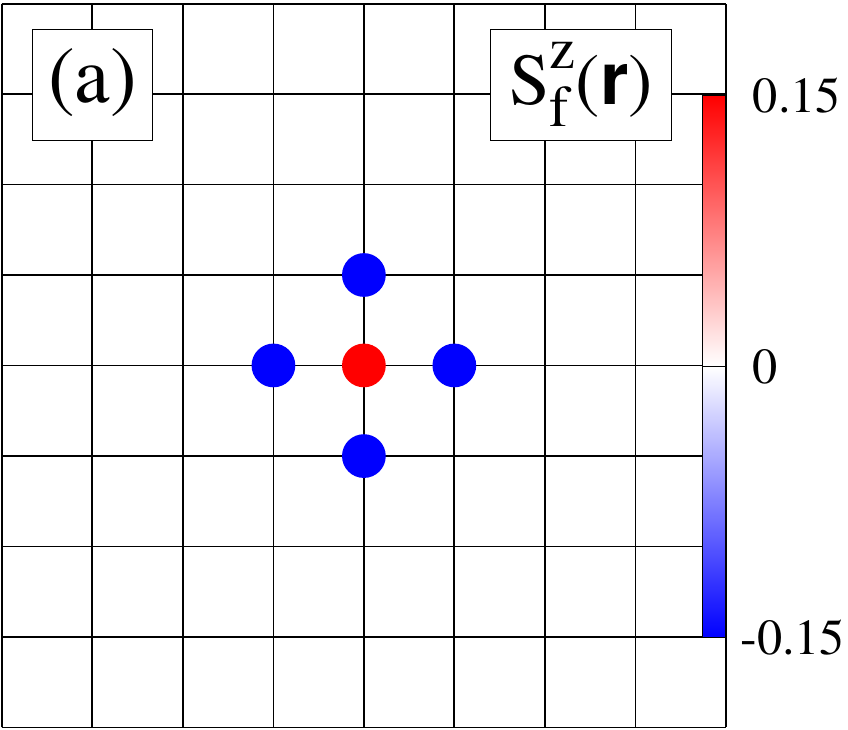}
\includegraphics[width=0.15\textwidth]{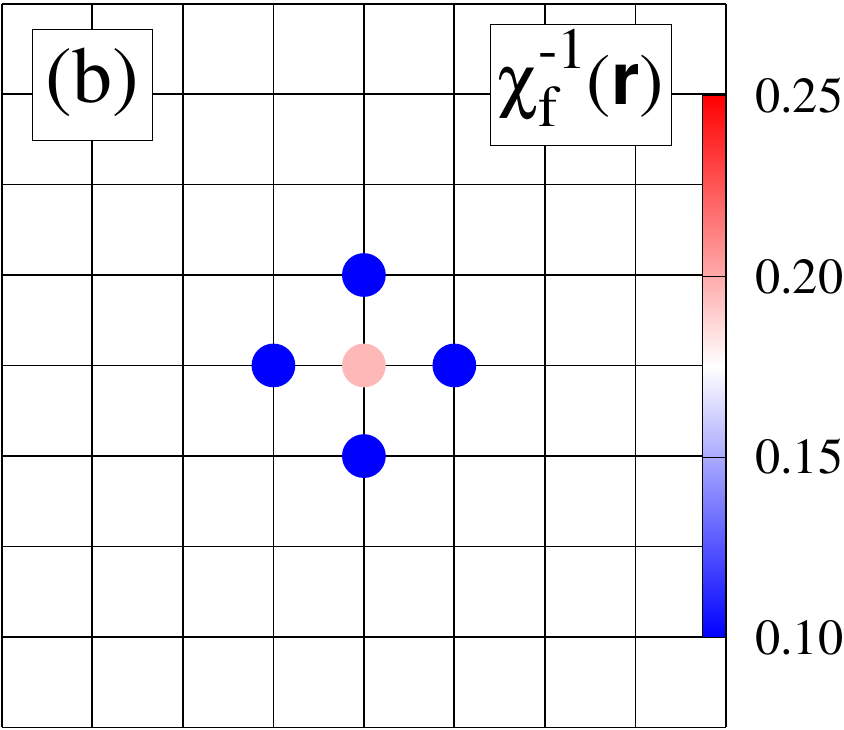}
\includegraphics[width=0.15\textwidth]{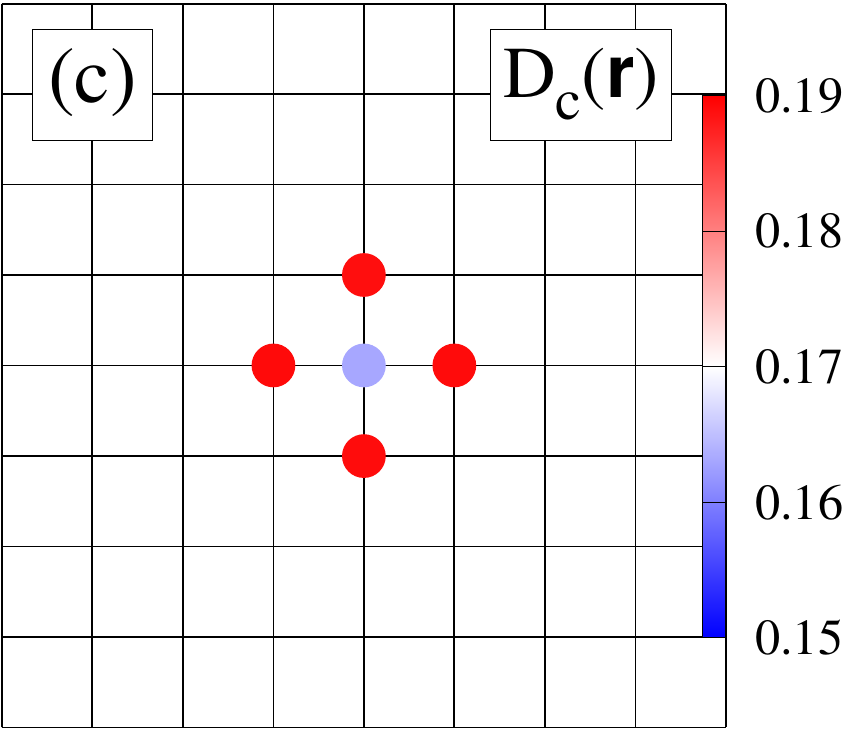}\\
\vspace*{0.1cm}
\includegraphics[width=0.15\textwidth]{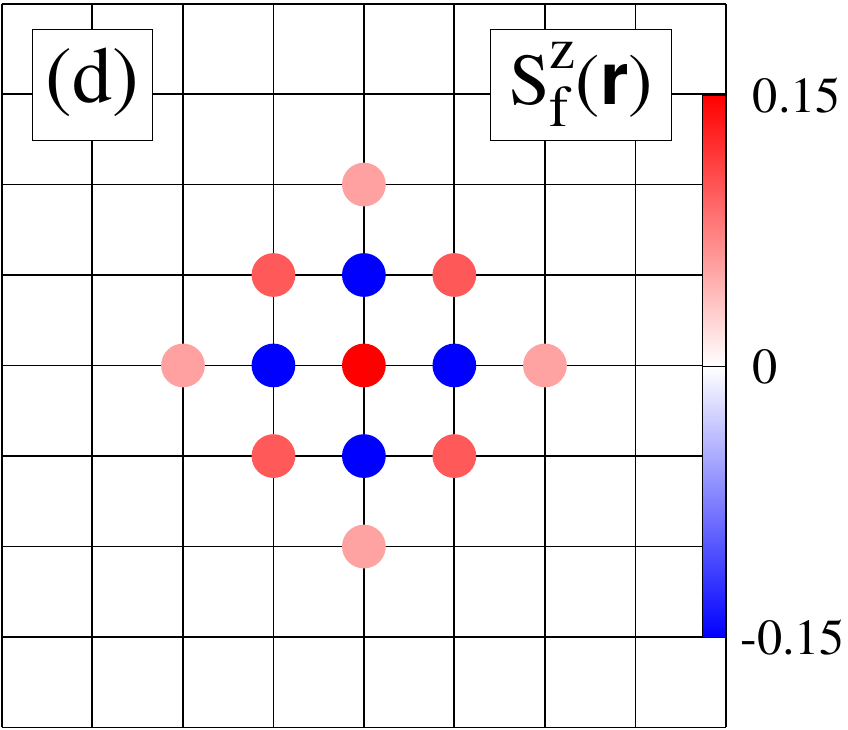}
\includegraphics[width=0.15\textwidth]{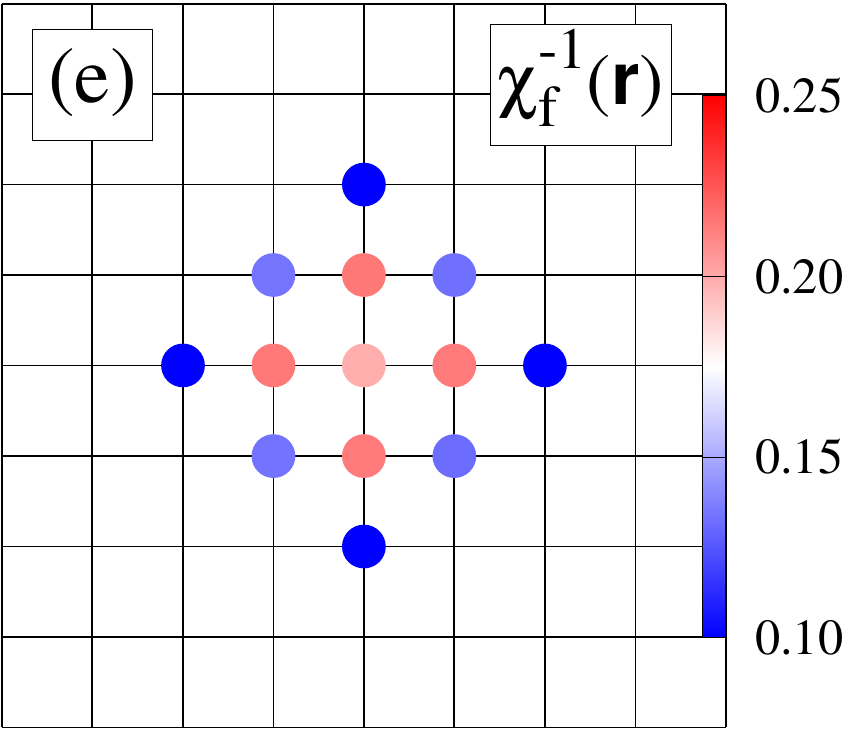}
\includegraphics[width=0.15\textwidth]{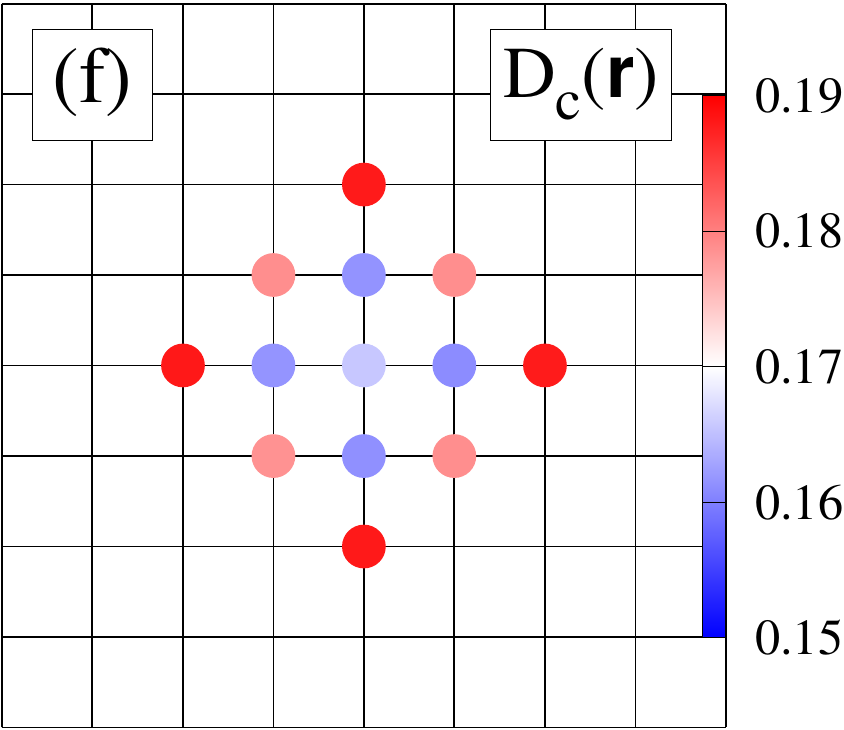}\\
\vspace*{0.1cm}
\includegraphics[width=0.15\textwidth]{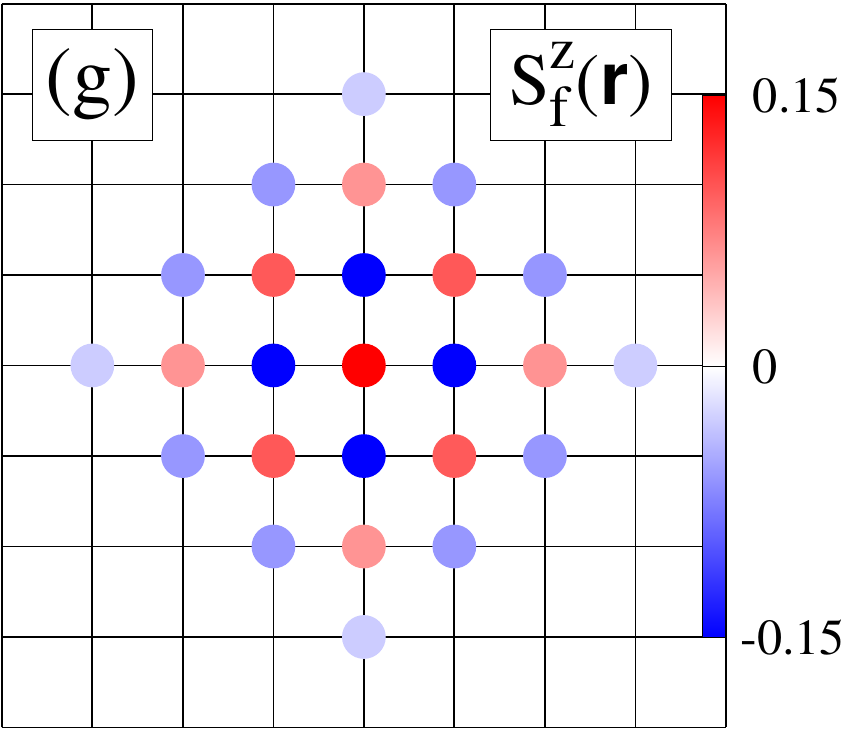}
\includegraphics[width=0.15\textwidth]{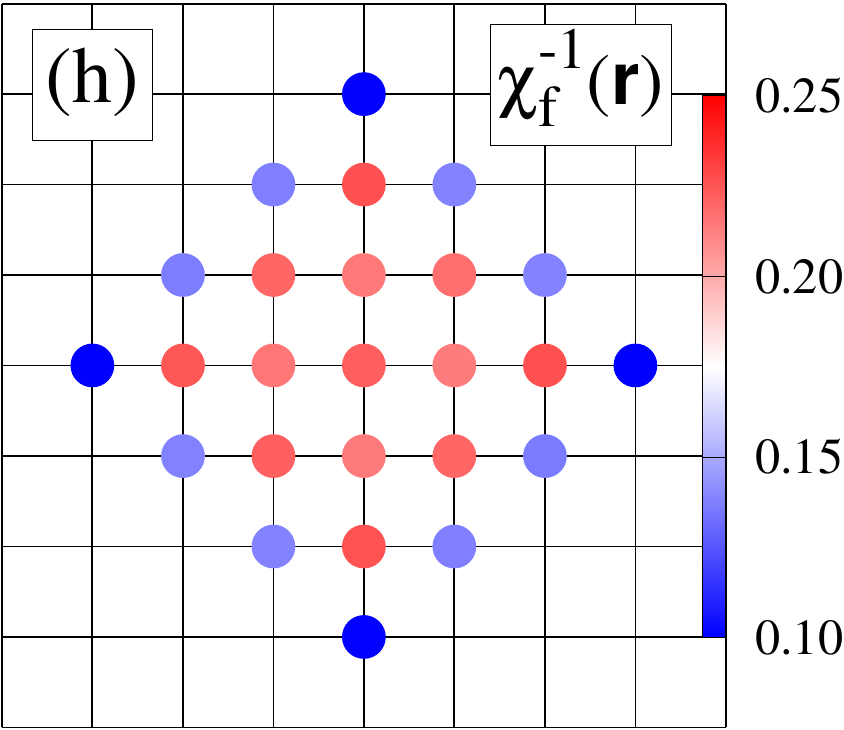}
\includegraphics[width=0.15\textwidth]{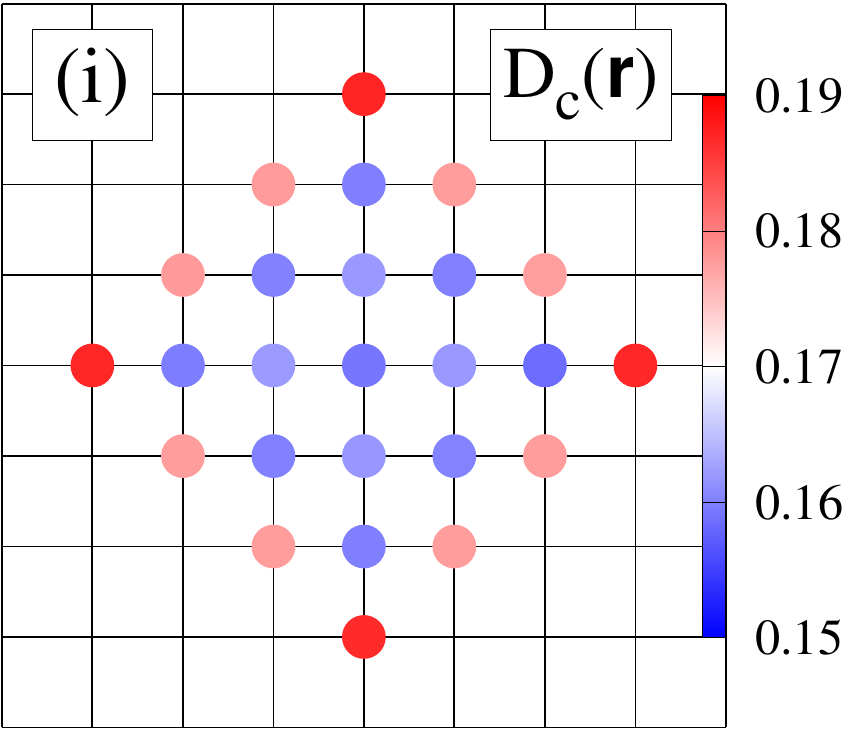}
\end{center}
\caption
{
Real-space spin correlations   $S_f^{z}(\pmb{r})$  relative to the central $\pmb r=(0,0)$  impurity (left) 
and spatial dependence of: inverse $f$-spin susceptibility $\chi_f^{-1}(\pmb{r})$ (middle) 
and $c$-electron double occupancy $D_c(\pmb{r})$ (right) from QMC simulations of the $n=1$ (a-c), 
$n=2$ (d-f), and $n=3$ (g-i) systems. Parameters: $L=16$ and $T=t/30$.
}
\label{Sz}
\end{figure}

Next, we consider the spatial properties of superlattices, summarized in Fig.~\ref{Sz}. 
The build up of intersite  AF  spin  correlations $S^z_f(\pmb{r})   = 4\langle S^z_{f,\pmb{r}} S^z_{f,\pmb{0}}\rangle$ 
measured relative to the central impurity gives rise to a collective-like screening of the $f$-impurities, seen as the enhancement 
of $\chi_f^{-1}(\pmb{r})$ at inner shells. 
It also assists the localization of $c$-electrons reflected in the reduction of $D_c(\pmb{r})$ in the center. 
In contrast, the outermost shell, and in particular the corner sites, clearly stands out. It stems from two effects: 
(i) a reduced number of nearest-neighbor impurities at the edge makes the RKKY interaction less important and 
(ii) since the edge is immersed in a conduction electron sea, a locally enhanced density of $c$-electron states 
$\rho_{c,\pmb{r}}$ available for Kondo quenching of the edge impurities introduces a site-dependent Kondo temperature 
$T_K(\pmb{r})\sim e^{-1/J\rho_{c,\pmb{r}}(\omega=0)}$. Furthermore, while the spatial characteristics of the $n=3$ 
and larger superlattices are very much alike, cf. Figs.~\ref{Sz}(g-i) and \ref{Sz2} in Ref.~\cite{SM}, 
one can discern a nonmonotonic evolution of both $\chi_f^{-1}(\pmb{r})$  and $D_c(\pmb{r})$ when moving from the center 
towards the corner atom of the $n=2$ system. Together with the initial oscillating behavior of both quantities 
upon increasing $n$ in Fig.~\ref{local}(a), it is indicative of a strong competition between the local Kondo physics at the edges 
and lattice effects in the center.

\begin{figure}[t!]
\begin{center}
\includegraphics[width=0.45\textwidth]{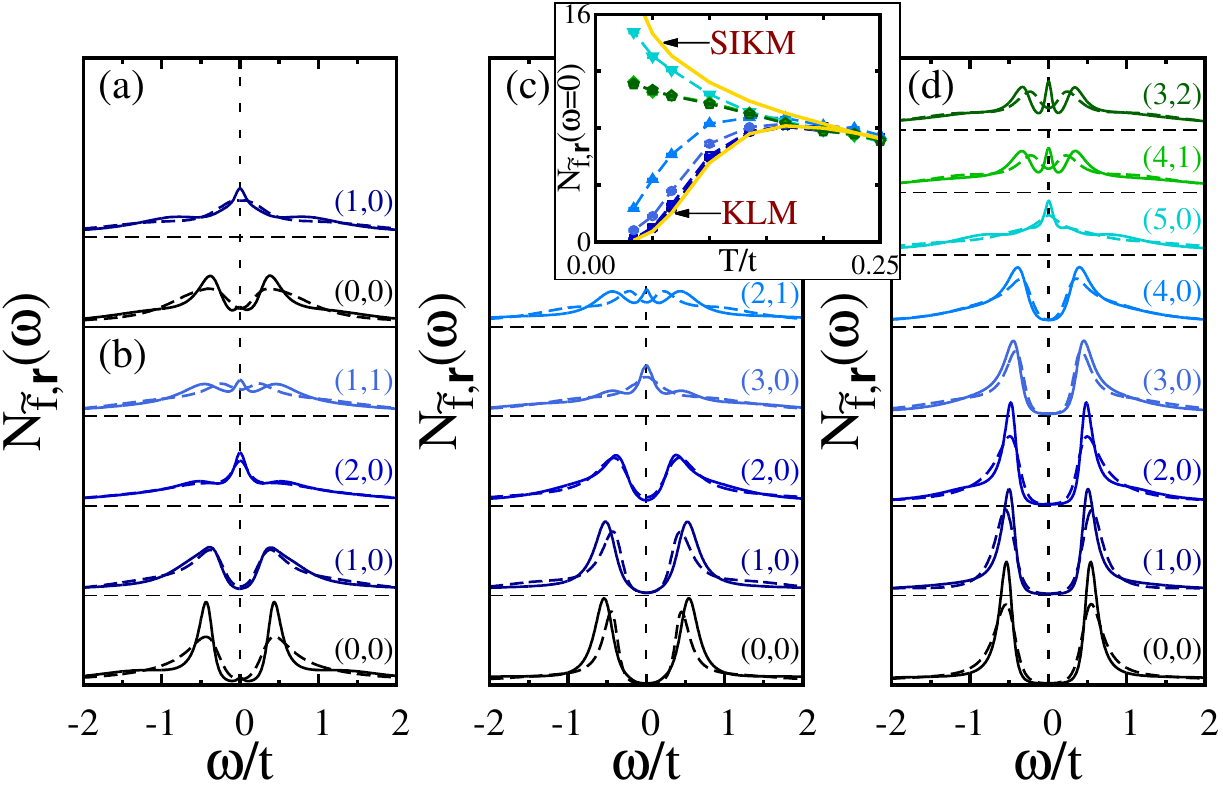}
\end{center}
\caption
{Spatial variation (from bottom to top) of the $\tilde{f}$-operator LDOS $N_{\tilde{f},\pmb{r}}(\omega)$ along 
the (0,0) $\to$ $(n,0)$ (corner) path  as well as at the corner-adjacent $(n-1,1)$ site of the outermost 
shell in the system with $n=1$ (a), $n=2$ (b), $n=3$ (c), and $n=5$ (d) shells. In (d) we also plot 
$N_{\tilde{f},\pmb{r}}(\omega)$ at the second inequivalent (3,2) site of the outermost shell. 
Parameters: $L=20$; $T=t/30$ (solid) and $T=t/15$ (dashed). Inset: $T$-dependence of $N_{\tilde{f},\pmb{r}}(\omega=0)$ 
in the $n=5$ case with color-coding of sites as in panel (d); solid lines show two limiting cases: 
2D KLM and the single-impurity Kondo model (SIKM).
}
\label{spatial_nf}
\end{figure}

Given the onset of a Kondo insulating phase in the core of systems with $n\ge 3$,  one would like to know what 
happens at the surface of the insulator immersed in a conduction electron sea, and in particular, whether and to 
what extent the insulator is penetrable to those electrons?
To get more insight into this issue, we show in Fig.~\ref{spatial_nf} how the $\tilde{f}$-operator LDOS 
$N_{\tilde{f},\pmb{r}}(\omega)$ evolves when moving from the central (0,0) impurity to the corner $(n,0)$ site 
for superlattices with different number of shells $n$.  As is apparent, independently of whether the gap at the 
central site is full or partial, the corner always develops the Kondo peak, followed by 
the other sites of the outermost shell upon lowering $T$, see Fig.~\ref{spatial_nf}.

To assess that the emergence of low-$T$ Kondo peaks at the edge is not an artifact of analytical 
continuation, we focus on the largest $n=5$ nanosystem and look at the $T$-dependence of 
$N_{\tilde{f},\pmb{r}}(\omega=0)$, see the inset in Fig.~\ref{spatial_nf}. 
It can be directly extracted from the Green's function $G_{\tilde{f},\pmb{r}}(\tau)$  by using  
$N_{\tilde{f},\pmb{r}}(\omega=0)\propto \lim\limits_{\beta\to\infty}\beta G_{\tilde{f},\pmb{r}}(\tau=\beta/2)$.
As expected, all the site-dependent curves are bounded from below (above) by those corresponding 
to the 2D KLM [single-impurity Kondo model (SIKM)], respectively.  The $T$-dependence of $N_{\tilde{f},\pmb{r}}(\omega=0)$ at 
the central impurity and at the two innermost $n\in\{1,2\}$ shells follows essentially that of the 2D KLM. 
Moreover, the $n\in\{3,4\}$ shells display a similar $T$-dependence indicative of the opening of full gap in the 
$T=0$ limit. In contrast, $N_{\tilde{f},\pmb{r}}(\omega=0)$  at the corner (5,0) site follows closely the    
SIKM behavior; similarly, $N_{\tilde{f},\pmb{r}}(\omega=0)$ at the other (4,1) and (3,2)  sites of the edge 
grows steadily with reducing $T$ lending further support for the emergent peaks at our lowest $T=t/30$, see Fig.~\ref{spatial_nf}(d).  
They signal the penetration of conduction electron gas into the correlated superlattice  via the Kondo effect and stem from 
the single-impurity Kondo physics, which prevails locally over intersite correlations. As such, they are specific to the 
geometry of a superlattice. Indeed,  interfacing  the Kondo lattice layer with a noninteracting metal leads merely to 
softening of the hybridization gap in the Kondo lattice layer~\cite{Peters13}.


\begin{figure}[t!]
\begin{center}
\includegraphics[width=0.15\textwidth]{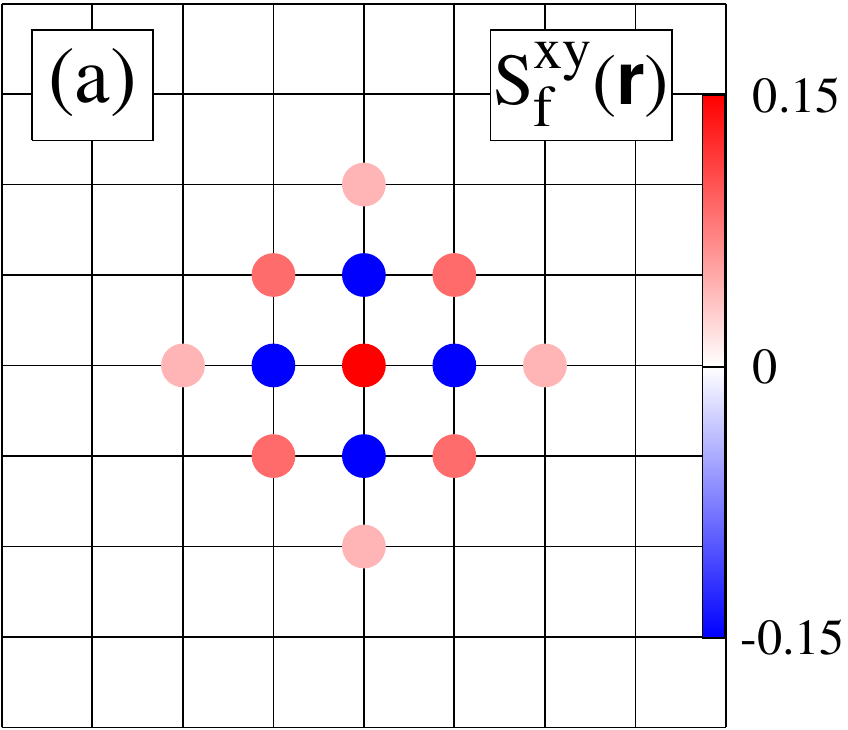}  
\includegraphics[width=0.15\textwidth]{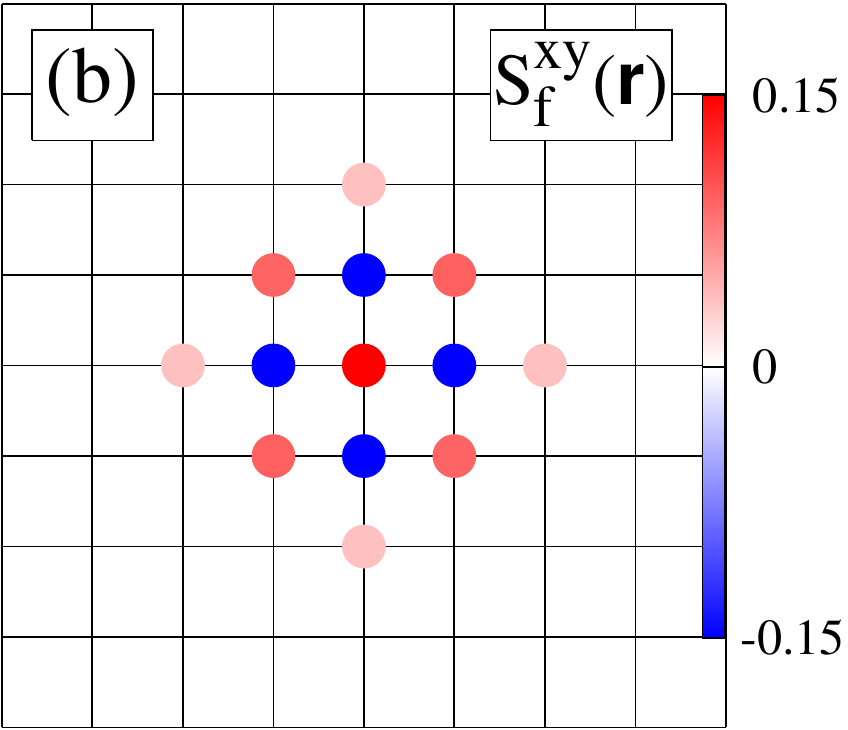}  
\includegraphics[width=0.15\textwidth]{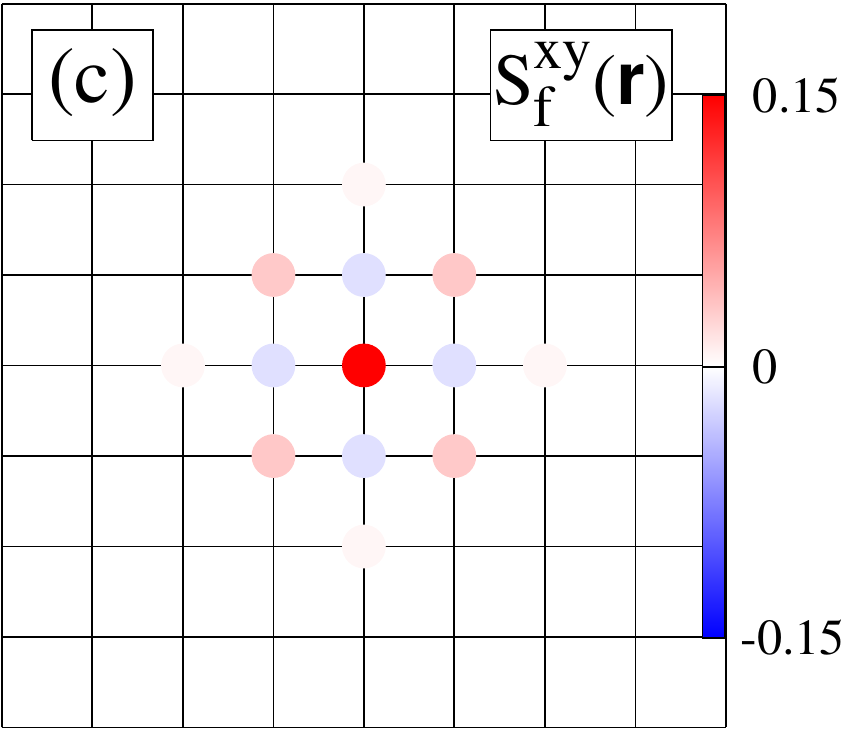}\\
\vspace*{0.1cm}
\includegraphics[width=0.15\textwidth]{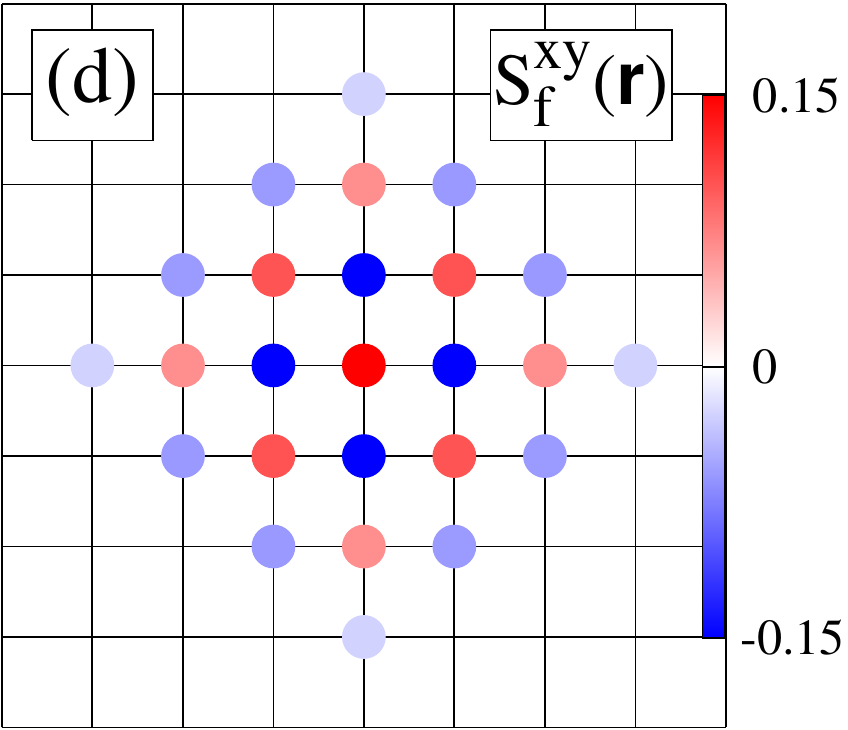}
\includegraphics[width=0.15\textwidth]{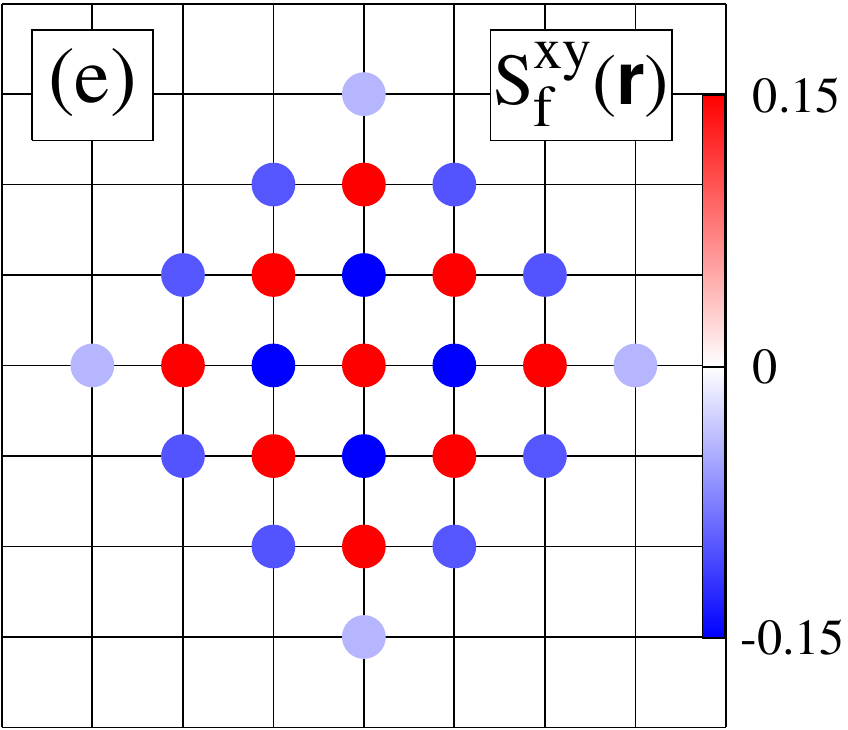}
\includegraphics[width=0.15\textwidth]{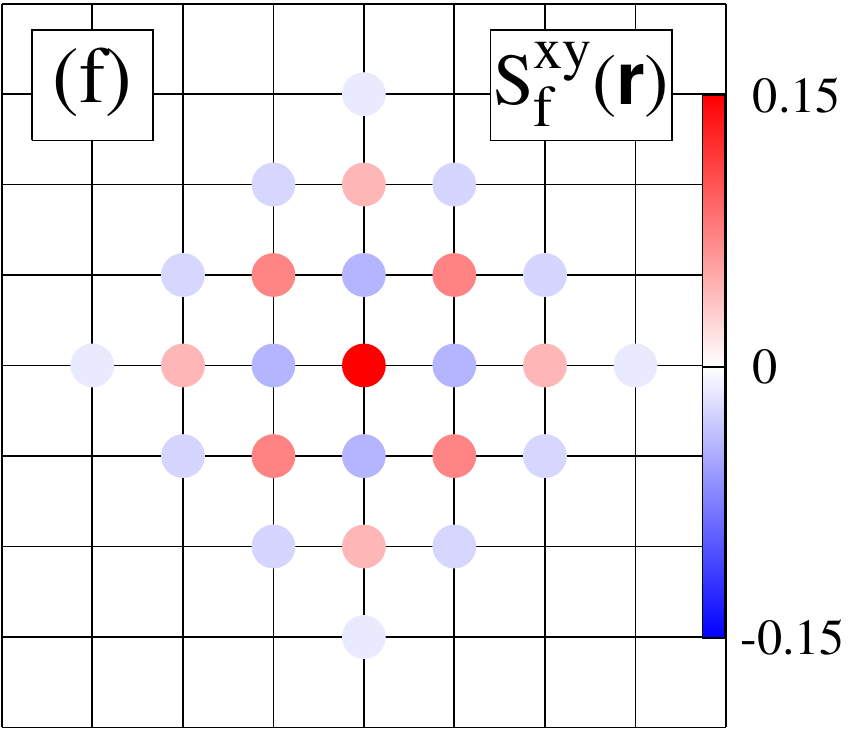}\\
\vspace*{0.1cm}
\includegraphics[width=0.15\textwidth]{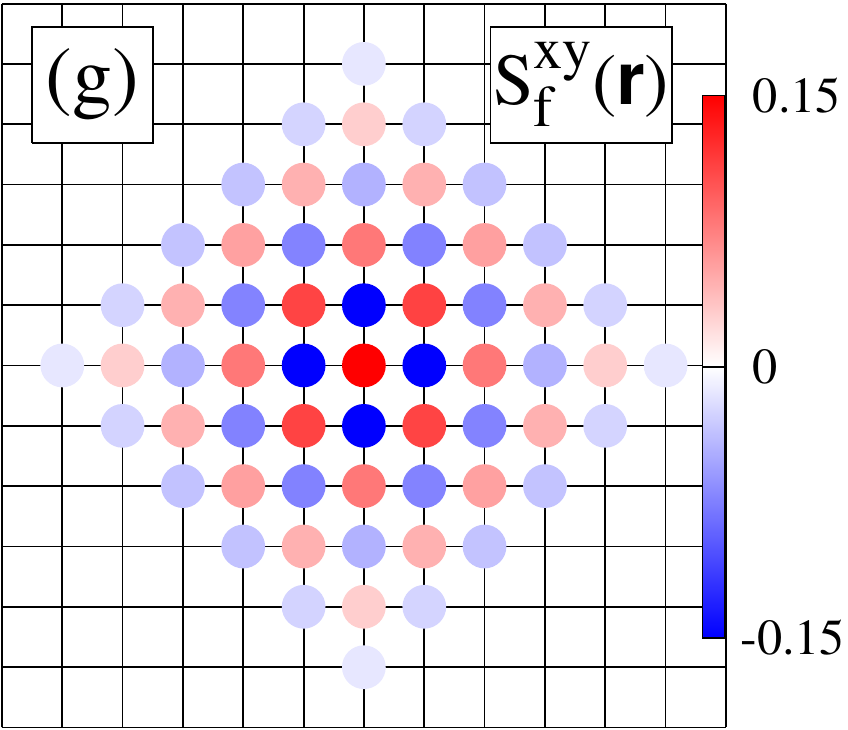}  
\includegraphics[width=0.15\textwidth]{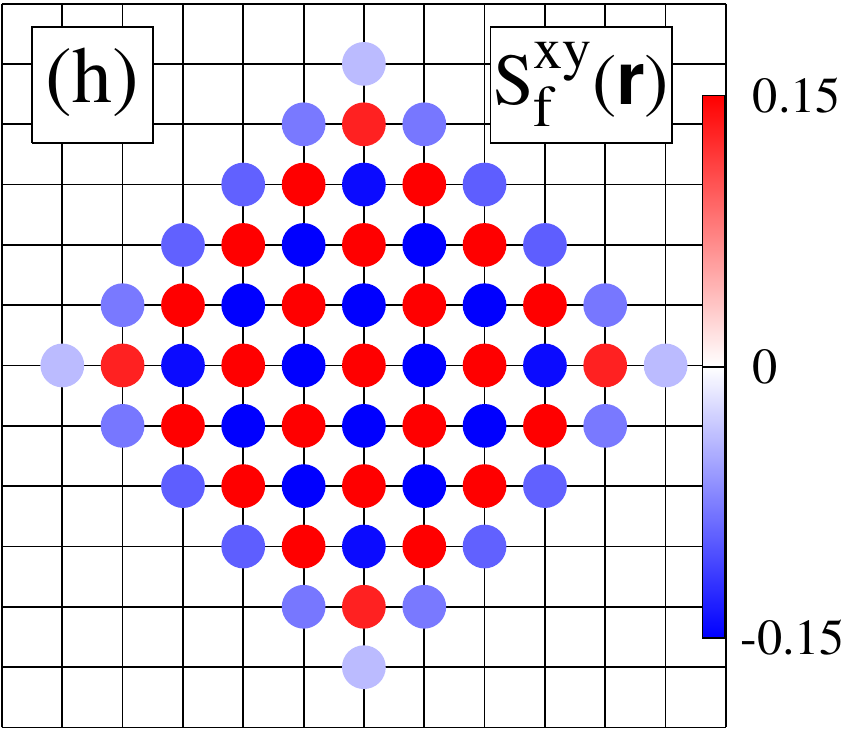}  
\includegraphics[width=0.15\textwidth]{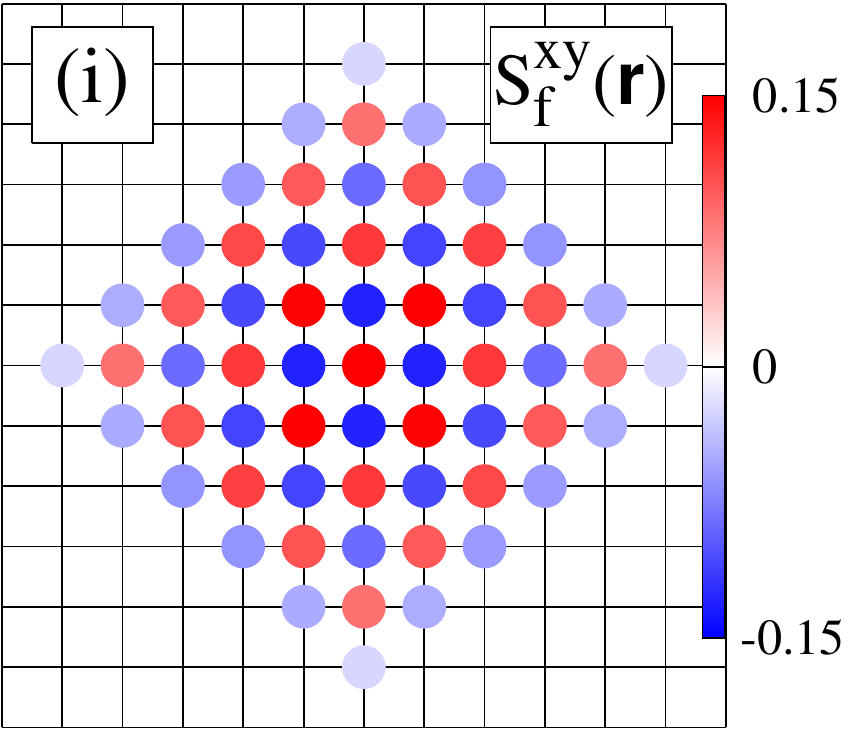}
\end{center}
\caption
{
Real-space transverse-spin correlations  $S_f^{xy}(\pmb{r})$  relative to the central impurity 
of the $n=2$ (a-c), $n=3$ (d-f), and $n=5$ (g-i) systems subject to increasing magnetic field 
$\mu_BB/t=0.06$ (left), 0.13 (middle), and 0.2 (right). Parameters: $L=16$ and $T=t/30$.
}
\label{Sxy}
\end{figure}

An important issue that one might be concerned about is whether the opening of a direct (optical) gap in the \emph{local} quantity 
$N_{\tilde{f},\pmb{r}}(\omega)$ can be considered as unambiguous evidence of coherence, which is a \emph{global} phenomenon 
in the system? Indeed,  in a translation-invariant system, the appearance of a hybridized HF band structure with a small indirect 
gap is usually inferred either from the momentum-dependent single-particle spectral function $
A(\pmb{k},\omega)$~\cite{Eder98,Shim07,Martin08,Otsuki09,Benlagra11,Klein11,Shen12} or from the $T$-dependence of transport 
properties~\cite{Logan05,Pruschke06}. 
Given that the translational symmetry is broken in our superlattices, one has to use an alternative strategy to assess the 
formation of HF bands. One possibility stems from the fact that the 2D particle-hole symmetric KLM 
subject to a magnetic field features a phase transition from the Kondo insulator to a canted AF state~\cite{Beach04,Milat04,Kawakami04}. 
The AF phase is understood as a spin-density-wave instability driven by perfect nesting of the particle and hole hybridized 
bands with opposite spin indices. Thus, it is legitimate to consider the field-induced transverse antiferromagnetism  as 
the \emph{criterion} of the coherent Kondo lattice behavior.

With this considerations in mind, we perform QMC simulations of the KLM (\ref{KLM}) augmented by a Zeeman term,
\begin{equation}
H_{B} = - g \mu_B B \sum_{ \pmb{i} } ( S_{c,\pmb{i}}^{z} + S_{f,\pmb{i}}^{z} ), 
\end{equation}
and we calculate real-space transverse-spin correlations 
$S^{xy}_f(\pmb{r}) = 2\biglb( \langle S^x_{f,\pmb{r}} S^x_{f,\pmb{0}}\rangle + 
                          \langle S^y_{f,\pmb{r}} S^y_{f,\pmb{0}}\rangle  \bigrb)$  
relative to the central impurity upon increasing an external magnetic field $B$, see Fig.~\ref{Sxy}.
The applied magnetic field quickly suppresses  $S_f^{xy}(\pmb{r})$  in the single-shell case, see Fig.~\ref{spatial_B} in Ref.~\cite{SM}.  
In the $n=2$ system, an enhanced range of finite $S_f^{xy}(\pmb{r})$ in the field, albeit without any strengthening 
in Figs.~\ref{Sxy}(a-c), is evocative of the compensation effect:   
on the one hand, opening of the hybridization gap is a prerequisite for the emergence of nesting between the upper and lower HF 
bands with the opposite spin indices; on the other hand, enhanced quasiparticle scattering off transverse AF spin fluctuations  
in the presence of an incomplete gap, as in the $n=2$ superlattice,  precludes the onset of coherence. 
In contrast, a full hybridization gap leads already in the $n=3$ system to a noticeable enhancement of transverse-spin correlations 
$S_f^{xy}(\pmb{r})$, cf. Figs.~\ref{Sxy}(d,e), analogous to that in the $n=5$ superlattice, see Figs.~\ref{Sxy}(g-i). 
Same critical cluster size is found at smaller $J/t=1.5$ in the close proximity to the quantum critical point 
[Figs.~\ref{spatial_B_ext}(a,b) in Ref.~\cite{SM}]  and at $J/t=1.8$ deep in the Kondo insulating phase 
[Figs.~\ref{spatial_B_ext}(c,d) in Ref.~\cite{SM}].

\begin{figure}[t!]
\begin{center}
\includegraphics[width=0.45\textwidth]{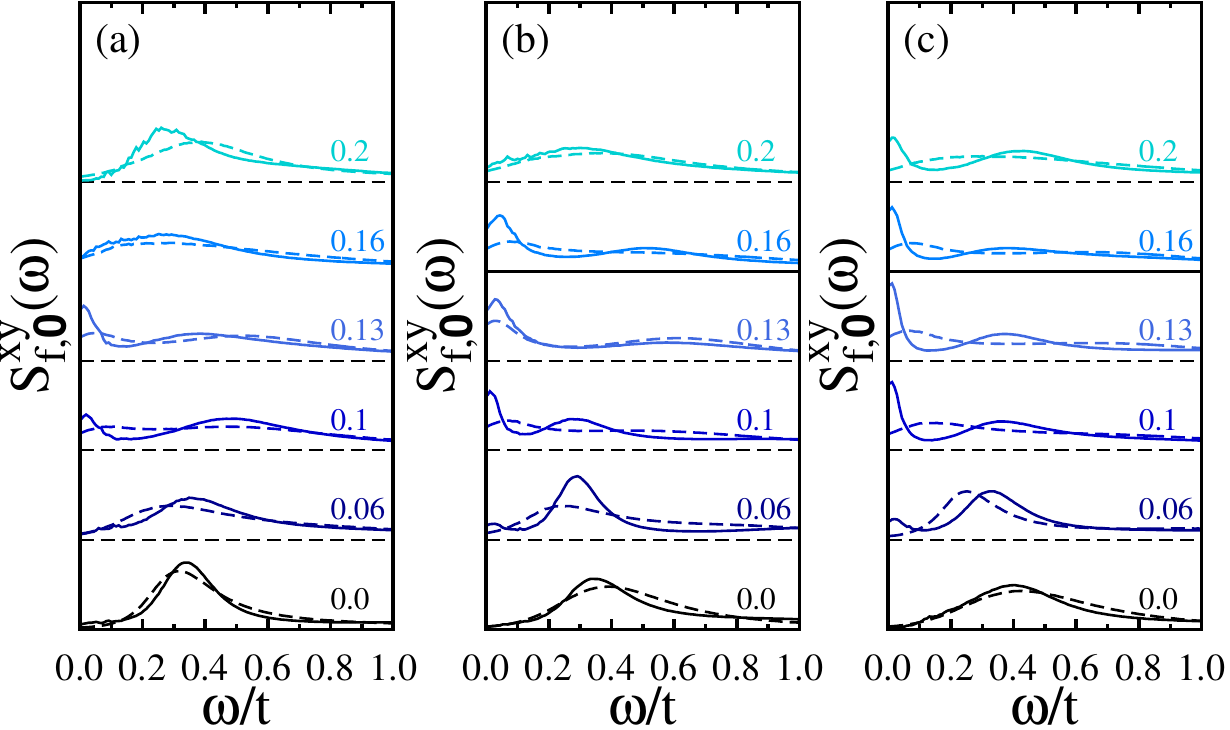}
\end{center}
\caption
{Dynamical transverse-spin structure factor   $S^{xy}_{f,\pmb{0}}(\omega)$ at the central impurity as a function of out-of-plane 
magnetic field $B$ for the system with $n=2$ (a), $n=3$ (b), and $n=5$ (c) shells. 
From bottom to top: $\mu_BB/t=0,\ldots,0.2$. Parameters: $L=16$; $T=t/30$ (solid) and
$T=t/15$ (dashed).
}
\label{local_B}
\end{figure}


Finally, Fig.~\ref{local_B}  displays how the corresponding  dynamical transverse-spin structure factor  
$S_{f,\pmb{0}}^{xy}(\omega)$ at the central site  evolves as a function of the magnetic-field strength $B$.
At $B=0$, all the spectra show a gap-like feature consistent with a collective screening of the $f$-impurities in the Kondo insulator.    
Increasing magnetic field results in a significant transfer of spectral weight to 
low-energy excitations. While the low-frequency spin-wave-like excitations are intimately connected to 
the field-induced transverse antiferromagnetism  in the $n\ge 3$ systems, see Figs.~\ref{local_B}(b,c), 
softening of spin excitations in the $n=2$ system in Fig.~\ref{local_B}(a) should be considered as 
a precursive feature of the magnetic order that emerges in larger superlattices.


In summary, 
we have provided the evidence for the emergent coherent lattice behavior in molecular Kondo systems: 
despite a broken-translation symmetry by the surface,  already a three-shell superlattice with $N_{\textrm{imp}}=25$ periodically 
arranged magnetic impurities is sufficient to recover in the bulk features of the 2D periodic KLM. 
Our predictions are consistent with the recent studies of diluted Kondo lattice systems where disconnected $f$-electron 
clusters continue to exhibit the coherence temperature comparable to the clean case~\cite{Costa18cm,Lawson19}. 
The opening of a hybridization gap in the presence of the spin gap (Kondo insulator regime) effectively 
decouples the bulk from the metallic surface. This can be contrasted with zigzag graphene nanoribbons 
where the bulk is always metallic which leads to boundary-critical phenomena~\cite{Affleck12}.
Our setup corresponds to a special band filling with exactly one conduction electron per impurity spin. 
It leads to the nesting-driven enhancement of the RKKY interaction, which assists the opening of the hybridization 
gap on top of that from coherent scattering off Kondo singlets. 
It motivates future QMC studies of the composite heavy quasiparticle formation in depleted Kondo nanosystems where 
some impurity spins are removed in a regular way, promoting HF metallicity~\cite{Assaad02,Costa18}.

\begin{acknowledgments}
We thank Dirk K. Morr for helpful discussions and T.~A. Costi for comments on the draft.
M.~R. was supported by the German Research Foundation (DFG) through Grant No. RA 2990/1-1. 
F.~F.~A. thanks the DFG for financial support through the SFB 1170 ToCoTronics (Project C01). 
The authors gratefully acknowledge the computing time granted by the John von Neumann Institute for 
Computing (NIC) and provided on the supercomputer JURECA~\cite{jureca} at J\"ulich Supercomputing Centre (JSC). 
 
\end{acknowledgments}

\paragraph*{Note --} A similar topic has recently been addressed in Ref.~\cite{Morr19}.


\begin{thebibliography}{94}%
\makeatletter
\providecommand \@ifxundefined [1]{%
 \@ifx{#1\undefined}
}%
\providecommand \@ifnum [1]{%
 \ifnum #1\expandafter \@firstoftwo
 \else \expandafter \@secondoftwo
 \fi
}%
\providecommand \@ifx [1]{%
 \ifx #1\expandafter \@firstoftwo
 \else \expandafter \@secondoftwo
 \fi
}%
\providecommand \natexlab [1]{#1}%
\providecommand \enquote  [1]{``#1''}%
\providecommand \bibnamefont  [1]{#1}%
\providecommand \bibfnamefont [1]{#1}%
\providecommand \citenamefont [1]{#1}%
\providecommand \href@noop [0]{\@secondoftwo}%
\providecommand \href [0]{\begingroup \@sanitize@url \@href}%
\providecommand \@href[1]{\@@startlink{#1}\@@href}%
\providecommand \@@href[1]{\endgroup#1\@@endlink}%
\providecommand \@sanitize@url [0]{\catcode `\\12\catcode `\$12\catcode
  `\&12\catcode `\#12\catcode `\^12\catcode `\_12\catcode `\%12\relax}%
\providecommand \@@startlink[1]{}%
\providecommand \@@endlink[0]{}%
\providecommand \url  [0]{\begingroup\@sanitize@url \@url }%
\providecommand \@url [1]{\endgroup\@href {#1}{\urlprefix }}%
\providecommand \urlprefix  [0]{URL }%
\providecommand \Eprint [0]{\href }%
\providecommand \doibase [0]{http://dx.doi.org/}%
\providecommand \selectlanguage [0]{\@gobble}%
\providecommand \bibinfo  [0]{\@secondoftwo}%
\providecommand \bibfield  [0]{\@secondoftwo}%
\providecommand \translation [1]{[#1]}%
\providecommand \BibitemOpen [0]{}%
\providecommand \bibitemStop [0]{}%
\providecommand \bibitemNoStop [0]{.\EOS\space}%
\providecommand \EOS [0]{\spacefactor3000\relax}%
\providecommand \BibitemShut  [1]{\csname bibitem#1\endcsname}%
\let\auto@bib@innerbib\@empty
\bibitem [{\citenamefont {Bloch}(1929)}]{Bloch29}%
  \BibitemOpen
  \bibfield  {author} {\bibinfo {author} {\bibfnamefont {F.}~\bibnamefont
  {Bloch}},\ }\bibfield  {title} {\textit {\bibinfo {title} {{{{\"U}ber die
  Quantenmechanik der Elektronen in Kristallgittern}}},\ }}\href {\doibase
  10.1007/BF01339455} {\bibfield  {journal} {\bibinfo  {journal} {Z. Phys.}\
  }\textbf {\bibinfo {volume} {52}},\ \bibinfo {pages} {555} (\bibinfo {year}
  {1929})}\BibitemShut {NoStop}%
\bibitem [{\citenamefont {Wirth}\ and\ \citenamefont
  {Steglich}(2016)}]{Wirth16}%
  \BibitemOpen
  \bibfield  {author} {\bibinfo {author} {\bibfnamefont {S.}~\bibnamefont
  {Wirth}}\ and\ \bibinfo {author} {\bibfnamefont {F.}~\bibnamefont
  {Steglich}},\ }\bibfield  {title} {\textit {\bibinfo {title} {{Exploring
  heavy fermions from macroscopic to microscopic length scales}},\ }}\href
  {\doibase 10.1038/natrevmats.2016.51} {\bibfield  {journal} {\bibinfo
  {journal} {Nat. Rev. Mat.}\ }\textbf {\bibinfo {volume} {1}},\ \bibinfo
  {pages} {16051} (\bibinfo {year} {2016})}\BibitemShut {NoStop}%
\bibitem [{\citenamefont {Hewson}(1993)}]{Hewson_book}%
  \BibitemOpen
  \bibfield  {author} {\bibinfo {author} {\bibfnamefont {A.~C.}\ \bibnamefont
  {Hewson}},\ }\href@noop {} {\emph {\bibinfo {title} {{The Kondo Problem to
  Heavy Fermions}}}}\ (\bibinfo  {publisher} {Cambridge University Press},\
  \bibinfo {address} {Cambridge},\ \bibinfo {year} {1993})\BibitemShut
  {NoStop}%
\bibitem [{\citenamefont {Costi}\ \emph {et~al.}(1988)\citenamefont {Costi},
  \citenamefont {M\"uller-Hartmann},\ and\ \citenamefont {Ulrich}}]{Costi88}%
  \BibitemOpen
  \bibfield  {author} {\bibinfo {author} {\bibfnamefont {T.}~\bibnamefont
  {Costi}}, \bibinfo {author} {\bibfnamefont {E.}~\bibnamefont
  {M\"uller-Hartmann}}, \ and\ \bibinfo {author} {\bibfnamefont
  {K.}~\bibnamefont {Ulrich}},\ }\bibfield  {title} {\textit {\bibinfo {title}
  {{Quantum Monte Carlo simulations of dilute and concentrated heavy-fermion
  systems}},\ }}\href {\doibase 10.1016/0038-1098(88)90853-8} {\bibfield
  {journal} {\bibinfo  {journal} {Solid State Commun.}\ }\textbf {\bibinfo
  {volume} {66}},\ \bibinfo {pages} {343 } (\bibinfo {year}
  {1988})}\BibitemShut {NoStop}%
\bibitem [{\citenamefont {Schlottmann}(1992)}]{Schlottmann92}%
  \BibitemOpen
  \bibfield  {author} {\bibinfo {author} {\bibfnamefont {P.}~\bibnamefont
  {Schlottmann}},\ }\bibfield  {title} {\textit {\bibinfo {title} {{Impurity
  bands in Kondo insulators}},\ }}\href {\doibase 10.1103/PhysRevB.46.998}
  {\bibfield  {journal} {\bibinfo  {journal} {Phys. Rev. B}\ }\textbf {\bibinfo
  {volume} {46}},\ \bibinfo {pages} {998} (\bibinfo {year} {1992})}\BibitemShut
  {NoStop}%
\bibitem [{\citenamefont {Miranda}\ \emph {et~al.}(1997)\citenamefont
  {Miranda}, \citenamefont {Dobrosavljevi\ifmmode~\acute{c}\else \'{c}\fi{}},\
  and\ \citenamefont {Kotliar}}]{Miranda97}%
  \BibitemOpen
  \bibfield  {author} {\bibinfo {author} {\bibfnamefont {E.}~\bibnamefont
  {Miranda}}, \bibinfo {author} {\bibfnamefont {V.}~\bibnamefont
  {Dobrosavljevi\ifmmode~\acute{c}\else \'{c}\fi{}}}, \ and\ \bibinfo {author}
  {\bibfnamefont {G.}~\bibnamefont {Kotliar}},\ }\bibfield  {title} {\textit
  {\bibinfo {title} {{Disorder-Driven Non-Fermi-Liquid Behavior in Kondo
  Alloys}},\ }}\href {\doibase 10.1103/PhysRevLett.78.290} {\bibfield
  {journal} {\bibinfo  {journal} {Phys. Rev. Lett.}\ }\textbf {\bibinfo
  {volume} {78}},\ \bibinfo {pages} {290} (\bibinfo {year} {1997})}\BibitemShut
  {NoStop}%
\bibitem [{\citenamefont {Barzykin}\ and\ \citenamefont
  {Affleck}(2000)}]{Affleck00}%
  \BibitemOpen
  \bibfield  {author} {\bibinfo {author} {\bibfnamefont {V.}~\bibnamefont
  {Barzykin}}\ and\ \bibinfo {author} {\bibfnamefont {I.}~\bibnamefont
  {Affleck}},\ }\bibfield  {title} {\textit {\bibinfo {title} {{Impurity
  correlations in dilute Kondo alloys}},\ }}\href {\doibase
  10.1103/PhysRevB.61.6170} {\bibfield  {journal} {\bibinfo  {journal} {Phys.
  Rev. B}\ }\textbf {\bibinfo {volume} {61}},\ \bibinfo {pages} {6170}
  (\bibinfo {year} {2000})}\BibitemShut {NoStop}%
\bibitem [{\citenamefont {Riseborough}(2003)}]{Riseboro03}%
  \BibitemOpen
  \bibfield  {author} {\bibinfo {author} {\bibfnamefont {P.~S.}\ \bibnamefont
  {Riseborough}},\ }\bibfield  {title} {\textit {\bibinfo {title} {{Collapse of
  the coherence gap in Kondo semiconductors}},\ }}\href {\doibase
  10.1103/PhysRevB.68.235213} {\bibfield  {journal} {\bibinfo  {journal} {Phys.
  Rev. B}\ }\textbf {\bibinfo {volume} {68}},\ \bibinfo {pages} {235213}
  (\bibinfo {year} {2003})}\BibitemShut {NoStop}%
\bibitem [{\citenamefont {Kaul}\ and\ \citenamefont {Vojta}(2007)}]{Vojta07}%
  \BibitemOpen
  \bibfield  {author} {\bibinfo {author} {\bibfnamefont {R.~K.}\ \bibnamefont
  {Kaul}}\ and\ \bibinfo {author} {\bibfnamefont {M.}~\bibnamefont {Vojta}},\
  }\bibfield  {title} {\textit {\bibinfo {title} {{Strongly inhomogeneous
  phases and non-Fermi-liquid behavior in randomly depleted Kondo lattices}},\
  }}\href {\doibase 10.1103/PhysRevB.75.132407} {\bibfield  {journal} {\bibinfo
   {journal} {Phys. Rev. B}\ }\textbf {\bibinfo {volume} {75}},\ \bibinfo
  {pages} {132407} (\bibinfo {year} {2007})}\BibitemShut {NoStop}%
\bibitem [{\citenamefont {Grenzebach}\ \emph {et~al.}(2008)\citenamefont
  {Grenzebach}, \citenamefont {Anders}, \citenamefont {Czycholl},\ and\
  \citenamefont {Pruschke}}]{Pruschke08}%
  \BibitemOpen
  \bibfield  {author} {\bibinfo {author} {\bibfnamefont {C.}~\bibnamefont
  {Grenzebach}}, \bibinfo {author} {\bibfnamefont {F.~B.}\ \bibnamefont
  {Anders}}, \bibinfo {author} {\bibfnamefont {G.}~\bibnamefont {Czycholl}}, \
  and\ \bibinfo {author} {\bibfnamefont {T.}~\bibnamefont {Pruschke}},\
  }\bibfield  {title} {\textit {\bibinfo {title} {{Influence of disorder on the
  transport properties of heavy-fermion systems}},\ }}\href {\doibase
  10.1103/PhysRevB.77.115125} {\bibfield  {journal} {\bibinfo  {journal} {Phys.
  Rev. B}\ }\textbf {\bibinfo {volume} {77}},\ \bibinfo {pages} {115125}
  (\bibinfo {year} {2008})}\BibitemShut {NoStop}%
\bibitem [{\citenamefont {Watanabe}\ and\ \citenamefont
  {Ogata}(2010)}]{Ogata10}%
  \BibitemOpen
  \bibfield  {author} {\bibinfo {author} {\bibfnamefont {H.}~\bibnamefont
  {Watanabe}}\ and\ \bibinfo {author} {\bibfnamefont {M.}~\bibnamefont
  {Ogata}},\ }\bibfield  {title} {\textit {\bibinfo {title} {{Crossover from
  dilute-Kondo system to heavy-fermion system}},\ }}\href {\doibase
  10.1103/PhysRevB.81.113111} {\bibfield  {journal} {\bibinfo  {journal} {Phys.
  Rev. B}\ }\textbf {\bibinfo {volume} {81}},\ \bibinfo {pages} {113111}
  (\bibinfo {year} {2010})}\BibitemShut {NoStop}%
\bibitem [{\citenamefont {Kumar}\ and\ \citenamefont
  {Vidhyadhiraja}(2014)}]{Kumar14}%
  \BibitemOpen
  \bibfield  {author} {\bibinfo {author} {\bibfnamefont {P.}~\bibnamefont
  {Kumar}}\ and\ \bibinfo {author} {\bibfnamefont {N.~S.}\ \bibnamefont
  {Vidhyadhiraja}},\ }\bibfield  {title} {\textit {\bibinfo {title}
  {{Kondo-hole substitution in heavy fermions: Dynamics and transport}},\
  }}\href {\doibase 10.1103/PhysRevB.90.235133} {\bibfield  {journal} {\bibinfo
   {journal} {Phys. Rev. B}\ }\textbf {\bibinfo {volume} {90}},\ \bibinfo
  {pages} {235133} (\bibinfo {year} {2014})}\BibitemShut {NoStop}%
\bibitem [{\citenamefont {Sen}\ \emph {et~al.}(2015)\citenamefont {Sen},
  \citenamefont {Moreno}, \citenamefont {Jarrell},\ and\ \citenamefont
  {Vidhyadhiraja}}]{Moreno15}%
  \BibitemOpen
  \bibfield  {author} {\bibinfo {author} {\bibfnamefont {S.}~\bibnamefont
  {Sen}}, \bibinfo {author} {\bibfnamefont {J.}~\bibnamefont {Moreno}},
  \bibinfo {author} {\bibfnamefont {M.}~\bibnamefont {Jarrell}}, \ and\
  \bibinfo {author} {\bibfnamefont {N.~S.}\ \bibnamefont {Vidhyadhiraja}},\
  }\bibfield  {title} {\textit {\bibinfo {title} {{Spectral changes in layered
  $f$-electron systems induced by Kondo hole substitution in the boundary
  layer}},\ }}\href {\doibase 10.1103/PhysRevB.91.155146} {\bibfield  {journal}
  {\bibinfo  {journal} {Phys. Rev. B}\ }\textbf {\bibinfo {volume} {91}},\
  \bibinfo {pages} {155146} (\bibinfo {year} {2015})}\BibitemShut {NoStop}%
\bibitem [{\citenamefont {Wei}\ and\ \citenamefont {Yang}(2017)}]{Wei17}%
  \BibitemOpen
  \bibfield  {author} {\bibinfo {author} {\bibfnamefont {L.-y.}\ \bibnamefont
  {Wei}}\ and\ \bibinfo {author} {\bibfnamefont {Y.-f.}\ \bibnamefont {Yang}},\
  }\bibfield  {title} {\textit {\bibinfo {title} {{Doping-induced perturbation
  and percolation in the two-dimensional Anderson lattice}},\ }}\href {\doibase
  10.1038/srep46089} {\bibfield  {journal} {\bibinfo  {journal} {Sci. Rep.}\
  }\textbf {\bibinfo {volume} {7}},\ \bibinfo {pages} {46089} (\bibinfo {year}
  {2017})}\BibitemShut {NoStop}%
\bibitem [{\citenamefont {Martin}(1982)}]{Martin82}%
  \BibitemOpen
  \bibfield  {author} {\bibinfo {author} {\bibfnamefont {R.~M.}\ \bibnamefont
  {Martin}},\ }\bibfield  {title} {\textit {\bibinfo {title} {{Fermi-Surfae Sum
  Rule and its Consequences for Periodic Kondo and Mixed-Valence Systems}},\
  }}\href {\doibase 10.1103/PhysRevLett.48.362} {\bibfield  {journal} {\bibinfo
   {journal} {Phys. Rev. Lett.}\ }\textbf {\bibinfo {volume} {48}},\ \bibinfo
  {pages} {362} (\bibinfo {year} {1982})}\BibitemShut {NoStop}%
\bibitem [{\citenamefont {Lacroix}(1986)}]{Lacroix86}%
  \BibitemOpen
  \bibfield  {author} {\bibinfo {author} {\bibfnamefont {C.}~\bibnamefont
  {Lacroix}},\ }\bibfield  {title} {\textit {\bibinfo {title} {{Coherence
  effects in the Kondo lattice}},\ }}\href {\doibase
  10.1016/0304-8853(86)90093-4} {\bibfield  {journal} {\bibinfo  {journal} {J.
  Magn. Magn. Mater.}\ }\textbf {\bibinfo {volume} {60}},\ \bibinfo {pages}
  {145} (\bibinfo {year} {1986})}\BibitemShut {NoStop}%
\bibitem [{\citenamefont {Tesanovic}\ and\ \citenamefont
  {Valls}(1986)}]{Tesa86}%
  \BibitemOpen
  \bibfield  {author} {\bibinfo {author} {\bibfnamefont {Z.}~\bibnamefont
  {Tesanovic}}\ and\ \bibinfo {author} {\bibfnamefont {O.~T.}\ \bibnamefont
  {Valls}},\ }\bibfield  {title} {\textit {\bibinfo {title} {{Kondo lattice and
  the formation of a heavy-fermion state}},\ }}\href {\doibase
  10.1103/PhysRevB.34.1918} {\bibfield  {journal} {\bibinfo  {journal} {Phys.
  Rev. B}\ }\textbf {\bibinfo {volume} {34}},\ \bibinfo {pages} {1918}
  (\bibinfo {year} {1986})}\BibitemShut {NoStop}%
\bibitem [{\citenamefont {Continentino}\ \emph {et~al.}(1989)\citenamefont
  {Continentino}, \citenamefont {Japiassu},\ and\ \citenamefont
  {Troper}}]{Continentino89}%
  \BibitemOpen
  \bibfield  {author} {\bibinfo {author} {\bibfnamefont {M.~A.}\ \bibnamefont
  {Continentino}}, \bibinfo {author} {\bibfnamefont {G.~M.}\ \bibnamefont
  {Japiassu}}, \ and\ \bibinfo {author} {\bibfnamefont {A.}~\bibnamefont
  {Troper}},\ }\bibfield  {title} {\textit {\bibinfo {title} {{Critical
  approach to the coherence transition in Kondo lattices}},\ }}\href {\doibase
  10.1103/PhysRevB.39.9734} {\bibfield  {journal} {\bibinfo  {journal} {Phys.
  Rev. B}\ }\textbf {\bibinfo {volume} {39}},\ \bibinfo {pages} {9734}
  (\bibinfo {year} {1989})}\BibitemShut {NoStop}%
\bibitem [{\citenamefont {Tahvildar-Zadeh}\ \emph {et~al.}(1998)\citenamefont
  {Tahvildar-Zadeh}, \citenamefont {Jarrell},\ and\ \citenamefont
  {Freericks}}]{Jarrell98}%
  \BibitemOpen
  \bibfield  {author} {\bibinfo {author} {\bibfnamefont {A.~N.}\ \bibnamefont
  {Tahvildar-Zadeh}}, \bibinfo {author} {\bibfnamefont {M.}~\bibnamefont
  {Jarrell}}, \ and\ \bibinfo {author} {\bibfnamefont {J.~K.}\ \bibnamefont
  {Freericks}},\ }\bibfield  {title} {\textit {\bibinfo {title}
  {{Low-Temperature Coherence in the Periodic Anderson Model: Predictions for
  Photoemission of Heavy Fermions}},\ }}\href {\doibase
  10.1103/PhysRevLett.80.5168} {\bibfield  {journal} {\bibinfo  {journal}
  {Phys. Rev. Lett.}\ }\textbf {\bibinfo {volume} {80}},\ \bibinfo {pages}
  {5168} (\bibinfo {year} {1998})}\BibitemShut {NoStop}%
\bibitem [{\citenamefont {Pruschke}\ \emph {et~al.}(2000)\citenamefont
  {Pruschke}, \citenamefont {Bulla},\ and\ \citenamefont
  {Jarrell}}]{Pruschke00}%
  \BibitemOpen
  \bibfield  {author} {\bibinfo {author} {\bibfnamefont {T.}~\bibnamefont
  {Pruschke}}, \bibinfo {author} {\bibfnamefont {R.}~\bibnamefont {Bulla}}, \
  and\ \bibinfo {author} {\bibfnamefont {M.}~\bibnamefont {Jarrell}},\
  }\bibfield  {title} {\textit {\bibinfo {title} {{Low-energy scale of the
  periodic Anderson model}},\ }}\href {\doibase 10.1103/PhysRevB.61.12799}
  {\bibfield  {journal} {\bibinfo  {journal} {Phys. Rev. B}\ }\textbf {\bibinfo
  {volume} {61}},\ \bibinfo {pages} {12799} (\bibinfo {year}
  {2000})}\BibitemShut {NoStop}%
\bibitem [{\citenamefont {Burdin}\ \emph {et~al.}(2000)\citenamefont {Burdin},
  \citenamefont {Georges},\ and\ \citenamefont {Grempel}}]{Burdin00}%
  \BibitemOpen
  \bibfield  {author} {\bibinfo {author} {\bibfnamefont {S.}~\bibnamefont
  {Burdin}}, \bibinfo {author} {\bibfnamefont {A.}~\bibnamefont {Georges}}, \
  and\ \bibinfo {author} {\bibfnamefont {D.~R.}\ \bibnamefont {Grempel}},\
  }\bibfield  {title} {\textit {\bibinfo {title} {{Coherence Scale of the Kondo
  Lattice}},\ }}\href {\doibase 10.1103/PhysRevLett.85.1048} {\bibfield
  {journal} {\bibinfo  {journal} {Phys. Rev. Lett.}\ }\textbf {\bibinfo
  {volume} {85}},\ \bibinfo {pages} {1048} (\bibinfo {year}
  {2000})}\BibitemShut {NoStop}%
\bibitem [{\citenamefont {Costi}\ and\ \citenamefont {Manini}(2002)}]{Costi02}%
  \BibitemOpen
  \bibfield  {author} {\bibinfo {author} {\bibfnamefont {T.~A.}\ \bibnamefont
  {Costi}}\ and\ \bibinfo {author} {\bibfnamefont {N.}~\bibnamefont {Manini}},\
  }\bibfield  {title} {\textit {\bibinfo {title} {{Low-Energy Scales and
  Temperature-Dependent Photoemission of Heavy Fermions}},\ }}\href {\doibase
  10.1023/A:1013882222679} {\bibfield  {journal} {\bibinfo  {journal} {J. Low
  Temp. Phys.}\ }\textbf {\bibinfo {volume} {126}},\ \bibinfo {pages} {835}
  (\bibinfo {year} {2002})}\BibitemShut {NoStop}%
\bibitem [{\citenamefont {Assaad}(2004)}]{Assaad04}%
  \BibitemOpen
  \bibfield  {author} {\bibinfo {author} {\bibfnamefont {F.~F.}\ \bibnamefont
  {Assaad}},\ }\bibfield  {title} {\textit {\bibinfo {title} {{Coherence scale
  of the two-dimensional Kondo lattice model}},\ }}\href {\doibase
  10.1103/PhysRevB.70.020402} {\bibfield  {journal} {\bibinfo  {journal} {Phys.
  Rev. B}\ }\textbf {\bibinfo {volume} {70}},\ \bibinfo {pages} {020402}
  (\bibinfo {year} {2004})}\BibitemShut {NoStop}%
\bibitem [{\citenamefont {Yang}\ \emph {et~al.}(2008)\citenamefont {Yang},
  \citenamefont {Fisk}, \citenamefont {Lee}, \citenamefont {Thompson},\ and\
  \citenamefont {Pines}}]{Fisk08}%
  \BibitemOpen
  \bibfield  {author} {\bibinfo {author} {\bibfnamefont {Y.-f.}\ \bibnamefont
  {Yang}}, \bibinfo {author} {\bibfnamefont {Z.}~\bibnamefont {Fisk}}, \bibinfo
  {author} {\bibfnamefont {H.-O.}\ \bibnamefont {Lee}}, \bibinfo {author}
  {\bibfnamefont {J.~D.}\ \bibnamefont {Thompson}}, \ and\ \bibinfo {author}
  {\bibfnamefont {D.}~\bibnamefont {Pines}},\ }\bibfield  {title} {\textit
  {\bibinfo {title} {{Scaling the Kondo lattice}},\ }}\href {\doibase
  10.1038/nature07157} {\bibfield  {journal} {\bibinfo  {journal} {Nature
  (London)}\ }\textbf {\bibinfo {volume} {454}},\ \bibinfo {pages} {611}
  (\bibinfo {year} {2008})}\BibitemShut {NoStop}%
\bibitem [{\citenamefont {Beach}\ and\ \citenamefont {Assaad}(2008)}]{Beach08}%
  \BibitemOpen
  \bibfield  {author} {\bibinfo {author} {\bibfnamefont {K.~S.~D.}\
  \bibnamefont {Beach}}\ and\ \bibinfo {author} {\bibfnamefont {F.~F.}\
  \bibnamefont {Assaad}},\ }\bibfield  {title} {\textit {\bibinfo {title}
  {{Coherence and metamagnetism in the two-dimensional Kondo lattice model}},\
  }}\href {\doibase 10.1103/PhysRevB.77.205123} {\bibfield  {journal} {\bibinfo
   {journal} {Phys. Rev. B}\ }\textbf {\bibinfo {volume} {77}},\ \bibinfo
  {pages} {205123} (\bibinfo {year} {2008})}\BibitemShut {NoStop}%
\bibitem [{\citenamefont {Raczkowski}\ \emph {et~al.}(2010)\citenamefont
  {Raczkowski}, \citenamefont {Zhang}, \citenamefont {Assaad}, \citenamefont
  {Pruschke},\ and\ \citenamefont {Jarrell}}]{Raczkowski10}%
  \BibitemOpen
  \bibfield  {author} {\bibinfo {author} {\bibfnamefont {M.}~\bibnamefont
  {Raczkowski}}, \bibinfo {author} {\bibfnamefont {P.}~\bibnamefont {Zhang}},
  \bibinfo {author} {\bibfnamefont {F.~F.}\ \bibnamefont {Assaad}}, \bibinfo
  {author} {\bibfnamefont {T.}~\bibnamefont {Pruschke}}, \ and\ \bibinfo
  {author} {\bibfnamefont {M.}~\bibnamefont {Jarrell}},\ }\bibfield  {title}
  {\textit {\bibinfo {title} {{Phonons and the coherence scale of models of
  heavy fermions}},\ }}\href {\doibase 10.1103/PhysRevB.81.054444} {\bibfield
  {journal} {\bibinfo  {journal} {Phys. Rev. B}\ }\textbf {\bibinfo {volume}
  {81}},\ \bibinfo {pages} {054444} (\bibinfo {year} {2010})}\BibitemShut
  {NoStop}%
\bibitem [{\citenamefont {Tanaskovi\ifmmode~\acute{c}\else \'{c}\fi{}}\ \emph
  {et~al.}(2011)\citenamefont {Tanaskovi\ifmmode~\acute{c}\else \'{c}\fi{}},
  \citenamefont {Haule}, \citenamefont {Kotliar},\ and\ \citenamefont
  {Dobrosavljevi\ifmmode~\acute{c}\else \'{c}\fi{}}}]{Tanaskovic11}%
  \BibitemOpen
  \bibfield  {author} {\bibinfo {author} {\bibfnamefont {D.}~\bibnamefont
  {Tanaskovi\ifmmode~\acute{c}\else \'{c}\fi{}}}, \bibinfo {author}
  {\bibfnamefont {K.}~\bibnamefont {Haule}}, \bibinfo {author} {\bibfnamefont
  {G.}~\bibnamefont {Kotliar}}, \ and\ \bibinfo {author} {\bibfnamefont
  {V.}~\bibnamefont {Dobrosavljevi\ifmmode~\acute{c}\else \'{c}\fi{}}},\
  }\bibfield  {title} {\textit {\bibinfo {title} {{Phase diagram, energy
  scales, and nonlocal correlations in the Anderson lattice model}},\ }}\href
  {\doibase 10.1103/PhysRevB.84.115105} {\bibfield  {journal} {\bibinfo
  {journal} {Phys. Rev. B}\ }\textbf {\bibinfo {volume} {84}},\ \bibinfo
  {pages} {115105} (\bibinfo {year} {2011})}\BibitemShut {NoStop}%
\bibitem [{\citenamefont {Bredl}\ \emph {et~al.}(1984)\citenamefont {Bredl},
  \citenamefont {Horn}, \citenamefont {Steglich}, \citenamefont {L\"uthi},\
  and\ \citenamefont {Martin}}]{Steglich84}%
  \BibitemOpen
  \bibfield  {author} {\bibinfo {author} {\bibfnamefont {C.~D.}\ \bibnamefont
  {Bredl}}, \bibinfo {author} {\bibfnamefont {S.}~\bibnamefont {Horn}},
  \bibinfo {author} {\bibfnamefont {F.}~\bibnamefont {Steglich}}, \bibinfo
  {author} {\bibfnamefont {B.}~\bibnamefont {L\"uthi}}, \ and\ \bibinfo
  {author} {\bibfnamefont {R.~M.}\ \bibnamefont {Martin}},\ }\bibfield  {title}
  {\textit {\bibinfo {title} {{Low-Temperature Specific Heat of
  $\mathrm{CeCu}_2\mathrm{Si}_2$ and $\mathrm{CeAl}_3$: Coherence Effects in
  Kondo Lattice Systems}},\ }}\href {\doibase 10.1103/PhysRevLett.52.1982}
  {\bibfield  {journal} {\bibinfo  {journal} {Phys. Rev. Lett.}\ }\textbf
  {\bibinfo {volume} {52}},\ \bibinfo {pages} {1982} (\bibinfo {year}
  {1984})}\BibitemShut {NoStop}%
\bibitem [{\citenamefont {\=Onuki}\ \emph {et~al.}(1985)\citenamefont
  {\=Onuki}, \citenamefont {Shimizu}, \citenamefont {Nishihara}, \citenamefont
  {Machii},\ and\ \citenamefont {Komatsubara}}]{Onuki85}%
  \BibitemOpen
  \bibfield  {author} {\bibinfo {author} {\bibfnamefont {Y.}~\bibnamefont
  {\=Onuki}}, \bibinfo {author} {\bibfnamefont {Y.}~\bibnamefont {Shimizu}},
  \bibinfo {author} {\bibfnamefont {M.}~\bibnamefont {Nishihara}}, \bibinfo
  {author} {\bibfnamefont {Y.}~\bibnamefont {Machii}}, \ and\ \bibinfo {author}
  {\bibfnamefont {T.}~\bibnamefont {Komatsubara}},\ }\bibfield  {title}
  {\textit {\bibinfo {title} {{Kondo Lattice Formation in
  $\mathrm{Ce}_x\mathrm{La}_{1-x}\mathrm{Cu}_6$ }},\ }}\href {\doibase
  10.1143/JPSJ.54.1964} {\bibfield  {journal} {\bibinfo  {journal} {J. Phys.
  Soc. Jpn.}\ }\textbf {\bibinfo {volume} {54}},\ \bibinfo {pages} {1964}
  (\bibinfo {year} {1985})}\BibitemShut {NoStop}%
\bibitem [{\citenamefont {Sumiyama}\ \emph {et~al.}(1986)\citenamefont
  {Sumiyama}, \citenamefont {Oda}, \citenamefont {Nagano}, \citenamefont
  {\=Onuki}, \citenamefont {Shibutani},\ and\ \citenamefont
  {Komatsubara}}]{Sumiyama86}%
  \BibitemOpen
  \bibfield  {author} {\bibinfo {author} {\bibfnamefont {A.}~\bibnamefont
  {Sumiyama}}, \bibinfo {author} {\bibfnamefont {Y.}~\bibnamefont {Oda}},
  \bibinfo {author} {\bibfnamefont {H.}~\bibnamefont {Nagano}}, \bibinfo
  {author} {\bibfnamefont {Y.}~\bibnamefont {\=Onuki}}, \bibinfo {author}
  {\bibfnamefont {K.}~\bibnamefont {Shibutani}}, \ and\ \bibinfo {author}
  {\bibfnamefont {T.}~\bibnamefont {Komatsubara}},\ }\bibfield  {title}
  {\textit {\bibinfo {title} {{Coherent Kondo State in a Dense Kondo Substance:
  $\mathrm{Ce}_x\mathrm{La}_{1-x}\mathrm{Cu}_6$ }},\ }}\href {\doibase
  10.1143/JPSJ.55.1294} {\bibfield  {journal} {\bibinfo  {journal} {J. Phys.
  Soc. Jpn.}\ }\textbf {\bibinfo {volume} {55}},\ \bibinfo {pages} {1294}
  (\bibinfo {year} {1986})}\BibitemShut {NoStop}%
\bibitem [{\citenamefont {Lin}\ \emph {et~al.}(1987)\citenamefont {Lin},
  \citenamefont {Wallash}, \citenamefont {Crow}, \citenamefont {Mihalisin},\
  and\ \citenamefont {Schlottmann}}]{Lin87}%
  \BibitemOpen
  \bibfield  {author} {\bibinfo {author} {\bibfnamefont {C.~L.}\ \bibnamefont
  {Lin}}, \bibinfo {author} {\bibfnamefont {A.}~\bibnamefont {Wallash}},
  \bibinfo {author} {\bibfnamefont {J.~E.}\ \bibnamefont {Crow}}, \bibinfo
  {author} {\bibfnamefont {T.}~\bibnamefont {Mihalisin}}, \ and\ \bibinfo
  {author} {\bibfnamefont {P.}~\bibnamefont {Schlottmann}},\ }\bibfield
  {title} {\textit {\bibinfo {title} {{Heavy-fermion behavior and the
  single-ion Kondo model}},\ }}\href {\doibase 10.1103/PhysRevLett.58.1232}
  {\bibfield  {journal} {\bibinfo  {journal} {Phys. Rev. Lett.}\ }\textbf
  {\bibinfo {volume} {58}},\ \bibinfo {pages} {1232} (\bibinfo {year}
  {1987})}\BibitemShut {NoStop}%
\bibitem [{\citenamefont {Bud'ko}\ \emph {et~al.}(1998)\citenamefont {Bud'ko},
  \citenamefont {Fontes},\ and\ \citenamefont {Baggio-Saitovitch}}]{Budko98}%
  \BibitemOpen
  \bibfield  {author} {\bibinfo {author} {\bibfnamefont {S.~L.}\ \bibnamefont
  {Bud'ko}}, \bibinfo {author} {\bibfnamefont {M.~B.}\ \bibnamefont {Fontes}},
  \ and\ \bibinfo {author} {\bibfnamefont {E.~M.}\ \bibnamefont
  {Baggio-Saitovitch}},\ }\bibfield  {title} {\textit {\bibinfo {title}
  {{Evolution of physical properties in the series
  $\mathrm{Ce}_{1-x}\mathrm{La}_x\mathrm{FeGe}_3$: crossover from Kondo lattice
  to single ion Kondo impurity systems}},\ }}\href {\doibase
  10.1088/0953-8984/10/39/017} {\bibfield  {journal} {\bibinfo  {journal} {J.
  Phys: Condens. Matter}\ }\textbf {\bibinfo {volume} {10}},\ \bibinfo {pages}
  {8815} (\bibinfo {year} {1998})}\BibitemShut {NoStop}%
\bibitem [{\citenamefont {Nakatsuji}\ \emph {et~al.}(2002)\citenamefont
  {Nakatsuji}, \citenamefont {Yeo}, \citenamefont {Balicas}, \citenamefont
  {Fisk}, \citenamefont {Schlottmann}, \citenamefont {Pagliuso}, \citenamefont
  {Moreno}, \citenamefont {Sarrao},\ and\ \citenamefont {Thompson}}]{Fisk02}%
  \BibitemOpen
  \bibfield  {author} {\bibinfo {author} {\bibfnamefont {S.}~\bibnamefont
  {Nakatsuji}}, \bibinfo {author} {\bibfnamefont {S.}~\bibnamefont {Yeo}},
  \bibinfo {author} {\bibfnamefont {L.}~\bibnamefont {Balicas}}, \bibinfo
  {author} {\bibfnamefont {Z.}~\bibnamefont {Fisk}}, \bibinfo {author}
  {\bibfnamefont {P.}~\bibnamefont {Schlottmann}}, \bibinfo {author}
  {\bibfnamefont {P.~G.}\ \bibnamefont {Pagliuso}}, \bibinfo {author}
  {\bibfnamefont {N.~O.}\ \bibnamefont {Moreno}}, \bibinfo {author}
  {\bibfnamefont {J.~L.}\ \bibnamefont {Sarrao}}, \ and\ \bibinfo {author}
  {\bibfnamefont {J.~D.}\ \bibnamefont {Thompson}},\ }\bibfield  {title}
  {\textit {\bibinfo {title} {{Intersite Coupling Effects in a Kondo
  Lattice}},\ }}\href {\doibase 10.1103/PhysRevLett.89.106402} {\bibfield
  {journal} {\bibinfo  {journal} {Phys. Rev. Lett.}\ }\textbf {\bibinfo
  {volume} {89}},\ \bibinfo {pages} {106402} (\bibinfo {year}
  {2002})}\BibitemShut {NoStop}%
\bibitem [{\citenamefont {Pikul}\ \emph {et~al.}(2012)\citenamefont {Pikul},
  \citenamefont {Stockert}, \citenamefont {Steppke}, \citenamefont {Cichorek},
  \citenamefont {Hartmann}, \citenamefont {Caroca-Canales}, \citenamefont
  {Oeschler}, \citenamefont {Brando}, \citenamefont {Geibel},\ and\
  \citenamefont {Steglich}}]{Pikul12}%
  \BibitemOpen
  \bibfield  {author} {\bibinfo {author} {\bibfnamefont {A.~P.}\ \bibnamefont
  {Pikul}}, \bibinfo {author} {\bibfnamefont {U.}~\bibnamefont {Stockert}},
  \bibinfo {author} {\bibfnamefont {A.}~\bibnamefont {Steppke}}, \bibinfo
  {author} {\bibfnamefont {T.}~\bibnamefont {Cichorek}}, \bibinfo {author}
  {\bibfnamefont {S.}~\bibnamefont {Hartmann}}, \bibinfo {author}
  {\bibfnamefont {N.}~\bibnamefont {Caroca-Canales}}, \bibinfo {author}
  {\bibfnamefont {N.}~\bibnamefont {Oeschler}}, \bibinfo {author}
  {\bibfnamefont {M.}~\bibnamefont {Brando}}, \bibinfo {author} {\bibfnamefont
  {C.}~\bibnamefont {Geibel}}, \ and\ \bibinfo {author} {\bibfnamefont
  {F.}~\bibnamefont {Steglich}},\ }\bibfield  {title} {\textit {\bibinfo
  {title} {{Single-Ion Kondo Scaling of the Coherent Fermi Liquid Regime in
  ${\mathrm{Ce}}_{1\ensuremath{-}x}{\mathrm{La}}_{x}{\mathrm{Ni}}_{2}{\mathrm{Ge}}_{2}$}},\
  }}\href {\doibase 10.1103/PhysRevLett.108.066405} {\bibfield  {journal}
  {\bibinfo  {journal} {Phys. Rev. Lett.}\ }\textbf {\bibinfo {volume} {108}},\
  \bibinfo {pages} {066405} (\bibinfo {year} {2012})}\BibitemShut {NoStop}%
\bibitem [{\citenamefont {Hodovanets}\ \emph {et~al.}(2015)\citenamefont
  {Hodovanets}, \citenamefont {Bud'ko}, \citenamefont {Straszheim},
  \citenamefont {Taufour}, \citenamefont {Mun}, \citenamefont {Kim},
  \citenamefont {Flint},\ and\ \citenamefont {Canfield}}]{Budko15}%
  \BibitemOpen
  \bibfield  {author} {\bibinfo {author} {\bibfnamefont {H.}~\bibnamefont
  {Hodovanets}}, \bibinfo {author} {\bibfnamefont {S.~L.}\ \bibnamefont
  {Bud'ko}}, \bibinfo {author} {\bibfnamefont {W.~E.}\ \bibnamefont
  {Straszheim}}, \bibinfo {author} {\bibfnamefont {V.}~\bibnamefont {Taufour}},
  \bibinfo {author} {\bibfnamefont {E.~D.}\ \bibnamefont {Mun}}, \bibinfo
  {author} {\bibfnamefont {H.}~\bibnamefont {Kim}}, \bibinfo {author}
  {\bibfnamefont {R.}~\bibnamefont {Flint}}, \ and\ \bibinfo {author}
  {\bibfnamefont {P.~C.}\ \bibnamefont {Canfield}},\ }\bibfield  {title}
  {\textit {\bibinfo {title} {{Remarkably Robust and Correlated Coherence and
  Antiferromagnetism in
  $({\mathrm{Ce}}_{1\ensuremath{-}x}{\mathrm{La}}_{x}){\mathrm{Cu}}_{2}{\mathrm{Ge}}_{2}$}},\
  }}\href {\doibase 10.1103/PhysRevLett.114.236601} {\bibfield  {journal}
  {\bibinfo  {journal} {Phys. Rev. Lett.}\ }\textbf {\bibinfo {volume} {114}},\
  \bibinfo {pages} {236601} (\bibinfo {year} {2015})}\BibitemShut {NoStop}%
\bibitem [{\citenamefont {Groten}\ \emph {et~al.}(2001)\citenamefont {Groten},
  \citenamefont {Baarle}, \citenamefont {Aarts}, \citenamefont {Nieuwenhuys},\
  and\ \citenamefont {Mydosh}}]{Groten01}%
  \BibitemOpen
  \bibfield  {author} {\bibinfo {author} {\bibfnamefont {D.}~\bibnamefont
  {Groten}}, \bibinfo {author} {\bibfnamefont {G.~J. C.~v.}\ \bibnamefont
  {Baarle}}, \bibinfo {author} {\bibfnamefont {J.}~\bibnamefont {Aarts}},
  \bibinfo {author} {\bibfnamefont {G.~J.}\ \bibnamefont {Nieuwenhuys}}, \ and\
  \bibinfo {author} {\bibfnamefont {J.~A.}\ \bibnamefont {Mydosh}},\ }\bibfield
   {title} {\textit {\bibinfo {title} {{Thickness dependence of the
  ground-state properties of thin films of the heavy-fermion compound
  ${\mathrm{CeCu}}_{6}$}},\ }}\href {\doibase 10.1103/PhysRevB.64.144425}
  {\bibfield  {journal} {\bibinfo  {journal} {Phys. Rev. B}\ }\textbf {\bibinfo
  {volume} {64}},\ \bibinfo {pages} {144425} (\bibinfo {year}
  {2001})}\BibitemShut {NoStop}%
\bibitem [{\citenamefont {Echevarria-Bonet}\ \emph {et~al.}(2018)\citenamefont
  {Echevarria-Bonet}, \citenamefont {Rojas}, \citenamefont {Espeso},
  \citenamefont {Rodriguez~Fern\'andez}, \citenamefont {Rodriguez~Fern\'andez},
  \citenamefont {Bauer}, \citenamefont {Burdin}, \citenamefont {Magalh\~aes},\
  and\ \citenamefont {Fern\'andez~Barquin}}]{Burdin18}%
  \BibitemOpen
  \bibfield  {author} {\bibinfo {author} {\bibfnamefont {C.}~\bibnamefont
  {Echevarria-Bonet}}, \bibinfo {author} {\bibfnamefont {D.~P.}\ \bibnamefont
  {Rojas}}, \bibinfo {author} {\bibfnamefont {J.~I.}\ \bibnamefont {Espeso}},
  \bibinfo {author} {\bibfnamefont {J.}~\bibnamefont {Rodriguez~Fern\'andez}},
  \bibinfo {author} {\bibfnamefont {L.}~\bibnamefont {Rodriguez~Fern\'andez}},
  \bibinfo {author} {\bibfnamefont {E.}~\bibnamefont {Bauer}}, \bibinfo
  {author} {\bibfnamefont {S.}~\bibnamefont {Burdin}}, \bibinfo {author}
  {\bibfnamefont {S.~G.}\ \bibnamefont {Magalh\~aes}}, \ and\ \bibinfo {author}
  {\bibfnamefont {L.}~\bibnamefont {Fern\'andez~Barquin}},\ }\bibfield  {title}
  {\textit {\bibinfo {title} {{Breakdown of the coherence effects and Fermi
  liquid behavior in $\mathrm{YbAl}_3$ nanoparticles}},\ }}\href {\doibase
  10.1088/1361-648X/aab0c7} {\bibfield  {journal} {\bibinfo  {journal} {J.
  Phys.: Condens. Matter}\ }\textbf {\bibinfo {volume} {30}},\ \bibinfo {pages}
  {135604} (\bibinfo {year} {2018})}\BibitemShut {NoStop}%
\bibitem [{\citenamefont {Shimozawa}\ \emph {et~al.}(2016)\citenamefont
  {Shimozawa}, \citenamefont {Goh}, \citenamefont {Shibauchi},\ and\
  \citenamefont {Matsuda}}]{superlatt16}%
  \BibitemOpen
  \bibfield  {author} {\bibinfo {author} {\bibfnamefont {M.}~\bibnamefont
  {Shimozawa}}, \bibinfo {author} {\bibfnamefont {S.~K.}\ \bibnamefont {Goh}},
  \bibinfo {author} {\bibfnamefont {T.}~\bibnamefont {Shibauchi}}, \ and\
  \bibinfo {author} {\bibfnamefont {Y.}~\bibnamefont {Matsuda}},\ }\bibfield
  {title} {\textit {\bibinfo {title} {{From Kondo lattices to Kondo
  superlattices}},\ }}\href {\doibase 10.1088/0034-4885/79/7/074503} {\bibfield
   {journal} {\bibinfo  {journal} {Rep. Prog. Phys.}\ }\textbf {\bibinfo
  {volume} {79}},\ \bibinfo {pages} {074503} (\bibinfo {year}
  {2016})}\BibitemShut {NoStop}%
\bibitem [{\citenamefont {Tada}\ \emph {et~al.}(2013)\citenamefont {Tada},
  \citenamefont {Peters},\ and\ \citenamefont {Oshikawa}}]{Tada13}%
  \BibitemOpen
  \bibfield  {author} {\bibinfo {author} {\bibfnamefont {Y.}~\bibnamefont
  {Tada}}, \bibinfo {author} {\bibfnamefont {R.}~\bibnamefont {Peters}}, \ and\
  \bibinfo {author} {\bibfnamefont {M.}~\bibnamefont {Oshikawa}},\ }\bibfield
  {title} {\textit {\bibinfo {title} {{Dimensional crossover in layered
  $f$-electron superlattices}},\ }}\href {\doibase 10.1103/PhysRevB.88.235121}
  {\bibfield  {journal} {\bibinfo  {journal} {Phys. Rev. B}\ }\textbf {\bibinfo
  {volume} {88}},\ \bibinfo {pages} {235121} (\bibinfo {year}
  {2013})}\BibitemShut {NoStop}%
\bibitem [{\citenamefont {Ternes}\ \emph {et~al.}(2009)\citenamefont {Ternes},
  \citenamefont {Heinrich},\ and\ \citenamefont {Schneider}}]{Ternes09}%
  \BibitemOpen
  \bibfield  {author} {\bibinfo {author} {\bibfnamefont {M.}~\bibnamefont
  {Ternes}}, \bibinfo {author} {\bibfnamefont {A.~J.}\ \bibnamefont
  {Heinrich}}, \ and\ \bibinfo {author} {\bibfnamefont {W.-D.}\ \bibnamefont
  {Schneider}},\ }\bibfield  {title} {\textit {\bibinfo {title} {{Spectroscopic
  manifestations of the Kondo effect on single adatoms}},\ }}\href {\doibase
  10.1088/0953-8984/21/5/053001} {\bibfield  {journal} {\bibinfo  {journal} {J.
  Phys.: Condens. Matter}\ }\textbf {\bibinfo {volume} {21}},\ \bibinfo {pages}
  {053001} (\bibinfo {year} {2009})}\BibitemShut {NoStop}%
\bibitem [{\citenamefont {Morr}(2017)}]{Morr17}%
  \BibitemOpen
  \bibfield  {author} {\bibinfo {author} {\bibfnamefont {D.~K.}\ \bibnamefont
  {Morr}},\ }\bibfield  {title} {\textit {\bibinfo {title} {{Theory of scanning
  tunneling spectroscopy: from Kondo impurities to heavy fermion materials}},\
  }}\href {\doibase 10.1088/0034-4885/80/1/014502} {\bibfield  {journal}
  {\bibinfo  {journal} {Rep. Prog. Phys.}\ }\textbf {\bibinfo {volume} {80}},\
  \bibinfo {pages} {014502} (\bibinfo {year} {2017})}\BibitemShut {NoStop}%
\bibitem [{\citenamefont {Madhavan}\ \emph {et~al.}(1998)\citenamefont
  {Madhavan}, \citenamefont {Chen}, \citenamefont {Jamneala}, \citenamefont
  {Crommie},\ and\ \citenamefont {Wingreen}}]{Madhavan98}%
  \BibitemOpen
  \bibfield  {author} {\bibinfo {author} {\bibfnamefont {V.}~\bibnamefont
  {Madhavan}}, \bibinfo {author} {\bibfnamefont {W.}~\bibnamefont {Chen}},
  \bibinfo {author} {\bibfnamefont {T.}~\bibnamefont {Jamneala}}, \bibinfo
  {author} {\bibfnamefont {M.~F.}\ \bibnamefont {Crommie}}, \ and\ \bibinfo
  {author} {\bibfnamefont {N.~S.}\ \bibnamefont {Wingreen}},\ }\bibfield
  {title} {\textit {\bibinfo {title} {{Tunneling into a Single Magnetic Atom:
  Spectroscopic Evidence of the Kondo Resonance}},\ }}\href {\doibase
  10.1126/science.280.5363.567} {\bibfield  {journal} {\bibinfo  {journal}
  {Science}\ }\textbf {\bibinfo {volume} {280}},\ \bibinfo {pages} {567}
  (\bibinfo {year} {1998})}\BibitemShut {NoStop}%
\bibitem [{\citenamefont {Li}\ \emph {et~al.}(1998)\citenamefont {Li},
  \citenamefont {Schneider}, \citenamefont {Berndt},\ and\ \citenamefont
  {Delley}}]{Li98}%
  \BibitemOpen
  \bibfield  {author} {\bibinfo {author} {\bibfnamefont {J.}~\bibnamefont
  {Li}}, \bibinfo {author} {\bibfnamefont {W.-D.}\ \bibnamefont {Schneider}},
  \bibinfo {author} {\bibfnamefont {R.}~\bibnamefont {Berndt}}, \ and\ \bibinfo
  {author} {\bibfnamefont {B.}~\bibnamefont {Delley}},\ }\bibfield  {title}
  {\textit {\bibinfo {title} {{Kondo Scattering Observed at a Single Magnetic
  Impurity}},\ }}\href {\doibase 10.1103/PhysRevLett.80.2893} {\bibfield
  {journal} {\bibinfo  {journal} {Phys. Rev. Lett.}\ }\textbf {\bibinfo
  {volume} {80}},\ \bibinfo {pages} {2893} (\bibinfo {year}
  {1998})}\BibitemShut {NoStop}%
\bibitem [{\citenamefont {Schmidt}\ \emph {et~al.}(2010)\citenamefont
  {Schmidt}, \citenamefont {Hamidian}, \citenamefont {Wahl}, \citenamefont
  {Meier}, \citenamefont {Balatsky}, \citenamefont {Garrett}, \citenamefont
  {Williams}, \citenamefont {Luke},\ and\ \citenamefont {Davis}}]{Schmidt10}%
  \BibitemOpen
  \bibfield  {author} {\bibinfo {author} {\bibfnamefont {A.~R.}\ \bibnamefont
  {Schmidt}}, \bibinfo {author} {\bibfnamefont {M.~H.}\ \bibnamefont
  {Hamidian}}, \bibinfo {author} {\bibfnamefont {P.}~\bibnamefont {Wahl}},
  \bibinfo {author} {\bibfnamefont {F.}~\bibnamefont {Meier}}, \bibinfo
  {author} {\bibfnamefont {A.~V.}\ \bibnamefont {Balatsky}}, \bibinfo {author}
  {\bibfnamefont {J.~D.}\ \bibnamefont {Garrett}}, \bibinfo {author}
  {\bibfnamefont {T.~J.}\ \bibnamefont {Williams}}, \bibinfo {author}
  {\bibfnamefont {G.~M.}\ \bibnamefont {Luke}}, \ and\ \bibinfo {author}
  {\bibfnamefont {J.~C.}\ \bibnamefont {Davis}},\ }\bibfield  {title} {\textit
  {\bibinfo {title} {{Imaging the Fano lattice to ‘hidden order’ transition
  in $\mathrm{URu}_2\mathrm{Si}_2$}},\ }}\href {\doibase 10.1038/nature09073}
  {\bibfield  {journal} {\bibinfo  {journal} {Nature (London)}\ }\textbf
  {\bibinfo {volume} {465}},\ \bibinfo {pages} {570} (\bibinfo {year}
  {2010})}\BibitemShut {NoStop}%
\bibitem [{\citenamefont {Ernst}\ \emph {et~al.}(2011)\citenamefont {Ernst},
  \citenamefont {Kirchner}, \citenamefont {Krellner}, \citenamefont {Geibel},
  \citenamefont {Zwicknagl}, \citenamefont {Steglich},\ and\ \citenamefont
  {Wirth}}]{Wirth11}%
  \BibitemOpen
  \bibfield  {author} {\bibinfo {author} {\bibfnamefont {S.}~\bibnamefont
  {Ernst}}, \bibinfo {author} {\bibfnamefont {S.}~\bibnamefont {Kirchner}},
  \bibinfo {author} {\bibfnamefont {C.}~\bibnamefont {Krellner}}, \bibinfo
  {author} {\bibfnamefont {C.}~\bibnamefont {Geibel}}, \bibinfo {author}
  {\bibfnamefont {G.}~\bibnamefont {Zwicknagl}}, \bibinfo {author}
  {\bibfnamefont {F.}~\bibnamefont {Steglich}}, \ and\ \bibinfo {author}
  {\bibfnamefont {S.}~\bibnamefont {Wirth}},\ }\bibfield  {title} {\textit
  {\bibinfo {title} {{Emerging local Kondo screening and spatial coherence in
  the heavy-fermion metal $\mathrm{YbRh}_2\mathrm{Si}_2$}},\ }}\href {\doibase
  10.1038/nature10148} {\bibfield  {journal} {\bibinfo  {journal} {Nature
  (London)}\ }\textbf {\bibinfo {volume} {474}},\ \bibinfo {pages} {362}
  (\bibinfo {year} {2011})}\BibitemShut {NoStop}%
\bibitem [{\citenamefont {Hamidian}\ \emph {et~al.}(2011)\citenamefont
  {Hamidian}, \citenamefont {Schmidt}, \citenamefont {Firmo}, \citenamefont
  {Allan}, \citenamefont {Bradley}, \citenamefont {Garrett}, \citenamefont
  {Williams}, \citenamefont {Luke}, \citenamefont {Dubi}, \citenamefont
  {Balatsky},\ and\ \citenamefont {Davis}}]{Hamidian11}%
  \BibitemOpen
  \bibfield  {author} {\bibinfo {author} {\bibfnamefont {M.~H.}\ \bibnamefont
  {Hamidian}}, \bibinfo {author} {\bibfnamefont {A.~R.}\ \bibnamefont
  {Schmidt}}, \bibinfo {author} {\bibfnamefont {I.~A.}\ \bibnamefont {Firmo}},
  \bibinfo {author} {\bibfnamefont {M.~P.}\ \bibnamefont {Allan}}, \bibinfo
  {author} {\bibfnamefont {P.}~\bibnamefont {Bradley}}, \bibinfo {author}
  {\bibfnamefont {J.~D.}\ \bibnamefont {Garrett}}, \bibinfo {author}
  {\bibfnamefont {T.~J.}\ \bibnamefont {Williams}}, \bibinfo {author}
  {\bibfnamefont {G.~M.}\ \bibnamefont {Luke}}, \bibinfo {author}
  {\bibfnamefont {Y.}~\bibnamefont {Dubi}}, \bibinfo {author} {\bibfnamefont
  {A.~V.}\ \bibnamefont {Balatsky}}, \ and\ \bibinfo {author} {\bibfnamefont
  {J.~C.}\ \bibnamefont {Davis}},\ }\bibfield  {title} {\textit {\bibinfo
  {title} {{How Kondo-holes create intense nanoscale heavy-fermion
  hybridization disorder}},\ }}\href {\doibase 10.1073/pnas.1115027108}
  {\bibfield  {journal} {\bibinfo  {journal} {Proc. Natl. Acad. Sci. U.S.A.}\
  }\textbf {\bibinfo {volume} {108}},\ \bibinfo {pages} {18233} (\bibinfo
  {year} {2011})}\BibitemShut {NoStop}%
\bibitem [{\citenamefont {Aynajian}\ \emph {et~al.}(2012)\citenamefont
  {Aynajian}, \citenamefont {da~Silva~Neto}, \citenamefont {Gyenis},
  \citenamefont {Baumbach}, \citenamefont {Thompson}, \citenamefont {Fisk},
  \citenamefont {Bauer},\ and\ \citenamefont {Yazdani}}]{Yazdani12}%
  \BibitemOpen
  \bibfield  {author} {\bibinfo {author} {\bibfnamefont {P.}~\bibnamefont
  {Aynajian}}, \bibinfo {author} {\bibfnamefont {E.~H.}\ \bibnamefont
  {da~Silva~Neto}}, \bibinfo {author} {\bibfnamefont {A.}~\bibnamefont
  {Gyenis}}, \bibinfo {author} {\bibfnamefont {R.~E.}\ \bibnamefont
  {Baumbach}}, \bibinfo {author} {\bibfnamefont {J.~D.}\ \bibnamefont
  {Thompson}}, \bibinfo {author} {\bibfnamefont {Z.}~\bibnamefont {Fisk}},
  \bibinfo {author} {\bibfnamefont {E.~D.}\ \bibnamefont {Bauer}}, \ and\
  \bibinfo {author} {\bibfnamefont {A.}~\bibnamefont {Yazdani}},\ }\bibfield
  {title} {\textit {\bibinfo {title} {{Visualizing heavy fermions emerging in a
  quantum critical Kondo lattice}},\ }}\href {\doibase 10.1038/nature11204}
  {\bibfield  {journal} {\bibinfo  {journal} {Nature (London)}\ }\textbf
  {\bibinfo {volume} {486}},\ \bibinfo {pages} {201} (\bibinfo {year}
  {2012})}\BibitemShut {NoStop}%
\bibitem [{\citenamefont {Zhou}\ \emph {et~al.}(2010)\citenamefont {Zhou},
  \citenamefont {Wiebe}, \citenamefont {Lounis}, \citenamefont {Vedmedenko},
  \citenamefont {Meier}, \citenamefont {Bl\"ugel}, \citenamefont {Dederichs},\
  and\ \citenamefont {Wiesendanger}}]{Wiebe10}%
  \BibitemOpen
  \bibfield  {author} {\bibinfo {author} {\bibfnamefont {L.}~\bibnamefont
  {Zhou}}, \bibinfo {author} {\bibfnamefont {J.}~\bibnamefont {Wiebe}},
  \bibinfo {author} {\bibfnamefont {S.}~\bibnamefont {Lounis}}, \bibinfo
  {author} {\bibfnamefont {E.}~\bibnamefont {Vedmedenko}}, \bibinfo {author}
  {\bibfnamefont {F.}~\bibnamefont {Meier}}, \bibinfo {author} {\bibfnamefont
  {S.}~\bibnamefont {Bl\"ugel}}, \bibinfo {author} {\bibfnamefont {P.~H.}\
  \bibnamefont {Dederichs}}, \ and\ \bibinfo {author} {\bibfnamefont
  {R.}~\bibnamefont {Wiesendanger}},\ }\bibfield  {title} {\textit {\bibinfo
  {title} {{Strength and directionality of surface
  Ruderman-Kittel-Kasuya-Yosida interaction mapped on the atomic scale}},\
  }}\href {\doibase 10.1038/nphys1514} {\bibfield  {journal} {\bibinfo
  {journal} {Nat. Phys.}\ }\textbf {\bibinfo {volume} {6}},\ \bibinfo {pages}
  {187} (\bibinfo {year} {2010})}\BibitemShut {NoStop}%
\bibitem [{\citenamefont {Pr\"user}\ \emph {et~al.}(2014)\citenamefont
  {Pr\"user}, \citenamefont {Dargel}, \citenamefont {Bouhassoune},
  \citenamefont {Ulbrich}, \citenamefont {Pruschke}, \citenamefont {Lounis},\
  and\ \citenamefont {Wenderoth}}]{Pruser14}%
  \BibitemOpen
  \bibfield  {author} {\bibinfo {author} {\bibfnamefont {H.}~\bibnamefont
  {Pr\"user}}, \bibinfo {author} {\bibfnamefont {P.~E.}\ \bibnamefont
  {Dargel}}, \bibinfo {author} {\bibfnamefont {M.}~\bibnamefont {Bouhassoune}},
  \bibinfo {author} {\bibfnamefont {R.~G.}\ \bibnamefont {Ulbrich}}, \bibinfo
  {author} {\bibfnamefont {T.}~\bibnamefont {Pruschke}}, \bibinfo {author}
  {\bibfnamefont {S.}~\bibnamefont {Lounis}}, \ and\ \bibinfo {author}
  {\bibfnamefont {M.}~\bibnamefont {Wenderoth}},\ }\bibfield  {title} {\textit
  {\bibinfo {title} {{Interplay between the Kondo effect and the
  Ruderman-Kittel-Kasuya-Yosida interaction}},\ }}\href {\doibase
  10.1038/ncomms6417} {\bibfield  {journal} {\bibinfo  {journal} {Nat.
  Commun.}\ }\textbf {\bibinfo {volume} {5}},\ \bibinfo {pages} {5417}
  (\bibinfo {year} {2014})}\BibitemShut {NoStop}%
\bibitem [{\citenamefont {Spinelli}\ \emph {et~al.}(2015)\citenamefont
  {Spinelli}, \citenamefont {Gerrits}, \citenamefont {Toskovic}, \citenamefont
  {Bryant}, \citenamefont {Ternes},\ and\ \citenamefont {Otte}}]{Spinelli15}%
  \BibitemOpen
  \bibfield  {author} {\bibinfo {author} {\bibfnamefont {A.}~\bibnamefont
  {Spinelli}}, \bibinfo {author} {\bibfnamefont {M.}~\bibnamefont {Gerrits}},
  \bibinfo {author} {\bibfnamefont {R.}~\bibnamefont {Toskovic}}, \bibinfo
  {author} {\bibfnamefont {B.}~\bibnamefont {Bryant}}, \bibinfo {author}
  {\bibfnamefont {M.}~\bibnamefont {Ternes}}, \ and\ \bibinfo {author}
  {\bibfnamefont {A.~F.}\ \bibnamefont {Otte}},\ }\bibfield  {title} {\textit
  {\bibinfo {title} {{Exploring the phase diagram of the two-impurity Kondo
  problem}},\ }}\href {\doibase 10.1038/ncomms10046} {\bibfield  {journal}
  {\bibinfo  {journal} {Nat. Commun.}\ }\textbf {\bibinfo {volume} {6}},\
  \bibinfo {pages} {10046} (\bibinfo {year} {2015})}\BibitemShut {NoStop}%
\bibitem [{\citenamefont {Allerdt}\ \emph {et~al.}(2015)\citenamefont
  {Allerdt}, \citenamefont {B\"usser}, \citenamefont {Martins},\ and\
  \citenamefont {Feiguin}}]{Carlos15}%
  \BibitemOpen
  \bibfield  {author} {\bibinfo {author} {\bibfnamefont {A.}~\bibnamefont
  {Allerdt}}, \bibinfo {author} {\bibfnamefont {C.~A.}\ \bibnamefont
  {B\"usser}}, \bibinfo {author} {\bibfnamefont {G.~B.}\ \bibnamefont
  {Martins}}, \ and\ \bibinfo {author} {\bibfnamefont {A.~E.}\ \bibnamefont
  {Feiguin}},\ }\bibfield  {title} {\textit {\bibinfo {title} {{Kondo versus
  indirect exchange: Role of lattice and actual range of RKKY interactions in
  real materials}},\ }}\href {\doibase 10.1103/PhysRevB.91.085101} {\bibfield
  {journal} {\bibinfo  {journal} {Phys. Rev. B}\ }\textbf {\bibinfo {volume}
  {91}},\ \bibinfo {pages} {085101} (\bibinfo {year} {2015})}\BibitemShut
  {NoStop}%
\bibitem [{\citenamefont {Mitchell}\ and\ \citenamefont
  {Bulla}(2015)}]{Bulla15}%
  \BibitemOpen
  \bibfield  {author} {\bibinfo {author} {\bibfnamefont {A.~K.}\ \bibnamefont
  {Mitchell}}\ and\ \bibinfo {author} {\bibfnamefont {R.}~\bibnamefont
  {Bulla}},\ }\bibfield  {title} {\textit {\bibinfo {title} {{Validity of the
  local self-energy approximation: Application to coupled quantum
  impurities}},\ }}\href {\doibase 10.1103/PhysRevB.92.155101} {\bibfield
  {journal} {\bibinfo  {journal} {Phys. Rev. B}\ }\textbf {\bibinfo {volume}
  {92}},\ \bibinfo {pages} {155101} (\bibinfo {year} {2015})}\BibitemShut
  {NoStop}%
\bibitem [{\citenamefont {Savkin}\ \emph {et~al.}(2005)\citenamefont {Savkin},
  \citenamefont {Rubtsov}, \citenamefont {Katsnelson},\ and\ \citenamefont
  {Lichtenstein}}]{Savkin05}%
  \BibitemOpen
  \bibfield  {author} {\bibinfo {author} {\bibfnamefont {V.~V.}\ \bibnamefont
  {Savkin}}, \bibinfo {author} {\bibfnamefont {A.~N.}\ \bibnamefont {Rubtsov}},
  \bibinfo {author} {\bibfnamefont {M.~I.}\ \bibnamefont {Katsnelson}}, \ and\
  \bibinfo {author} {\bibfnamefont {A.~I.}\ \bibnamefont {Lichtenstein}},\
  }\bibfield  {title} {\textit {\bibinfo {title} {{Correlated Adatom Trimer on
  a Metal Surface: A Continuous-Time Quantum Monte Carlo Study}},\ }}\href
  {\doibase 10.1103/PhysRevLett.94.026402} {\bibfield  {journal} {\bibinfo
  {journal} {Phys. Rev. Lett.}\ }\textbf {\bibinfo {volume} {94}},\ \bibinfo
  {pages} {026402} (\bibinfo {year} {2005})}\BibitemShut {NoStop}%
\bibitem [{\citenamefont {DiLullo}\ \emph {et~al.}(2012)\citenamefont
  {DiLullo}, \citenamefont {Chang}, \citenamefont {Baadji}, \citenamefont
  {Clark}, \citenamefont {Klöckner}, \citenamefont {Prosenc}, \citenamefont
  {Sanvito}, \citenamefont {Wiesendanger}, \citenamefont {Hoffmann},\ and\
  \citenamefont {Hla}}]{DiLullo12}%
  \BibitemOpen
  \bibfield  {author} {\bibinfo {author} {\bibfnamefont {A.}~\bibnamefont
  {DiLullo}}, \bibinfo {author} {\bibfnamefont {S.-H.}\ \bibnamefont {Chang}},
  \bibinfo {author} {\bibfnamefont {N.}~\bibnamefont {Baadji}}, \bibinfo
  {author} {\bibfnamefont {K.}~\bibnamefont {Clark}}, \bibinfo {author}
  {\bibfnamefont {J.-P.}\ \bibnamefont {Klöckner}}, \bibinfo {author}
  {\bibfnamefont {M.-H.}\ \bibnamefont {Prosenc}}, \bibinfo {author}
  {\bibfnamefont {S.}~\bibnamefont {Sanvito}}, \bibinfo {author} {\bibfnamefont
  {R.}~\bibnamefont {Wiesendanger}}, \bibinfo {author} {\bibfnamefont
  {G.}~\bibnamefont {Hoffmann}}, \ and\ \bibinfo {author} {\bibfnamefont
  {S.-W.}\ \bibnamefont {Hla}},\ }\bibfield  {title} {\textit {\bibinfo {title}
  {{Molecular Kondo Chain}},\ }}\href {\doibase 10.1021/nl301149d} {\bibfield
  {journal} {\bibinfo  {journal} {Nano Lett.}\ }\textbf {\bibinfo {volume}
  {12}},\ \bibinfo {pages} {3174} (\bibinfo {year} {2012})}\BibitemShut
  {NoStop}%
\bibitem [{\citenamefont {Lobos}\ \emph {et~al.}(2012)\citenamefont {Lobos},
  \citenamefont {Cazalilla},\ and\ \citenamefont {Chudzinski}}]{Chudzinski12}%
  \BibitemOpen
  \bibfield  {author} {\bibinfo {author} {\bibfnamefont {A.~M.}\ \bibnamefont
  {Lobos}}, \bibinfo {author} {\bibfnamefont {M.~A.}\ \bibnamefont
  {Cazalilla}}, \ and\ \bibinfo {author} {\bibfnamefont {P.}~\bibnamefont
  {Chudzinski}},\ }\bibfield  {title} {\textit {\bibinfo {title} {{Magnetic
  phases in the one-dimensional Kondo chain on a metallic surface}},\ }}\href
  {\doibase 10.1103/PhysRevB.86.035455} {\bibfield  {journal} {\bibinfo
  {journal} {Phys. Rev. B}\ }\textbf {\bibinfo {volume} {86}},\ \bibinfo
  {pages} {035455} (\bibinfo {year} {2012})}\BibitemShut {NoStop}%
\bibitem [{\citenamefont {Mitchell}\ \emph {et~al.}(2015)\citenamefont
  {Mitchell}, \citenamefont {Derry},\ and\ \citenamefont {Logan}}]{Mitchell15}%
  \BibitemOpen
  \bibfield  {author} {\bibinfo {author} {\bibfnamefont {A.~K.}\ \bibnamefont
  {Mitchell}}, \bibinfo {author} {\bibfnamefont {P.~G.}\ \bibnamefont {Derry}},
  \ and\ \bibinfo {author} {\bibfnamefont {D.~E.}\ \bibnamefont {Logan}},\
  }\bibfield  {title} {\textit {\bibinfo {title} {{Multiple magnetic impurities
  on surfaces: Scattering and quasiparticle interference}},\ }}\href {\doibase
  10.1103/PhysRevB.91.235127} {\bibfield  {journal} {\bibinfo  {journal} {Phys.
  Rev. B}\ }\textbf {\bibinfo {volume} {91}},\ \bibinfo {pages} {235127}
  (\bibinfo {year} {2015})}\BibitemShut {NoStop}%
\bibitem [{\citenamefont {Tsunetsugu}\ \emph {et~al.}(1997)\citenamefont
  {Tsunetsugu}, \citenamefont {Sigrist},\ and\ \citenamefont {Ueda}}]{Ueda97}%
  \BibitemOpen
  \bibfield  {author} {\bibinfo {author} {\bibfnamefont {H.}~\bibnamefont
  {Tsunetsugu}}, \bibinfo {author} {\bibfnamefont {M.}~\bibnamefont {Sigrist}},
  \ and\ \bibinfo {author} {\bibfnamefont {K.}~\bibnamefont {Ueda}},\
  }\bibfield  {title} {\textit {\bibinfo {title} {{The ground-state phase
  diagram of the one-dimensional Kondo lattice model}},\ }}\href {\doibase
  10.1103/RevModPhys.69.809} {\bibfield  {journal} {\bibinfo  {journal} {Rev.
  Mod. Phys.}\ }\textbf {\bibinfo {volume} {69}},\ \bibinfo {pages} {809}
  (\bibinfo {year} {1997})}\BibitemShut {NoStop}%
\bibitem [{\citenamefont {Assaad}(1999)}]{Assaad99}%
  \BibitemOpen
  \bibfield  {author} {\bibinfo {author} {\bibfnamefont {F.~F.}\ \bibnamefont
  {Assaad}},\ }\bibfield  {title} {\textit {\bibinfo {title} {{Quantum Monte
  Carlo Simulations of the Half-Filled Two-Dimensional Kondo Lattice Model}},\
  }}\href {\doibase 10.1103/PhysRevLett.83.796} {\bibfield  {journal} {\bibinfo
   {journal} {Phys. Rev. Lett.}\ }\textbf {\bibinfo {volume} {83}},\ \bibinfo
  {pages} {796} (\bibinfo {year} {1999})}\BibitemShut {NoStop}%
\bibitem [{\citenamefont {Bercx}\ \emph {et~al.}(2017)\citenamefont {Bercx},
  \citenamefont {Goth}, \citenamefont {Hofmann},\ and\ \citenamefont
  {Assaad}}]{alf}%
  \BibitemOpen
  \bibfield  {author} {\bibinfo {author} {\bibfnamefont {M.}~\bibnamefont
  {Bercx}}, \bibinfo {author} {\bibfnamefont {F.}~\bibnamefont {Goth}},
  \bibinfo {author} {\bibfnamefont {J.~S.}\ \bibnamefont {Hofmann}}, \ and\
  \bibinfo {author} {\bibfnamefont {F.~F.}\ \bibnamefont {Assaad}},\ }\bibfield
   {title} {\textit {\bibinfo {title} {{The ALF (Algorithms for Lattice
  Fermions) project release 1.0. Documentation for the auxiliary field quantum
  Monte Carlo code}},\ }}\href {\doibase 10.21468/SciPostPhys.3.2.013}
  {\bibfield  {journal} {\bibinfo  {journal} {SciPost Phys.}\ }\textbf
  {\bibinfo {volume} {3}},\ \bibinfo {pages} {013} (\bibinfo {year}
  {2017})}\BibitemShut {NoStop}%
\bibitem [{\citenamefont {Tsukahara}\ \emph {et~al.}(2011)\citenamefont
  {Tsukahara}, \citenamefont {Shiraki}, \citenamefont {Itou}, \citenamefont
  {Ohta}, \citenamefont {Takagi},\ and\ \citenamefont {Kawai}}]{Tsukahara11}%
  \BibitemOpen
  \bibfield  {author} {\bibinfo {author} {\bibfnamefont {N.}~\bibnamefont
  {Tsukahara}}, \bibinfo {author} {\bibfnamefont {S.}~\bibnamefont {Shiraki}},
  \bibinfo {author} {\bibfnamefont {S.}~\bibnamefont {Itou}}, \bibinfo {author}
  {\bibfnamefont {N.}~\bibnamefont {Ohta}}, \bibinfo {author} {\bibfnamefont
  {N.}~\bibnamefont {Takagi}}, \ and\ \bibinfo {author} {\bibfnamefont
  {M.}~\bibnamefont {Kawai}},\ }\bibfield  {title} {\textit {\bibinfo {title}
  {{Evolution of Kondo Resonance from a Single Impurity Molecule to the
  Two-Dimensional Lattice}},\ }}\href {\doibase 10.1103/PhysRevLett.106.187201}
  {\bibfield  {journal} {\bibinfo  {journal} {Phys. Rev. Lett.}\ }\textbf
  {\bibinfo {volume} {106}},\ \bibinfo {pages} {187201} (\bibinfo {year}
  {2011})}\BibitemShut {NoStop}%
\bibitem [{\citenamefont {Lobos}\ \emph {et~al.}(2014)\citenamefont {Lobos},
  \citenamefont {Romero},\ and\ \citenamefont {Aligia}}]{Lobos14}%
  \BibitemOpen
  \bibfield  {author} {\bibinfo {author} {\bibfnamefont {A.~M.}\ \bibnamefont
  {Lobos}}, \bibinfo {author} {\bibfnamefont {M.}~\bibnamefont {Romero}}, \
  and\ \bibinfo {author} {\bibfnamefont {A.~A.}\ \bibnamefont {Aligia}},\
  }\bibfield  {title} {\textit {\bibinfo {title} {{Spectral evolution of the
  SU(4) Kondo effect from the single impurity to the two-dimensional limit}},\
  }}\href {\doibase 10.1103/PhysRevB.89.121406} {\bibfield  {journal} {\bibinfo
   {journal} {Phys. Rev. B}\ }\textbf {\bibinfo {volume} {89}},\ \bibinfo
  {pages} {121406} (\bibinfo {year} {2014})}\BibitemShut {NoStop}%
\bibitem [{\citenamefont {{Beach}}(2004)}]{Beach04a}%
  \BibitemOpen
  \bibfield  {author} {\bibinfo {author} {\bibfnamefont {K.~S.~D.}\
  \bibnamefont {{Beach}}},\ }\bibfield  {title} {\textit {\bibinfo {title}
  {{{Identifying the maximum entropy method as a special limit of stochastic
  analytic continuation}}},\ }}\href@noop {} {\bibfield  {journal} {\bibinfo
  {journal} {arXiv e-prints}\ } (\bibinfo {year} {2004})},\ \Eprint
  {http://arxiv.org/abs/cond-mat/0403055} {cond-mat/0403055} \BibitemShut
  {NoStop}%
\bibitem [{\citenamefont {Capponi}\ and\ \citenamefont
  {Assaad}(2001)}]{Capponi01}%
  \BibitemOpen
  \bibfield  {author} {\bibinfo {author} {\bibfnamefont {S.}~\bibnamefont
  {Capponi}}\ and\ \bibinfo {author} {\bibfnamefont {F.~F.}\ \bibnamefont
  {Assaad}},\ }\bibfield  {title} {\textit {\bibinfo {title} {{Spin and charge
  dynamics of the ferromagnetic and antiferromagnetic two-dimensional
  half-filled Kondo lattice model}},\ }}\href {\doibase
  10.1103/PhysRevB.63.155114} {\bibfield  {journal} {\bibinfo  {journal} {Phys.
  Rev. B}\ }\textbf {\bibinfo {volume} {63}},\ \bibinfo {pages} {155114}
  (\bibinfo {year} {2001})}\BibitemShut {NoStop}%
\bibitem [{\citenamefont {Schrieffer}\ and\ \citenamefont
  {Wolff}(1966)}]{SW_transf}%
  \BibitemOpen
  \bibfield  {author} {\bibinfo {author} {\bibfnamefont {J.~R.}\ \bibnamefont
  {Schrieffer}}\ and\ \bibinfo {author} {\bibfnamefont {P.~A.}\ \bibnamefont
  {Wolff}},\ }\bibfield  {title} {\textit {\bibinfo {title} {{Relation between
  the Anderson and Kondo Hamiltonians}},\ }}\href {\doibase
  10.1103/PhysRev.149.491} {\bibfield  {journal} {\bibinfo  {journal} {Phys.
  Rev.}\ }\textbf {\bibinfo {volume} {149}},\ \bibinfo {pages} {491} (\bibinfo
  {year} {1966})}\BibitemShut {NoStop}%
\bibitem [{SM()}]{SM}%
  \BibitemOpen
  \href@noop {} {\bibinfo  {journal} {See Supplemental Material, which includes
  Ref.~\cite{Costi00}, for a derivation of the $\tilde{f}$-operator and the
  additional numerical evidence of the emergent coherence in Kondo
  superlattices}\ }\BibitemShut {NoStop}%
\bibitem [{\citenamefont {Maltseva}\ \emph {et~al.}(2009)\citenamefont
  {Maltseva}, \citenamefont {Dzero},\ and\ \citenamefont {Coleman}}]{Dzero09}%
  \BibitemOpen
\bibfield  {journal} {  }\bibfield  {author} {\bibinfo {author} {\bibfnamefont
  {M.}~\bibnamefont {Maltseva}}, \bibinfo {author} {\bibfnamefont
  {M.}~\bibnamefont {Dzero}}, \ and\ \bibinfo {author} {\bibfnamefont
  {P.}~\bibnamefont {Coleman}},\ }\bibfield  {title} {\textit {\bibinfo {title}
  {{Electron Cotunneling into a Kondo Lattice}},\ }}\href {\doibase
  10.1103/PhysRevLett.103.206402} {\bibfield  {journal} {\bibinfo  {journal}
  {Phys. Rev. Lett.}\ }\textbf {\bibinfo {volume} {103}},\ \bibinfo {pages}
  {206402} (\bibinfo {year} {2009})}\BibitemShut {NoStop}%
\bibitem [{\citenamefont {Figgins}\ and\ \citenamefont {Morr}(2010)}]{Morr10}%
  \BibitemOpen
  \bibfield  {author} {\bibinfo {author} {\bibfnamefont {J.}~\bibnamefont
  {Figgins}}\ and\ \bibinfo {author} {\bibfnamefont {D.~K.}\ \bibnamefont
  {Morr}},\ }\bibfield  {title} {\textit {\bibinfo {title} {{Differential
  Conductance and Quantum Interference in Kondo Systems}},\ }}\href {\doibase
  10.1103/PhysRevLett.104.187202} {\bibfield  {journal} {\bibinfo  {journal}
  {Phys. Rev. Lett.}\ }\textbf {\bibinfo {volume} {104}},\ \bibinfo {pages}
  {187202} (\bibinfo {year} {2010})}\BibitemShut {NoStop}%
\bibitem [{\citenamefont {W\"olfle}\ \emph {et~al.}(2010)\citenamefont
  {W\"olfle}, \citenamefont {Dubi},\ and\ \citenamefont {Balatsky}}]{Wolfle10}%
  \BibitemOpen
  \bibfield  {author} {\bibinfo {author} {\bibfnamefont {P.}~\bibnamefont
  {W\"olfle}}, \bibinfo {author} {\bibfnamefont {Y.}~\bibnamefont {Dubi}}, \
  and\ \bibinfo {author} {\bibfnamefont {A.~V.}\ \bibnamefont {Balatsky}},\
  }\bibfield  {title} {\textit {\bibinfo {title} {{Tunneling into Clean Heavy
  Fermion Compounds: Origin of the Fano Line Shape}},\ }}\href {\doibase
  10.1103/PhysRevLett.105.246401} {\bibfield  {journal} {\bibinfo  {journal}
  {Phys. Rev. Lett.}\ }\textbf {\bibinfo {volume} {105}},\ \bibinfo {pages}
  {246401} (\bibinfo {year} {2010})}\BibitemShut {NoStop}%
\bibitem [{\citenamefont {Zhu}\ and\ \citenamefont {Zhu}(2011)}]{Zhu11}%
  \BibitemOpen
  \bibfield  {author} {\bibinfo {author} {\bibfnamefont {L.}~\bibnamefont
  {Zhu}}\ and\ \bibinfo {author} {\bibfnamefont {J.-X.}\ \bibnamefont {Zhu}},\
  }\bibfield  {title} {\textit {\bibinfo {title} {{Coherence scale of coupled
  Anderson impurities}},\ }}\href {\doibase 10.1103/PhysRevB.83.195103}
  {\bibfield  {journal} {\bibinfo  {journal} {Phys. Rev. B}\ }\textbf {\bibinfo
  {volume} {83}},\ \bibinfo {pages} {195103} (\bibinfo {year}
  {2011})}\BibitemShut {NoStop}%
\bibitem [{\citenamefont {Jabben}\ \emph {et~al.}(2012)\citenamefont {Jabben},
  \citenamefont {Grewe},\ and\ \citenamefont {Schmitt}}]{Grewe12}%
  \BibitemOpen
  \bibfield  {author} {\bibinfo {author} {\bibfnamefont {T.}~\bibnamefont
  {Jabben}}, \bibinfo {author} {\bibfnamefont {N.}~\bibnamefont {Grewe}}, \
  and\ \bibinfo {author} {\bibfnamefont {S.}~\bibnamefont {Schmitt}},\
  }\bibfield  {title} {\textit {\bibinfo {title} {{Spectral properties of the
  two-impurity Anderson model with varying distance and various
  interactions}},\ }}\href {\doibase 10.1103/PhysRevB.85.045133} {\bibfield
  {journal} {\bibinfo  {journal} {Phys. Rev. B}\ }\textbf {\bibinfo {volume}
  {85}},\ \bibinfo {pages} {045133} (\bibinfo {year} {2012})}\BibitemShut
  {NoStop}%
\bibitem [{\citenamefont {Wang}\ \emph {et~al.}(2015)\citenamefont {Wang},
  \citenamefont {Shinaoka},\ and\ \citenamefont {Troyer}}]{Wang15}%
  \BibitemOpen
  \bibfield  {author} {\bibinfo {author} {\bibfnamefont {L.}~\bibnamefont
  {Wang}}, \bibinfo {author} {\bibfnamefont {H.}~\bibnamefont {Shinaoka}}, \
  and\ \bibinfo {author} {\bibfnamefont {M.}~\bibnamefont {Troyer}},\
  }\bibfield  {title} {\textit {\bibinfo {title} {{Fidelity Susceptibility
  Perspective on the Kondo Effect and Impurity Quantum Phase Transitions}},\
  }}\href {\doibase 10.1103/PhysRevLett.115.236601} {\bibfield  {journal}
  {\bibinfo  {journal} {Phys. Rev. Lett.}\ }\textbf {\bibinfo {volume} {115}},\
  \bibinfo {pages} {236601} (\bibinfo {year} {2015})}\BibitemShut {NoStop}%
\bibitem [{\citenamefont {Veki\ifmmode~\acute{c}\else \'{c}\fi{}}\ \emph
  {et~al.}(1995)\citenamefont {Veki\ifmmode~\acute{c}\else \'{c}\fi{}},
  \citenamefont {Cannon}, \citenamefont {Scalapino}, \citenamefont
  {Scalettar},\ and\ \citenamefont {Sugar}}]{Vekic95}%
  \BibitemOpen
  \bibfield  {author} {\bibinfo {author} {\bibfnamefont {M.}~\bibnamefont
  {Veki\ifmmode~\acute{c}\else \'{c}\fi{}}}, \bibinfo {author} {\bibfnamefont
  {J.~W.}\ \bibnamefont {Cannon}}, \bibinfo {author} {\bibfnamefont {D.~J.}\
  \bibnamefont {Scalapino}}, \bibinfo {author} {\bibfnamefont {R.~T.}\
  \bibnamefont {Scalettar}}, \ and\ \bibinfo {author} {\bibfnamefont {R.~L.}\
  \bibnamefont {Sugar}},\ }\bibfield  {title} {\textit {\bibinfo {title}
  {{Competition between Antiferromagnetic Order and Spin-Liquid Behavior in the
  Two-Dimensional Periodic Anderson Model at Half Filling}},\ }}\href {\doibase
  10.1103/PhysRevLett.74.2367} {\bibfield  {journal} {\bibinfo  {journal}
  {Phys. Rev. Lett.}\ }\textbf {\bibinfo {volume} {74}},\ \bibinfo {pages}
  {2367} (\bibinfo {year} {1995})}\BibitemShut {NoStop}%
\bibitem [{\citenamefont {Vidhyadhiraja}\ \emph {et~al.}(2003)\citenamefont
  {Vidhyadhiraja}, \citenamefont {Smith}, \citenamefont {Logan},\ and\
  \citenamefont {Krishnamurthy}}]{Logan03}%
  \BibitemOpen
  \bibfield  {author} {\bibinfo {author} {\bibfnamefont {N.~S.}\ \bibnamefont
  {Vidhyadhiraja}}, \bibinfo {author} {\bibfnamefont {V.~E.}\ \bibnamefont
  {Smith}}, \bibinfo {author} {\bibfnamefont {D.~E.}\ \bibnamefont {Logan}}, \
  and\ \bibinfo {author} {\bibfnamefont {H.~R.}\ \bibnamefont
  {Krishnamurthy}},\ }\bibfield  {title} {\textit {\bibinfo {title} {{Dynamics
  and transport properties of Kondo insulators}},\ }}\href {\doibase
  10.1088/0953-8984/15/24/301} {\bibfield  {journal} {\bibinfo  {journal} {J.
  Phys.: Condens. Matter}\ }\textbf {\bibinfo {volume} {15}},\ \bibinfo {pages}
  {4045} (\bibinfo {year} {2003})}\BibitemShut {NoStop}%
\bibitem [{\citenamefont {Peters}\ \emph {et~al.}(2013)\citenamefont {Peters},
  \citenamefont {Tada},\ and\ \citenamefont {Kawakami}}]{Peters13}%
  \BibitemOpen
  \bibfield  {author} {\bibinfo {author} {\bibfnamefont {R.}~\bibnamefont
  {Peters}}, \bibinfo {author} {\bibfnamefont {Y.}~\bibnamefont {Tada}}, \ and\
  \bibinfo {author} {\bibfnamefont {N.}~\bibnamefont {Kawakami}},\ }\bibfield
  {title} {\textit {\bibinfo {title} {{Kondo effect in $f$-electron
  superlattices}},\ }}\href {\doibase 10.1103/PhysRevB.88.155134} {\bibfield
  {journal} {\bibinfo  {journal} {Phys. Rev. B}\ }\textbf {\bibinfo {volume}
  {88}},\ \bibinfo {pages} {155134} (\bibinfo {year} {2013})}\BibitemShut
  {NoStop}%
\bibitem [{\citenamefont {Gr\"ober}\ and\ \citenamefont {Eder}(1998)}]{Eder98}%
  \BibitemOpen
  \bibfield  {author} {\bibinfo {author} {\bibfnamefont {C.}~\bibnamefont
  {Gr\"ober}}\ and\ \bibinfo {author} {\bibfnamefont {R.}~\bibnamefont
  {Eder}},\ }\bibfield  {title} {\textit {\bibinfo {title}
  {{Temperature-dependent band structure of the Kondo insulator}},\ }}\href
  {\doibase 10.1103/PhysRevB.57.R12659} {\bibfield  {journal} {\bibinfo
  {journal} {Phys. Rev. B}\ }\textbf {\bibinfo {volume} {57}},\ \bibinfo
  {pages} {R12659} (\bibinfo {year} {1998})}\BibitemShut {NoStop}%
\bibitem [{\citenamefont {Shim}\ \emph {et~al.}(2007)\citenamefont {Shim},
  \citenamefont {Haule},\ and\ \citenamefont {Kotliar}}]{Shim07}%
  \BibitemOpen
  \bibfield  {author} {\bibinfo {author} {\bibfnamefont {J.~H.}\ \bibnamefont
  {Shim}}, \bibinfo {author} {\bibfnamefont {K.}~\bibnamefont {Haule}}, \ and\
  \bibinfo {author} {\bibfnamefont {G.}~\bibnamefont {Kotliar}},\ }\bibfield
  {title} {\textit {\bibinfo {title} {{Modeling the Localized-to-Itinerant
  Electronic Transition in the Heavy Fermion System $\mathrm{CeIrIn}_5$ }},\
  }}\href {\doibase 10.1126/science.1149064} {\bibfield  {journal} {\bibinfo
  {journal} {Science}\ }\textbf {\bibinfo {volume} {318}},\ \bibinfo {pages}
  {1615} (\bibinfo {year} {2007})}\BibitemShut {NoStop}%
\bibitem [{\citenamefont {Martin}\ and\ \citenamefont
  {Assaad}(2008)}]{Martin08}%
  \BibitemOpen
  \bibfield  {author} {\bibinfo {author} {\bibfnamefont {L.~C.}\ \bibnamefont
  {Martin}}\ and\ \bibinfo {author} {\bibfnamefont {F.~F.}\ \bibnamefont
  {Assaad}},\ }\bibfield  {title} {\textit {\bibinfo {title} {{Evolution of the
  Fermi Surface across a Magnetic Order-Disorder Transition in the
  Two-Dimensional Kondo Lattice Model: A Dynamical Cluster Approach}},\ }}\href
  {\doibase 10.1103/PhysRevLett.101.066404} {\bibfield  {journal} {\bibinfo
  {journal} {Phys. Rev. Lett.}\ }\textbf {\bibinfo {volume} {101}},\ \bibinfo
  {pages} {066404} (\bibinfo {year} {2008})}\BibitemShut {NoStop}%
\bibitem [{\citenamefont {Otsuki}\ \emph {et~al.}(2009)\citenamefont {Otsuki},
  \citenamefont {Kusunose},\ and\ \citenamefont {Kuramoto}}]{Otsuki09}%
  \BibitemOpen
  \bibfield  {author} {\bibinfo {author} {\bibfnamefont {J.}~\bibnamefont
  {Otsuki}}, \bibinfo {author} {\bibfnamefont {H.}~\bibnamefont {Kusunose}}, \
  and\ \bibinfo {author} {\bibfnamefont {Y.}~\bibnamefont {Kuramoto}},\
  }\bibfield  {title} {\textit {\bibinfo {title} {{Evolution of a Large Fermi
  Surface in the Kondo Lattice}},\ }}\href {\doibase
  10.1103/PhysRevLett.102.017202} {\bibfield  {journal} {\bibinfo  {journal}
  {Phys. Rev. Lett.}\ }\textbf {\bibinfo {volume} {102}},\ \bibinfo {pages}
  {017202} (\bibinfo {year} {2009})}\BibitemShut {NoStop}%
\bibitem [{\citenamefont {Benlagra}\ \emph {et~al.}(2011)\citenamefont
  {Benlagra}, \citenamefont {Pruschke},\ and\ \citenamefont
  {Vojta}}]{Benlagra11}%
  \BibitemOpen
  \bibfield  {author} {\bibinfo {author} {\bibfnamefont {A.}~\bibnamefont
  {Benlagra}}, \bibinfo {author} {\bibfnamefont {T.}~\bibnamefont {Pruschke}},
  \ and\ \bibinfo {author} {\bibfnamefont {M.}~\bibnamefont {Vojta}},\
  }\bibfield  {title} {\textit {\bibinfo {title} {{Finite-temperature spectra
  and quasiparticle interference in Kondo lattices: From light electrons to
  coherent heavy quasiparticles}},\ }}\href {\doibase
  10.1103/PhysRevB.84.195141} {\bibfield  {journal} {\bibinfo  {journal} {Phys.
  Rev. B}\ }\textbf {\bibinfo {volume} {84}},\ \bibinfo {pages} {195141}
  (\bibinfo {year} {2011})}\BibitemShut {NoStop}%
\bibitem [{\citenamefont {Klein}\ \emph {et~al.}(2011)\citenamefont {Klein},
  \citenamefont {Nuber}, \citenamefont {Schwab}, \citenamefont {Albers},
  \citenamefont {Tobita}, \citenamefont {Higashiguchi}, \citenamefont {Jiang},
  \citenamefont {Fukuda}, \citenamefont {Tanaka}, \citenamefont {Shimada},
  \citenamefont {Mulazzi}, \citenamefont {Assaad},\ and\ \citenamefont
  {Reinert}}]{Klein11}%
  \BibitemOpen
  \bibfield  {author} {\bibinfo {author} {\bibfnamefont {M.}~\bibnamefont
  {Klein}}, \bibinfo {author} {\bibfnamefont {A.}~\bibnamefont {Nuber}},
  \bibinfo {author} {\bibfnamefont {H.}~\bibnamefont {Schwab}}, \bibinfo
  {author} {\bibfnamefont {C.}~\bibnamefont {Albers}}, \bibinfo {author}
  {\bibfnamefont {N.}~\bibnamefont {Tobita}}, \bibinfo {author} {\bibfnamefont
  {M.}~\bibnamefont {Higashiguchi}}, \bibinfo {author} {\bibfnamefont
  {J.}~\bibnamefont {Jiang}}, \bibinfo {author} {\bibfnamefont
  {S.}~\bibnamefont {Fukuda}}, \bibinfo {author} {\bibfnamefont
  {K.}~\bibnamefont {Tanaka}}, \bibinfo {author} {\bibfnamefont
  {K.}~\bibnamefont {Shimada}}, \bibinfo {author} {\bibfnamefont
  {M.}~\bibnamefont {Mulazzi}}, \bibinfo {author} {\bibfnamefont {F.~F.}\
  \bibnamefont {Assaad}}, \ and\ \bibinfo {author} {\bibfnamefont
  {F.}~\bibnamefont {Reinert}},\ }\bibfield  {title} {\textit {\bibinfo {title}
  {{Coherent Heavy Quasiparticles in a $\mathrm{Ce}{\mathrm{Pt}}_{5}$ Surface
  Alloy}},\ }}\href {\doibase 10.1103/PhysRevLett.106.186407} {\bibfield
  {journal} {\bibinfo  {journal} {Phys. Rev. Lett.}\ }\textbf {\bibinfo
  {volume} {106}},\ \bibinfo {pages} {186407} (\bibinfo {year}
  {2011})}\BibitemShut {NoStop}%
\bibitem [{\citenamefont {Mo}\ \emph {et~al.}(2012)\citenamefont {Mo},
  \citenamefont {Lee}, \citenamefont {Schmitt}, \citenamefont {Chen},
  \citenamefont {Lu}, \citenamefont {Capan}, \citenamefont {Kim}, \citenamefont
  {Fisk}, \citenamefont {Zhang}, \citenamefont {Hussain},\ and\ \citenamefont
  {Shen}}]{Shen12}%
  \BibitemOpen
  \bibfield  {author} {\bibinfo {author} {\bibfnamefont {S.-K.}\ \bibnamefont
  {Mo}}, \bibinfo {author} {\bibfnamefont {W.~S.}\ \bibnamefont {Lee}},
  \bibinfo {author} {\bibfnamefont {F.}~\bibnamefont {Schmitt}}, \bibinfo
  {author} {\bibfnamefont {Y.~L.}\ \bibnamefont {Chen}}, \bibinfo {author}
  {\bibfnamefont {D.~H.}\ \bibnamefont {Lu}}, \bibinfo {author} {\bibfnamefont
  {C.}~\bibnamefont {Capan}}, \bibinfo {author} {\bibfnamefont {D.~J.}\
  \bibnamefont {Kim}}, \bibinfo {author} {\bibfnamefont {Z.}~\bibnamefont
  {Fisk}}, \bibinfo {author} {\bibfnamefont {C.-Q.}\ \bibnamefont {Zhang}},
  \bibinfo {author} {\bibfnamefont {Z.}~\bibnamefont {Hussain}}, \ and\
  \bibinfo {author} {\bibfnamefont {Z.-X.}\ \bibnamefont {Shen}},\ }\bibfield
  {title} {\textit {\bibinfo {title} {{Emerging coherence with unified energy,
  temperature, and lifetime scale in heavy fermion
  $\mathrm{YbRh}_{2}\mathrm{Si}_2$}},\ }}\href {\doibase
  10.1103/PhysRevB.85.241103} {\bibfield  {journal} {\bibinfo  {journal} {Phys.
  Rev. B}\ }\textbf {\bibinfo {volume} {85}},\ \bibinfo {pages} {241103}
  (\bibinfo {year} {2012})}\BibitemShut {NoStop}%
\bibitem [{\citenamefont {Logan}\ and\ \citenamefont
  {Vidhyadhiraja}(2005)}]{Logan05}%
  \BibitemOpen
  \bibfield  {author} {\bibinfo {author} {\bibfnamefont {D.~E.}\ \bibnamefont
  {Logan}}\ and\ \bibinfo {author} {\bibfnamefont {N.~S.}\ \bibnamefont
  {Vidhyadhiraja}},\ }\bibfield  {title} {\textit {\bibinfo {title} {{Dynamics
  and transport properties of heavy fermions: theory}},\ }}\href {\doibase
  10.1088/0953-8984/17/19/009} {\bibfield  {journal} {\bibinfo  {journal} {J.
  Phys.: Condens. Matter}\ }\textbf {\bibinfo {volume} {17}},\ \bibinfo {pages}
  {2935} (\bibinfo {year} {2005})}\BibitemShut {NoStop}%
\bibitem [{\citenamefont {Grenzebach}\ \emph {et~al.}(2006)\citenamefont
  {Grenzebach}, \citenamefont {Anders}, \citenamefont {Czycholl},\ and\
  \citenamefont {Pruschke}}]{Pruschke06}%
  \BibitemOpen
  \bibfield  {author} {\bibinfo {author} {\bibfnamefont {C.}~\bibnamefont
  {Grenzebach}}, \bibinfo {author} {\bibfnamefont {F.~B.}\ \bibnamefont
  {Anders}}, \bibinfo {author} {\bibfnamefont {G.}~\bibnamefont {Czycholl}}, \
  and\ \bibinfo {author} {\bibfnamefont {T.}~\bibnamefont {Pruschke}},\
  }\bibfield  {title} {\textit {\bibinfo {title} {{Transport properties of
  heavy-fermion systems}},\ }}\href {\doibase 10.1103/PhysRevB.74.195119}
  {\bibfield  {journal} {\bibinfo  {journal} {Phys. Rev. B}\ }\textbf {\bibinfo
  {volume} {74}},\ \bibinfo {pages} {195119} (\bibinfo {year}
  {2006})}\BibitemShut {NoStop}%
\bibitem [{\citenamefont {Beach}\ \emph {et~al.}(2004)\citenamefont {Beach},
  \citenamefont {Lee},\ and\ \citenamefont {Monthoux}}]{Beach04}%
  \BibitemOpen
  \bibfield  {author} {\bibinfo {author} {\bibfnamefont {K.~S.~D.}\
  \bibnamefont {Beach}}, \bibinfo {author} {\bibfnamefont {P.~A.}\ \bibnamefont
  {Lee}}, \ and\ \bibinfo {author} {\bibfnamefont {P.}~\bibnamefont
  {Monthoux}},\ }\bibfield  {title} {\textit {\bibinfo {title} {{Field-Induced
  Antiferromagnetism in the Kondo Insulator}},\ }}\href {\doibase
  10.1103/PhysRevLett.92.026401} {\bibfield  {journal} {\bibinfo  {journal}
  {Phys. Rev. Lett.}\ }\textbf {\bibinfo {volume} {92}},\ \bibinfo {pages}
  {026401} (\bibinfo {year} {2004})}\BibitemShut {NoStop}%
\bibitem [{\citenamefont {{Milat, I.}}\ \emph {et~al.}(2004)\citenamefont
  {{Milat, I.}}, \citenamefont {{Assaad, F.}},\ and\ \citenamefont {{Sigrist,
  M.}}}]{Milat04}%
  \BibitemOpen
  \bibfield  {author} {\bibinfo {author} {\bibnamefont {{Milat, I.}}}, \bibinfo
  {author} {\bibnamefont {{Assaad, F.}}}, \ and\ \bibinfo {author}
  {\bibnamefont {{Sigrist, M.}}},\ }\bibfield  {title} {\textit {\bibinfo
  {title} {{Field induced magnetic ordering transition in Kondo insulators}},\
  }}\href {\doibase 10.1140/epjb/e2004-00154-5} {\bibfield  {journal} {\bibinfo
   {journal} {Eur. Phys. J. B}\ }\textbf {\bibinfo {volume} {38}},\ \bibinfo
  {pages} {571} (\bibinfo {year} {2004})}\BibitemShut {NoStop}%
\bibitem [{\citenamefont {Ohashi}\ \emph {et~al.}(2004)\citenamefont {Ohashi},
  \citenamefont {Koga}, \citenamefont {Suga},\ and\ \citenamefont
  {Kawakami}}]{Kawakami04}%
  \BibitemOpen
  \bibfield  {author} {\bibinfo {author} {\bibfnamefont {T.}~\bibnamefont
  {Ohashi}}, \bibinfo {author} {\bibfnamefont {A.}~\bibnamefont {Koga}},
  \bibinfo {author} {\bibfnamefont {S.-i.}\ \bibnamefont {Suga}}, \ and\
  \bibinfo {author} {\bibfnamefont {N.}~\bibnamefont {Kawakami}},\ }\bibfield
  {title} {\textit {\bibinfo {title} {{Field-induced phase transitions in a
  Kondo insulator}},\ }}\href {\doibase 10.1103/PhysRevB.70.245104} {\bibfield
  {journal} {\bibinfo  {journal} {Phys. Rev. B}\ }\textbf {\bibinfo {volume}
  {70}},\ \bibinfo {pages} {245104} (\bibinfo {year} {2004})}\BibitemShut
  {NoStop}%
\bibitem [{\citenamefont {{Costa}}\ \emph {et~al.}(2018)\citenamefont
  {{Costa}}, \citenamefont {{Mendes-Santos}}, \citenamefont {{Paiva}},
  \citenamefont {{Curro}}, \citenamefont {{dos Santos}},\ and\ \citenamefont
  {{Scalettar}}}]{Costa18cm}%
  \BibitemOpen
  \bibfield  {author} {\bibinfo {author} {\bibfnamefont {N.~C.}\ \bibnamefont
  {{Costa}}}, \bibinfo {author} {\bibfnamefont {T.}~\bibnamefont
  {{Mendes-Santos}}}, \bibinfo {author} {\bibfnamefont {T.}~\bibnamefont
  {{Paiva}}}, \bibinfo {author} {\bibfnamefont {N.~J.}\ \bibnamefont
  {{Curro}}}, \bibinfo {author} {\bibfnamefont {R.~R.}\ \bibnamefont {{dos
  Santos}}}, \ and\ \bibinfo {author} {\bibfnamefont {R.~T.}\ \bibnamefont
  {{Scalettar}}},\ }\bibfield  {title} {\textit {\bibinfo {title} {{{The
  Coherence Temperature in the Diluted Periodic Anderson Model}}},\
  }}\href@noop {} {\bibfield  {journal} {\bibinfo  {journal} {arXiv e-prints}\
  } (\bibinfo {year} {2018})},\ \Eprint {http://arxiv.org/abs/1812.09426}
  {arXiv:1812.09426} \BibitemShut {NoStop}%
\bibitem [{\citenamefont {{Lawson}}\ \emph {et~al.}(2019)\citenamefont
  {{Lawson}}, \citenamefont {{Bush}}, \citenamefont {{Shockley}}, \citenamefont
  {{Capan}}, \citenamefont {{Fisk}},\ and\ \citenamefont {{Curro}}}]{Lawson19}%
  \BibitemOpen
  \bibfield  {author} {\bibinfo {author} {\bibfnamefont {M.}~\bibnamefont
  {{Lawson}}}, \bibinfo {author} {\bibfnamefont {B.~T.}\ \bibnamefont
  {{Bush}}}, \bibinfo {author} {\bibfnamefont {A.~C.}\ \bibnamefont
  {{Shockley}}}, \bibinfo {author} {\bibfnamefont {C.}~\bibnamefont {{Capan}}},
  \bibinfo {author} {\bibfnamefont {Z.}~\bibnamefont {{Fisk}}}, \ and\ \bibinfo
  {author} {\bibfnamefont {N.~J.}\ \bibnamefont {{Curro}}},\ }\bibfield
  {title} {\textit {\bibinfo {title} {{Site Specific Knight Shift Measurements
  of the Dilute Kondo lattice System
  $\mathrm{Ce}_{1-x}\mathrm{La}_x\mathrm{CoIn}_5$}},\ }}\href@noop {}
  {\bibfield  {journal} {\bibinfo  {journal} {arXiv e-prints}\ } (\bibinfo
  {year} {2019})},\ \Eprint {http://arxiv.org/abs/1901.01333}
  {arXiv:1901.01333} \BibitemShut {NoStop}%
\bibitem [{\citenamefont {Karimi}\ and\ \citenamefont
  {Affleck}(2012)}]{Affleck12}%
  \BibitemOpen
  \bibfield  {author} {\bibinfo {author} {\bibfnamefont {H.}~\bibnamefont
  {Karimi}}\ and\ \bibinfo {author} {\bibfnamefont {I.}~\bibnamefont
  {Affleck}},\ }\bibfield  {title} {\textit {\bibinfo {title} {{Towards a
  rigorous proof of magnetism on the edges of graphene nanoribbons}},\ }}\href
  {\doibase 10.1103/PhysRevB.86.115446} {\bibfield  {journal} {\bibinfo
  {journal} {Phys. Rev. B}\ }\textbf {\bibinfo {volume} {86}},\ \bibinfo
  {pages} {115446} (\bibinfo {year} {2012})}\BibitemShut {NoStop}%
\bibitem [{\citenamefont {Assaad}(2002)}]{Assaad02}%
  \BibitemOpen
  \bibfield  {author} {\bibinfo {author} {\bibfnamefont {F.~F.}\ \bibnamefont
  {Assaad}},\ }\bibfield  {title} {\textit {\bibinfo {title} {{Depleted Kondo
  lattices: Quantum Monte Carlo and mean-field calculations}},\ }}\href
  {\doibase 10.1103/PhysRevB.65.115104} {\bibfield  {journal} {\bibinfo
  {journal} {Phys. Rev. B}\ }\textbf {\bibinfo {volume} {65}},\ \bibinfo
  {pages} {115104} (\bibinfo {year} {2002})}\BibitemShut {NoStop}%
\bibitem [{\citenamefont {Costa}\ \emph {et~al.}(2018)\citenamefont {Costa},
  \citenamefont {Ara\'ujo}, \citenamefont {Lima}, \citenamefont {Paiva},
  \citenamefont {dos Santos},\ and\ \citenamefont {Scalettar}}]{Costa18}%
  \BibitemOpen
  \bibfield  {author} {\bibinfo {author} {\bibfnamefont {N.~C.}\ \bibnamefont
  {Costa}}, \bibinfo {author} {\bibfnamefont {M.~V.}\ \bibnamefont {Ara\'ujo}},
  \bibinfo {author} {\bibfnamefont {J.~P.}\ \bibnamefont {Lima}}, \bibinfo
  {author} {\bibfnamefont {T.}~\bibnamefont {Paiva}}, \bibinfo {author}
  {\bibfnamefont {R.~R.}\ \bibnamefont {dos Santos}}, \ and\ \bibinfo {author}
  {\bibfnamefont {R.~T.}\ \bibnamefont {Scalettar}},\ }\bibfield  {title}
  {\textit {\bibinfo {title} {{Compressible ferrimagnetism in the depleted
  periodic Anderson model}},\ }}\href {\doibase 10.1103/PhysRevB.97.085123}
  {\bibfield  {journal} {\bibinfo  {journal} {Phys. Rev. B}\ }\textbf {\bibinfo
  {volume} {97}},\ \bibinfo {pages} {085123} (\bibinfo {year}
  {2018})}\BibitemShut {NoStop}%
\bibitem [{\citenamefont {{J\"{u}lich Supercomputing Centre}}(2016)}]{jureca}%
  \BibitemOpen
  \bibfield  {author} {\bibinfo {author} {\bibnamefont {{J\"{u}lich
  Supercomputing Centre}}},\ }\bibfield  {title} {\textit {\bibinfo {title}
  {{JURECA: General-purpose supercomputer at J\"{u}lich Supercomputing
  Centre}},\ }}\href {\doibase 10.17815/jlsrf-2-121} {\bibfield  {journal}
  {\bibinfo  {journal} {{J. Large-Scale Res. Facil.}}\ }\textbf {\bibinfo
  {volume} {2}},\ \bibinfo {pages} {A62} (\bibinfo {year} {2016})}\BibitemShut
  {NoStop}%
\bibitem [{\citenamefont {{Figgins}}\ \emph {et~al.}(2019)\citenamefont
  {{Figgins}}, \citenamefont {{Mattos}}, \citenamefont {{Mar}}, \citenamefont
  {{Chen}}, \citenamefont {{Manoharan}},\ and\ \citenamefont
  {{Morr}}}]{Morr19}%
  \BibitemOpen
  \bibfield  {author} {\bibinfo {author} {\bibfnamefont {J.}~\bibnamefont
  {{Figgins}}}, \bibinfo {author} {\bibfnamefont {L.~S.}\ \bibnamefont
  {{Mattos}}}, \bibinfo {author} {\bibfnamefont {W.}~\bibnamefont {{Mar}}},
  \bibinfo {author} {\bibfnamefont {Y.-T.}\ \bibnamefont {{Chen}}}, \bibinfo
  {author} {\bibfnamefont {H.~C.}\ \bibnamefont {{Manoharan}}}, \ and\ \bibinfo
  {author} {\bibfnamefont {D.~K.}\ \bibnamefont {{Morr}}},\ }\bibfield  {title}
  {\textit {\bibinfo {title} {{Quantum Engineered Kondo Lattices}},\
  }}\href@noop {} {\bibfield  {journal} {\bibinfo  {journal} {arXiv e-prints}\
  } (\bibinfo {year} {2019})},\ \Eprint {http://arxiv.org/abs/1902.04680}
  {arXiv:1902.04680} \BibitemShut {NoStop}%
\bibitem [{\citenamefont {Costi}(2000)}]{Costi00}%
  \BibitemOpen
  \bibfield  {author} {\bibinfo {author} {\bibfnamefont {T.~A.}\ \bibnamefont
  {Costi}},\ }\bibfield  {title} {\textit {\bibinfo {title} {{Kondo Effect in a
  Magnetic Field and the Magnetoresistivity of Kondo Alloys}},\ }}\href
  {\doibase 10.1103/PhysRevLett.85.1504} {\bibfield  {journal} {\bibinfo
  {journal} {Phys. Rev. Lett.}\ }\textbf {\bibinfo {volume} {85}},\ \bibinfo
  {pages} {1504} (\bibinfo {year} {2000})}\BibitemShut {NoStop}%
\end{thebibliography}

%


\clearpage 
\setcounter{figure}{0}
\renewcommand{\thefigure}{\arabic{figure}S}

\makeatletter
     \@addtoreset{figure}{section}
\makeatother

\onecolumngrid
\begin{center}
\Large{\bf Supplemental Material for: Emergent Coherent Lattice Behavior in Kondo Nanosystems }
\end{center}
\twocolumngrid

\subsection{Schrieffer-Wolff transformation and electron  cotunneling} 

For simplicity we  consider  a symmetric single-impurity Anderson model (SIAM) defined as: 
\begin{align}
 H_{\rm SIAM} &=  \underbrace{\sum_{\pmb{k},\sigma}   
\varepsilon(\pmb{k}) c^{\dagger}_{\pmb{k},\sigma}  c^{\phantom\dagger}_{\pmb{k},\sigma}        +
  U \left( f^{\dagger}_{\uparrow} f^{\phantom\dagger}_{\uparrow} -  1/2 \right)  
\left(  f^{\dagger}_{\downarrow} f^{\phantom\dagger}_{\downarrow} -  1/2   \right)}_{H_0}  \nonumber \\ 
&+  \underbrace{ \frac{V}{\sqrt{N}} \sum_{\pmb{k},\sigma} 
\left(  c^{\dagger}_{\pmb{k},\sigma} f^{\phantom\dagger}_{\sigma}  +   f^{\dagger}_{\sigma}c^{\phantom\dagger}_{\pmb{k},\sigma}\right) }_{H_1}.
\end{align}
The Abrikosov-Suhl resonance of the SIAM  emerges in the single-particle spectral function of the $f$-electrons: 
\begin{equation}
        G_f(\tau) \equiv \sum_{\sigma}\langle f_{\sigma}(\tau)  f^{\dagger}_{\sigma} (0) \rangle    = \int  d \omega   
        K(\tau,\omega) \text{ Im } G^{f}_{ret}(\omega),  
\end{equation}
where, 
\begin{align}
        \text{Im} G^{f}_{ret}(\omega)   &=  \frac{\pi}{Z}  \sum_{n,m,\sigma}  \left( e^{-\beta E_m }  + e^{-\beta E_n }  \right) \nonumber \\
                                        &\times | \langle n | f^{\dagger}_{\sigma} | m\rangle |^2 \delta \left(  \omega + E_n - E_m \right),
\end{align}
and  the Kernel is given by: 
\begin{equation}
        K(\tau, \omega) =  \frac{1}{\pi }\frac{e^{ \tau \omega} }{1 + e^{\beta \omega}}.
\end{equation}
We now carry out the Schrieffer-Wolff canonical transformation  required  to eliminate the hybridization term in first order and 
to  derive the Kondo model.
Let 
\begin{equation}
    S^{\dagger}   = - S,    \text{ and  }   \left[ S, H_0 \right]  = - H_1,   
\end{equation}
then 
\begin{equation}
          e^{S} H e^{-S}   = H_0  +   \frac{1}{2} \left[ S, H_1 \right]   + { \cal  O } \left(  \epsilon^3   \right). 
\end{equation}
Here, we have formally assumed that  $H_1$ and the generator $S$ are of order $ \epsilon $. Under this canonical transformation, 
the  imaginary-time-displaced Green's function: 
\begin{equation}
G_{\tilde{f}}(\tau) \equiv \sum_{\sigma}
  \langle \underbrace{e^{S}f_{\sigma} e^{-S}}_{\tilde{f}_{\sigma}} (\tau)   
          \underbrace{e^{S}f^{\dagger}_{\sigma}e^{-S} }_{\tilde{f}^{\dagger}_{\sigma}} (\tau=0) \rangle,
\end{equation}
should reproduce the Abrikosov-Suhl resonance of the SIAM  in the Kondo model with a single impurity. The  calculation gives: 
\begin{align}
        S  &=   \frac{V}{ \sqrt{N} } \sum_{\pmb{k},\sigma} \nonumber \\
           &\left(  \frac{ c^{\dagger}_{\pmb{k},\sigma} f^{}_\sigma P_o} {\varepsilon(\pmb{k}) + U/2 - i 0^+ }   
           +   \frac{ c^{\dagger}_{\pmb{k},\sigma} f^{}_\sigma P_e } {\varepsilon(\pmb{k}) - U/2 - i 0^+ }     -  H.c.\right), 
\end{align}
where  $P_e$ ($P_o$)  is the projection of the even, $(-1)^{\sum_\sigma f^{\dagger}_\sigma f^{}_\sigma} = 1$ 
(odd, $(-1)^{\sum_\sigma f^{\dagger}_\sigma f^{}_\sigma} = -1$), respectively,  parity sector of the $f$-electron.  
We can now transform the creation operator for $f$-electrons,  and constrain the Hilbert space to $P_o$ so as to obtain: 
\begin{align}
        \tilde{f}^{\dagger}_{\sigma'}   &\equiv e^{S} f^{\dagger}_{\sigma'} e^{-S} \nonumber \\  
                                        &\simeq 
        \frac{V}{\sqrt{N}}  \sum_{\pmb{k},\sigma} \left(  \frac{ c^{\dagger}_{\pmb{k},\sigma} f^{}_\sigma f^{\dagger}_{\sigma'}} 
       {\varepsilon(\pmb{k}) - U/2 - i 0^+ }   -   
        \frac{   f^{\dagger}_{\sigma'}c^{\dagger}_{\pmb{k},\sigma} f^{}_\sigma } {\varepsilon(\pmb{k}) + U/2 - i 0^+ }  \right). 
\end{align}
The first (second) term involves a  virtual doubly (empty) occupied $f$-site. Assuming that  $U$ is the largest scale so that we can set 
$\varepsilon(\pmb{k})$ to zero gives the final result: 
\begin{equation}
 \tilde{f}^{\dagger}_{\sigma'}     \simeq   -\frac{V}{U} \sum_{\sigma}  \left(   
         c^{\dagger}_{\pmb{r}=\pmb{0},\sigma} f^{}_\sigma f^{\dagger}_{\sigma'}   +    
f^{\dagger}_{\sigma'}c^{\dagger}_{\pmb{r}=\pmb{0},\sigma} f^{}_\sigma  \right).
\end{equation}
Using the relation, $2f^{\dagger}_{\sigma'}  f^{}_{\sigma'}  = n_f + \sigma' 2S^{z}_f$, 
where $\sigma'$ takes the value $1$ ($-1$) for up  (down) spin degrees of freedom, 
$ S^{z}_f  = \frac{1}{2} \sum_{\sigma'}  \sigma' f^{\dagger}_{\sigma'} f^{}_{\sigma'}$, and $n_f=\sum_{\sigma'}  f^{\dagger}_{\sigma'}  f^{}_{\sigma'}$,  
the above operator can be rewritten as:   
\begin{equation}
         \tilde{f}^{\dagger}_{\sigma'}     \simeq  \frac{2V}{U}  
     \left( c^{\dagger}_{\pmb{r}=\pmb{0},-\sigma'}  S^{\sigma'}_{f} +  \sigma'   c^{\dagger}_{\pmb{r}=\pmb{0},\sigma'} S^{z}_f   \right).
\end{equation}
Here $ S^{\sigma'}_{f}  =  f^{\dagger}_{\sigma'} f^{}_{-\sigma'} $ and we have used the fact that in the Kondo regime $n_f  = 1$. 
The corresponding impurity spectral function for the Kondo model matches that derived in Ref.~\cite{Costi00} using the equation of motion 
for the $c$-electron Green's function.

\subsection{Supplemental data}

\begin{figure}[h!]
\centering
\begin{minipage}{.475\textwidth}
\centering
\includegraphics[width=0.32\textwidth]{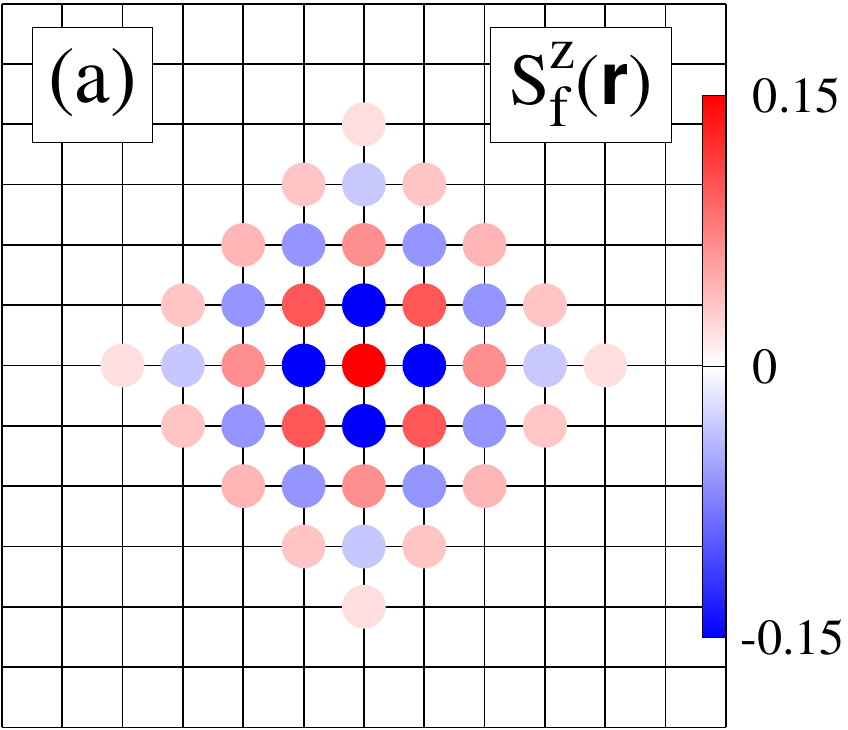}
\includegraphics[width=0.32\textwidth]{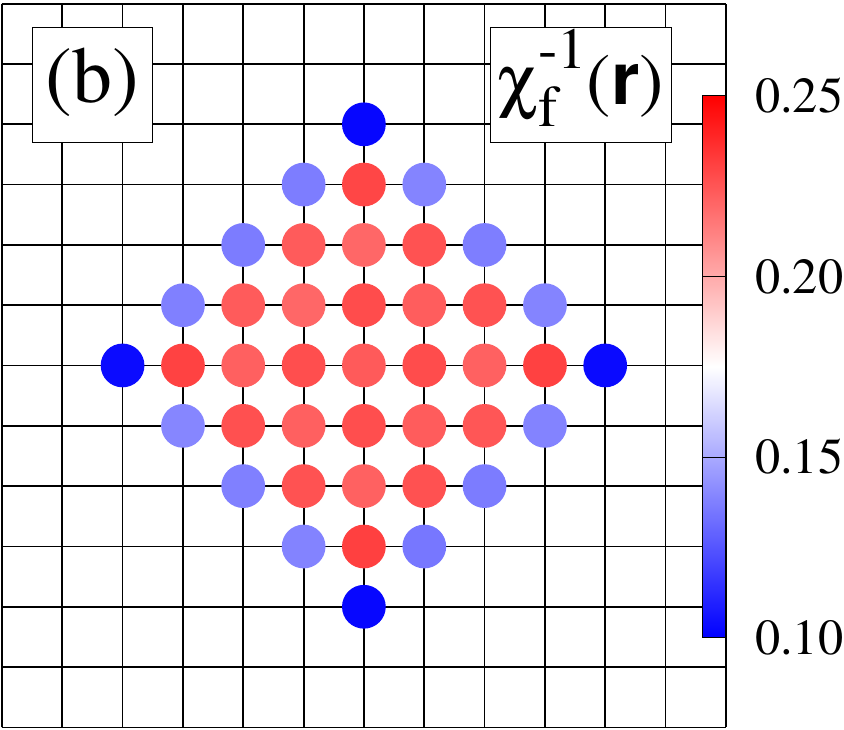}
\includegraphics[width=0.32\textwidth]{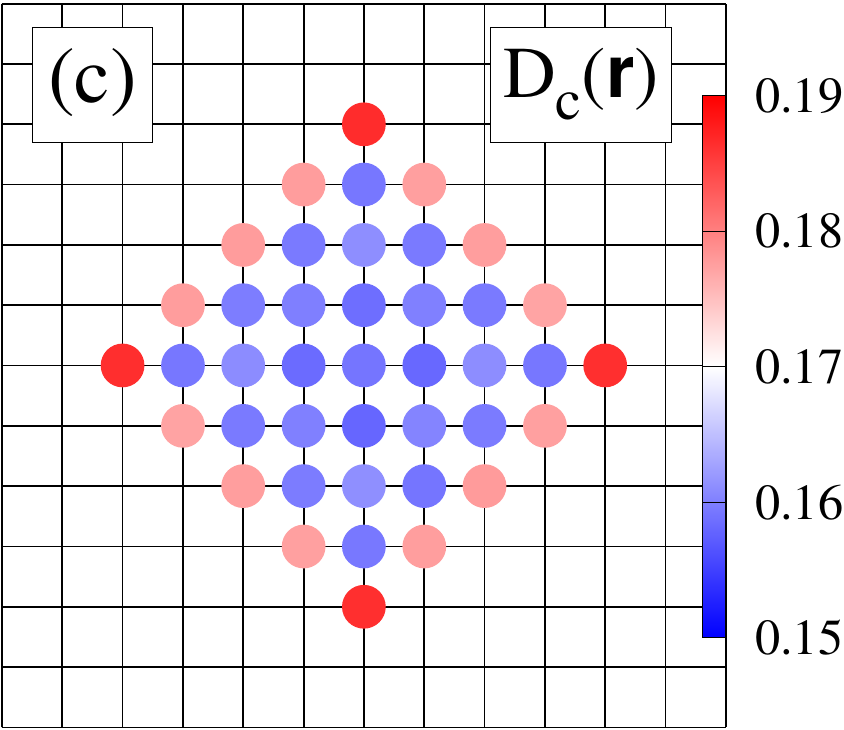}\\
\vspace*{0.1cm}
\includegraphics[width=0.32\textwidth]{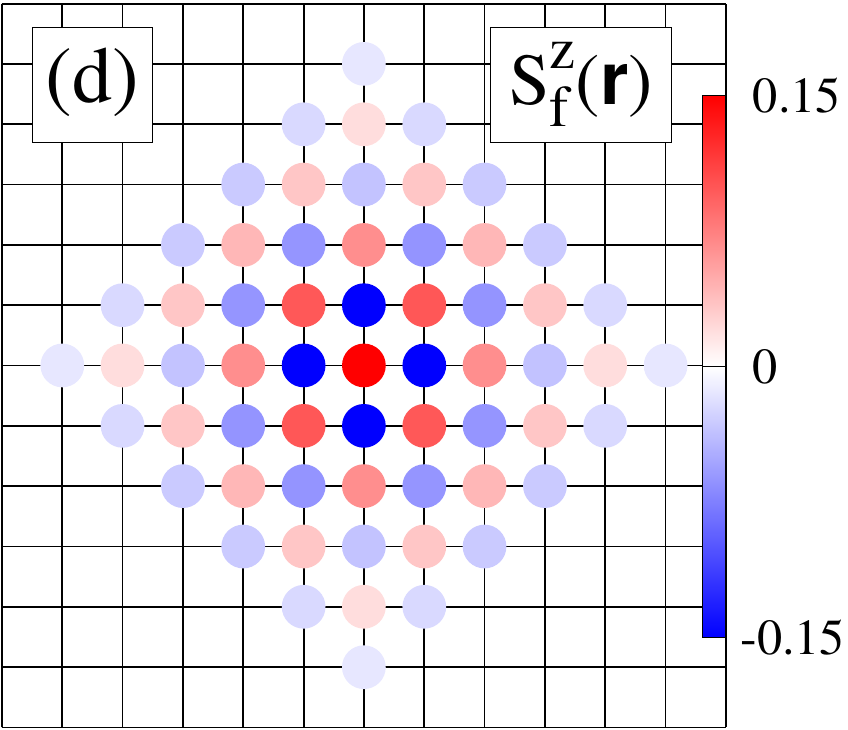}
\includegraphics[width=0.32\textwidth]{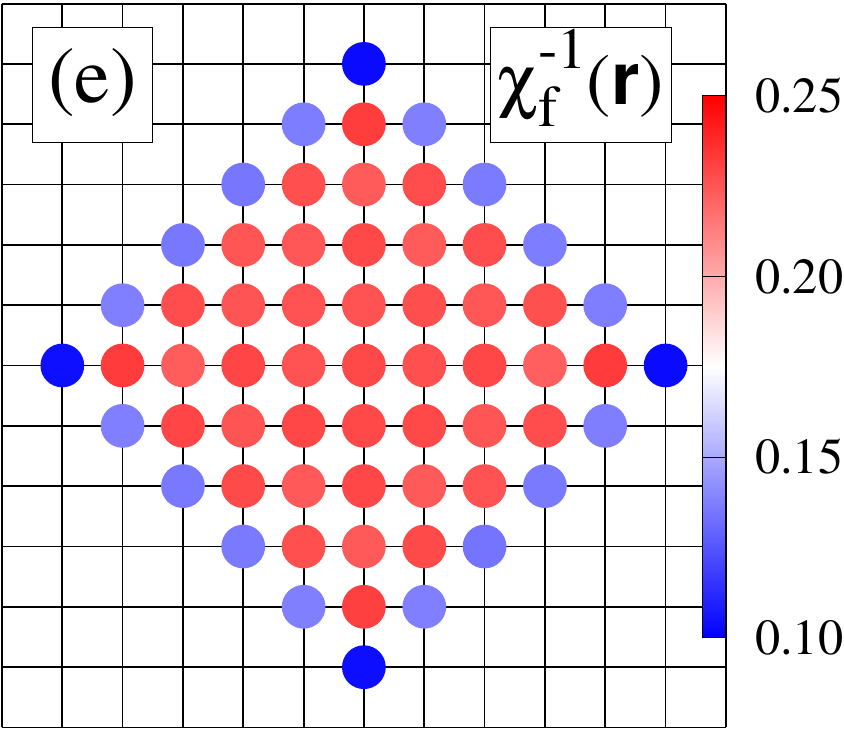}
\includegraphics[width=0.32\textwidth]{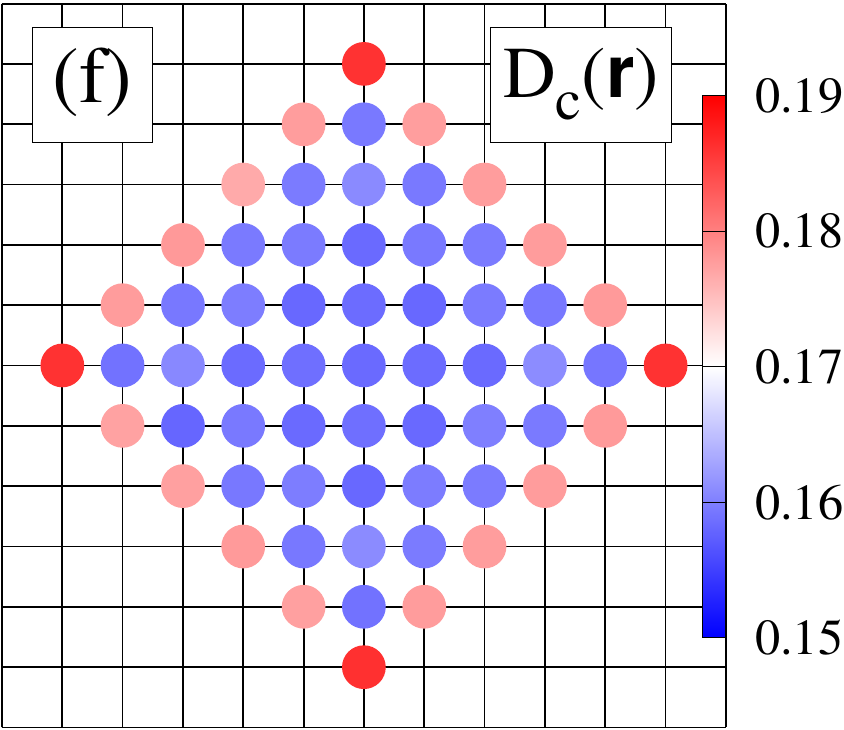}
\caption
{
Real-space spin correlations $S_f^{z}(\pmb{r})$  relative to the central impurity (left)
and spatial dependence of: inverse $f$-spin susceptibility $\chi_f^{-1}(\pmb{r})$ (middle)
and $c$-electron double occupancy $D_c(\pmb{r})$ (right) in the $n=4$ (a-c) and $n=5$ (d-f) systems.
Parameters: $J/t=1.6$, $T=t/30$, and $L=16$.
}
\label{Sz2}
\end{minipage}
\hfill 
\begin{minipage}{.475\textwidth}
\centering
\includegraphics[width=0.95\textwidth]{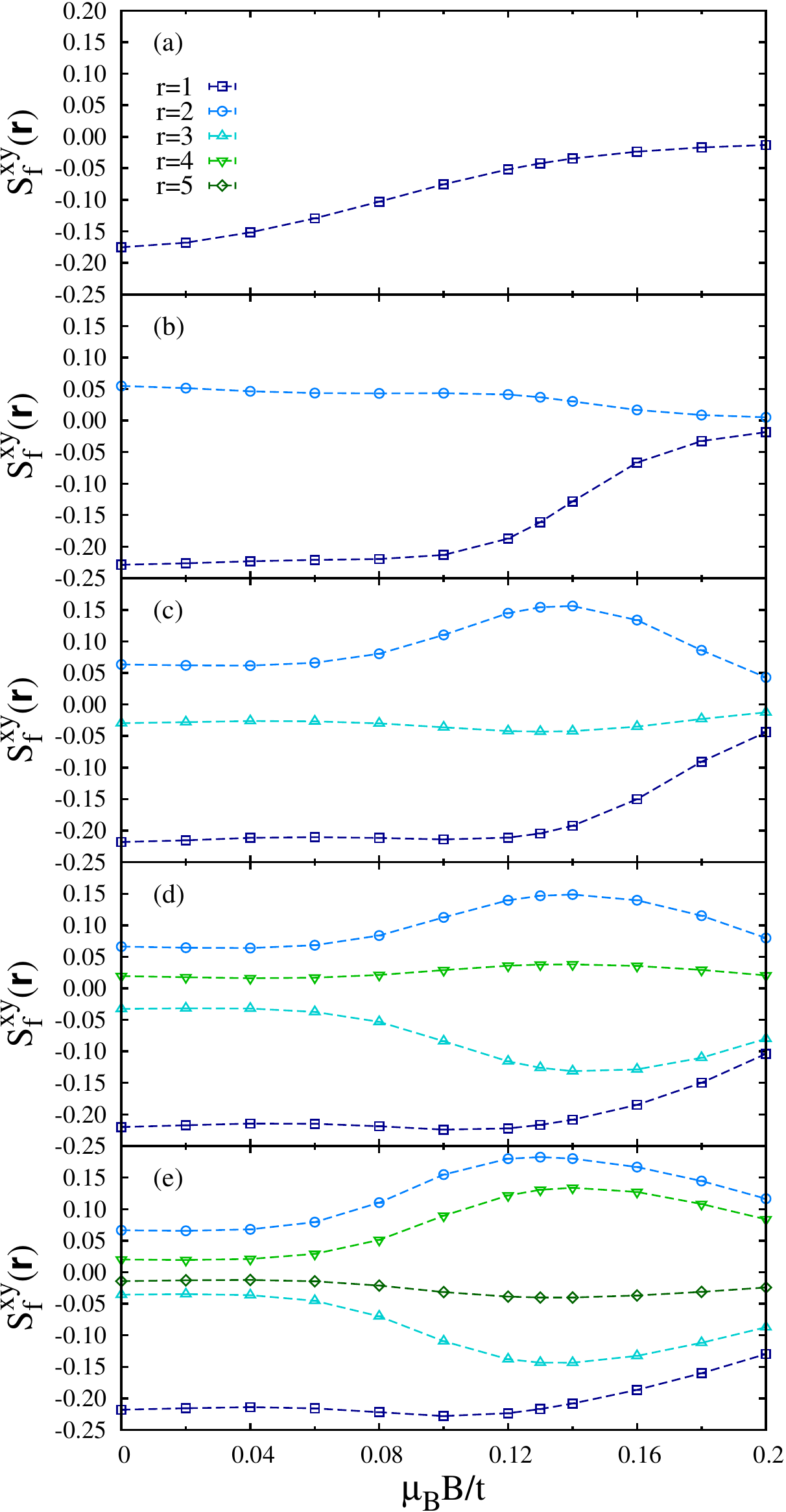}
\caption
{
Transverse-spin correlations $S_f^{xy}(\pmb{r})$  between the $(i,0)$ site with $i=1,\ldots,n$ and the central (0,0) impurity 
as a function of ouf-of-plane magnetic field $B$ for the system with $n=1$ (a), $n=2$ (b), $n=3$ (c), $n=4$ (d), and $n=5$ (e) 
shells. Parameters as in Fig.~\ref{Sz2}.
}
\label{spatial_B}
\end{minipage}
\end{figure}

\begin{figure}[t!]
\begin{center}
\includegraphics[width=0.45\textwidth]{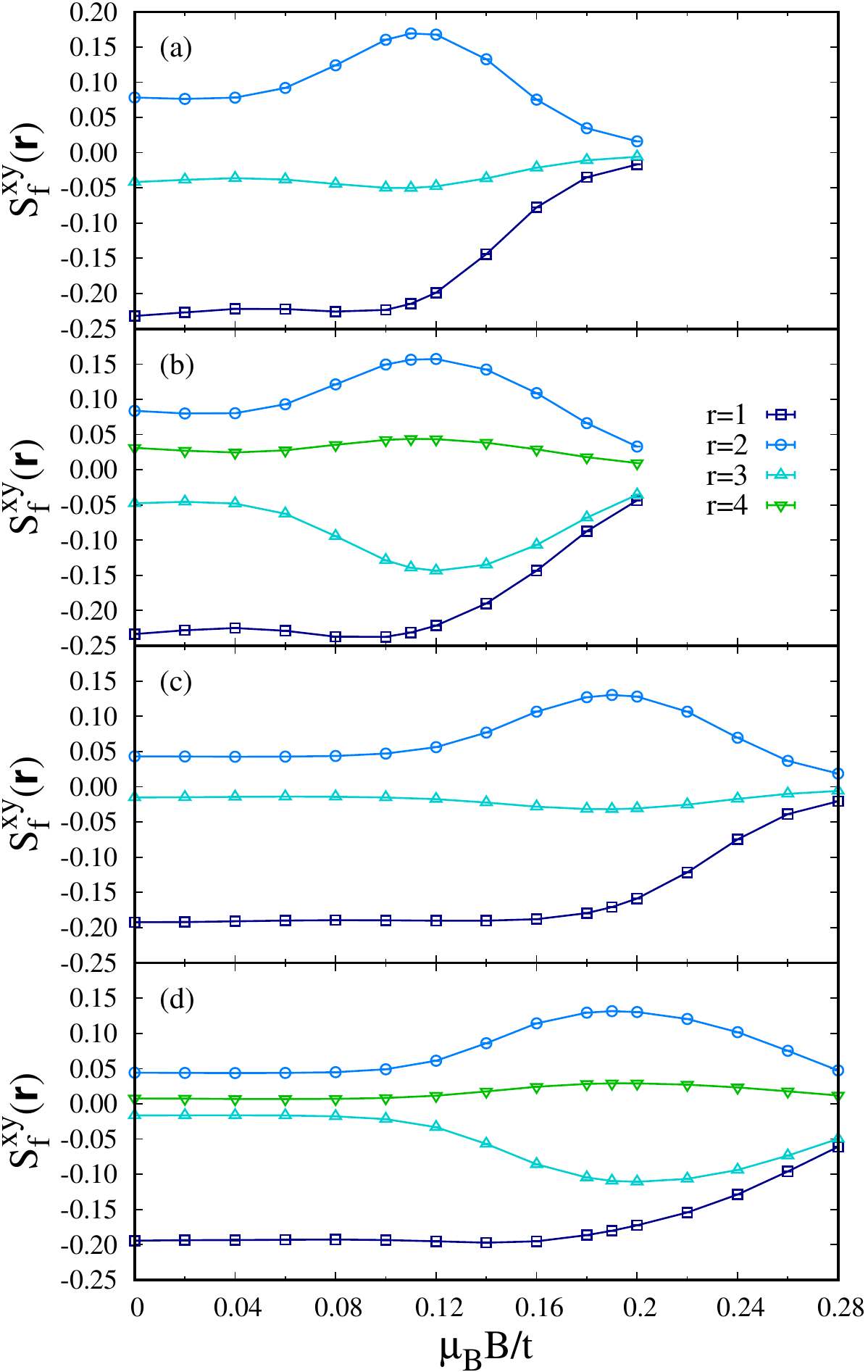}
\end{center}
\caption
{
Same as in Fig.~\ref{spatial_B} but for $J/t=1.5$ with $n=3$ (a) and $n=4$ (b) and for $J/t=1.8$ with 
$n=3$ (c) and $n=4$ (d).
}
\label{spatial_B_ext}
\end{figure}

\begin{table}[h!]
\caption{
Total number of magnetic impurities $N_{\textrm{imp}}$ in a nanosystem with $n$ shells.
}
\begin{ruledtabular}
\begin{tabular}{lcccccc}
 $n$                  &     0 &     1 &    2  &  3  &  4  &  5  \cr
 $N_{\textrm{imp}}$   &     1 &     5 &    13 &  25 &  41 &  61 \cr
\end{tabular}
\end{ruledtabular}
\label{table}
\end{table}
The relation between the number of shells $n$ and the corresponding total number of impurities $N_{\textrm{imp}}$ in
a superlattice is summarized in Table~\ref{table}. Figure~\ref{Sz2} shows the spatial characteristics of the $n\in\{4,5\}$ superlattices.
The enhancement  of transverse-spin correlations $S^{xy}_f(\pmb{r})$ for $n\ge 3$ upon increasing the external magnetic field $B$
at $J/t=1.6$ is documented more quantitatively in Fig.~\ref{spatial_B}.
Same critical cluster size required to observe the enhancement of $S^{xy}_f(\pmb{r})$ is found at smaller $J/t=1.5$ in the close proximity
to the quantum critical point [Figs.~\ref{spatial_B_ext}(a,b)]  and at $J/t=1.8$ deep in the Kondo insulating phase
[Figs.~\ref{spatial_B_ext}(c,d)].

\onecolumngrid

\end{document}